\documentclass[%
 reprint,
 superscriptaddress,
 showpacs,
 amsmath,amssymb,
 aps,
 pra,
 11pt,
 onecolumn,
 showkeys
]{revtex4-1}
\usepackage[position=t,singlelinecheck=off,
caption=false]{subfig}
\usepackage{xcolor}
\usepackage{pdfcomment}
\usepackage{mathrsfs}
\usepackage{graphicx}
\usepackage{dcolumn}
\usepackage{bm}
\usepackage{hyperref}
\usepackage{algorithm,algpseudocode}
\usepackage{soul}

\usepackage{amsmath,amsfonts,amssymb,color}
\usepackage{amsthm}
\usepackage{leftidx}
\usepackage{graphicx}
\usepackage{xcolor}
\usepackage{dcolumn}
\usepackage{bm}
\usepackage{epstopdf}
\usepackage{epsfig}
\setstcolor{red}

\usepackage[a4paper,top=1.8cm,bottom=1.8cm,left=1.5cm,right=1.5cm]{geometry}

\newcommand{\Hc}{\mathrm{h.c.}}
\newcommand{\cc}{\mathrm{c.c.}}
\newcommand{\ee}{\mathrm{e}}
\newcommand{\ii}{\mathrm{i}}
\newcommand{\hatx}{\bm{\hat{\mathrm{x}}}}
\newcommand{\haty}{\bm{\hat{\mathrm{y}}}}
\newcommand{\TrSite}{\mathrm{Tr}_{\Lambda}}
\newcommand{\fermi}{f}
\newcommand{\nH}{n_{\mathrm{H}}}
\newcommand{\nF}{n_{\mathrm{F}}}
\newcommand{\period}{\mathscr{T}}

\begin{document}

\preprint{APS/123-QED}

\title{A computational study of two-terminal transport of Floquet quantum Hall insulators}

\author{Han Hoe \surname{Yap}}
\affiliation{NUS Graduate School for Integrative Sciences and Engineering, Singapore 117456, Republic of Singapore}
\affiliation{Department of Physics, National University of Singapore, Singapore 117551, Republic of Singapore}
\author{Longwen \surname{Zhou}}
\affiliation{Department of Physics, National University of Singapore, Singapore 117551, Republic of Singapore}
\author{Jian-Sheng \surname{Wang}}
\affiliation{Department of Physics, National University of Singapore, Singapore 117551, Republic of Singapore}
\author{Jiangbin \surname{Gong}} \email{phygj@nus.edu.sg}
\affiliation{Department of Physics, National University of Singapore, Singapore 117551, Republic of Singapore}
\affiliation{NUS Graduate School for Integrative Sciences and Engineering, Singapore 117456, Republic of Singapore}

\date{\today}

\begin{abstract}
Periodic driving fields can induce topological phase transitions, resulting in Floquet topological phases with intriguing properties such as very large Chern numbers and unusual edge states. 
Whether such Floquet topological phases could generate robust edge state conductance much larger than their static counterparts is an interesting question. In this paper, working under the Keldysh formalism, we study two-lead transport via the edge states of irradiated quantum Hall insulators using the method of recursive Floquet-Green's functions. Focusing on a harmonically-driven Hofstadter model, we show that quantized Hall conductance as large as $8e^2/h$ can be realized, but only after applying the so-called Floquet sum rule. To assess the robustness of edge state transport, we analyze the DC conductance, time-averaged current profile and local density of states. It is found that co-propagating chiral edge modes are more robust against disorder and defects as compared with the remarkable counter-propagating edge modes, as well as certain symmetry-restricted Floquet edge modes. Furthermore, we go beyond the wide-band limit, which is often assumed for the leads, to study how the conductance quantization (after applying the Floquet sum rule) of Floquet edge states can be affected if the leads have finite bandwidths. These results may be useful for the design of transport devices based on Floquet topological matter. 

\end{abstract}

\maketitle


\section{\label{sec:level1}Introduction}
Under time-periodic modulations, the long-time behavior of a quantum system is governed by its Floquet states, which may carry topological features absent in its equilibrium predecessor. Originally anticipated in quantum chaos and nonlinear dynamics~\cite{Leboeuf1990}, Floquet topological phases are under intensive studies in recent years. Theoretically, a large class of topological Floquet states has been proposed, including Floquet topological insulators~\cite{Oka2009,Kitagawa2011,Lindner2011,GongPRL2012,Gomezleon2013,Rudner2013,Cayssol2013}, Floquet Anderson insulators~\cite{Titum2015,Titum2016}, Floquet Majorana Fermions~\cite{Jiang2011,Kundu2013,Gong2013,Klinovaja2016,Thakurathi2017}, Floquet topological semimetals~\cite{WangFWSM2014,Raditya2016,Zhou2016,Raditya2016b}, etc. New classification schemes for these states have also been introduced, which go beyond their static counterparts~\cite{Harper2016a,Harper2016b}. Experimentally, Floquet techniques have been implemented successfully to realize the long-awaited Haldane Chern insulator in cold-atom systems~\cite{Jotzu2014}. The flexibility of driving fields in manipulating topological surface states have also been demonstrated~\cite{Wang453}. Meanwhile, various photonic~\cite{Rechtsman2013,Hu2015,Maczewsky2017,Mukherjee2017}, phononic and acoustic~\cite{Xiao2015,Susstrunk2015,Susstrunk2017} analogs of Floquet topological phases with robust transport along their edges were observed, revealing a new strategy to harness the propagation of light and sound~\cite{Peano2015}.

One important reason why Floquet topological phases are attractive is that their topological invariants
can be very large \cite{GongPRL2012, Gong2013,Derek2014,Zhou2014,Xiong2016}. This hints that Floquet topological matter may be able to exploit a considerable number of edge state channels to realize robust and appreciable edge state transport.  However, very few studies have investigated this possibility from the perspective of two-terminal transport.  In conventional electronic devices, it is well known that in the quantum Hall regime, the Hall conductance is quantized and equal to $e^2/h$ times the number of chiral edge modes on the Fermi surface lying between Landau levels~\cite{Buttiker1988,Datta1995}. Consider now a sample subject to a periodic driving field. { Stroboscopically, the system is described by the Floquet operator and its eigenstates, with new features absent in the instantaneous eigenstates of the system Hamiltonian.}  Its transport properties may then dramatically change {, because the electrons can now absorb or emit photons from the driving field, resulting in a photon-dressed steady state which may explore conduction channels inaccessible in the static case. Therefore} studies of the transport of Floquet edge modes in a two-terminal device are necessary but not straightforward.  We also mention that
before the advent of Floquet topological phases, many efforts have been devoted to understanding driven transport at the nanoscale~\cite{Tien1963,Platero2004,Kohler2005}. Various theoretical approaches, such as nonequilibrium Green's functions~(NEGF)~\cite{Jauho1994,Arrachea2005,Tsuji2008}, master equations~\cite{Kohler2005,Bruder1994}, and Floquet scattering matrix~\cite{Moskalets2002,Kim2004,Arrachea2006} have been adopted to study this problem from different perspectives.

Regarding the potential applications of Floquet topological phases in transport devices, the key questions to be answered are then as follows: (i) Are Floquet topological phases indeed useful in achieving tunable, appreciable and robust conductance? (ii) How is the Floquet edge state conductance related to the symmetry and topological invariants of the system? (iii) Is the conductance of Floquet edge states quantized? If not, why and when?   Answers to these questions are of both theoretical and experimental interest.

We note that a couple of recent studies have focused on edge state transport in Floquet topological insulators. In Refs.~\cite{Gu2011,Foatorres2014,Kundu2014}, it was shown that the DC conductance of a Floquet topological insulator is not linked to the topological invariants of the full Floquet bands, and therefore not quantized in general. Yet, by employing a Floquet sum rule proposed in Ref.~\cite{Kundu2013}, the authors of Refs.~\cite{Farrell2015,Farrell2016} found a way to recover the conductance quantization of Floquet edge states in a quantum spin Hall insulator, which was further confirmed in a Floquet-Chern insulator model~\cite{Fruchart2016}. In Ref.~\cite{Fulga2016}, a scattering matrix invariant is introduced to describe the conductance of Floquet edge states at a given quasienergy, which assumes only integer values. {Using a boundary-condition matching approach generalized to Floquet systems, the authors in Ref.~\cite{Gu2011} studied the scaling of conductance in an irradiated graphene nanoribbon.} In the meantime, a couple of complementary works reveal some anomalous transport properties of Floquet edge states in larger systems due to the interplay between driving fields and reservoir-induced cooling~\cite{Dehghani2015a,Dehghani2015b,Dehghani2016}. These interesting debates and discoveries, together with the lack of experiments on Floquet edge states transport in electronic systems, call for more efforts to understand the conductance property of Floquet topological edge states and their relevant experimental signatures.

In this work, we apply the Keldysh nonequilibrium Green's function (NEGF) approach and the recursive Floquet-Green's function method to explore a number of aspects of Floquet edge state transport in a two-lead setup. As an example, we focus on a quantum well heterostructure, which is described by the Hofstadter lattice model subject to a harmonic driving and coupled to two metallic leads. The effective magnetic flux in the Hofstadter model breaks time reversal symmetry, giving rise to a model system even simpler than the driven quantum spin Hall system studied in Refs.~\cite{Farrell2015,Farrell2016}. By calculating the DC conductance, DC profile and time-averaged local density of states (LDOS) in both pristine and disordered samples, we present intriguing transport properties for different types of Floquet edge states. In particular, we indeed find robust and tunable edge current as we change the system parameters and the characteristics of the driving field. In certain parameter regimes, quantized DC conductance as large as $8e^2/h$ is found after applying the Floquet sum rule. We also discuss possible experimental observations of such theoretical observations. Furthermore, by introducing disorder and defects to the driven quantum well heterostructure, the robustness of three types of Floquet edge states are compared. It is found that co-propagating Floquet chiral edge states are more robust than counter-propagating and symmetry-restricted Floquet edge states. Finally, we study how the transport property of Floquet edge modes is modified if the leads have a finite bandwidth.   All these detailed results constitute a necessary guide towards the potential use of Floquet topological matter in designing novel transport devices.

This paper is organized as follows. In Sec.~\ref{sec:theory} we outline the theoretical framework and introduce the transport quantities studied in this work. Readers not interested in technical details may skip this section. In Sec.~\ref{sec:model}, we introduce the harmonically-driven Hofstadter model in a two-terminal setup. In Sec.~\ref{subsec:large}, we analyze the two-terminal transport via Floquet edge states in pristine samples. In Sec.~\ref{subsec:robust}, we study the response of Floquet edge states to disorder and defect. We then investigate the implications of finite-bandwidth leads on the quantization of DC conductance in Sec.~\ref{subsec:finiteBW}. This paper ends with conclusion and outlook in Sec.~\ref{sec:conclusion}.

\section{\label{sec:theory}Theory}

In this section, we first outline the Keldysh framework for quantum transport. {Next, we introduce the Floquet representation of Green's functions~\cite{AokiRMP2014,GomezLeon_Keldysh}.} This then leads us to three transport quantities: ($1$) DC conductance, ($2$) local DC profile, and ($3$) time-averaged local density of states (T-LDOS), which we use to investigate edge state transport. Lastly, we discuss how to compute the Floquet-Green's functions of interest. Throughout this paper, we use the following notations: $\period$ for the period of the driving field, $\Omega=2\pi/\period$ for the driving frequency, $\epsilon$ for quasienergy and $E$ for energy.  Note that although the central system is under periodic driving and hence what matters is its quasienergy, the leads are however not driven and its energy must be specified in a theory of quantum transport.
\subsection{\label{subsec:transport}Transport within the Keldysh-NEGF framework}
The Hamiltonian of noninteracting electrons in a tight-binding square lattice subject to driving fields can be generally written as {$H(t)=\sum_{i,j}J_{ij}(t)c_i^\dag c_j$,
where $c^{\dagger}_i$~($c_i$) is the creation (annihilation) operator on lattice site $i$, $J_{ij}(t)=J^*_{ji}(t)$ is a time-dependent hopping amplitude~($i\neq j$) or onsite potential~($i=j$).}
Following Ref.~\cite{Jauho1994}, we consider the driving fields to be switched on in the distant past, but not switched off throughout the duration of our interest. In a two-terminal device, the central system $H(t)$ is connected to a left and a right electronic lead, which are kept in equilibrium with chemical potentials $\mu_L$ and $\mu_R$, respectively. The electronic current from the left lead to the central system is given by $I^L(t)=e\langle\mathrm{d}N_L/\mathrm{d}t\rangle$, where the electron charge is $-e$, and $N_L$ is the particle number operator of the left lead. The explicit time dependence of the current is due to the driving field applied to the center.

In the Keldysh-NEGF framework, the contour-ordered Green's function is defined as $G(\tau,\tau')=-\frac{\ii}{\hbar}\mathcal{T}\langle c(\tau)c^\dag(\tau')\rangle$, where $\cal{T}$ orders contour time variables $\tau,\tau'$ along the Keldysh contour~\cite{Haug2008,Wang2008,Jishi2014}. Following the Meir-Wingreen prescription, the expression of the current reads~\cite{Jauho1994}:
\begin{equation}
I^L(t) = e \left[\int_\mathcal{C} \TrSite\Big[\Sigma_L(\tau,\tilde{\tau})G(\tilde{\tau},\tau')\Big]\mathrm{d}\tilde{\tau} \right]^<_{\tau=\tau'\mapsto t} + \cc, \label{current3}
\end{equation}
where $\Sigma_L(\tau,\tilde{\tau})$ is the self-energy of the left lead. $\int_{\mathcal{C}}\mathrm{d}\tilde{\tau}\cdots$ is an integral over the Keldysh contour~\cite{Wang2008,Wang2014}$.\
  \TrSite[\cdots]$ is a trace over the central lattice system. $[\cdots]^<$ means resolving a contour function into its lesser component, which can be done following the Langreth's rules~\cite{Haug2008}. The notation $\cdots|_{\tau=\tau'\mapsto t}$ means setting the two contour times $\tau,\tau'$ at equal time $t$.

\subsection{Green's functions in Floquet representation}
{The Green's functions of a periodically-driven system depend on two independent time variables, even in a steady state. In the Wigner representation, a two-point correlation function $g(t,t')$ is given by $\bar{g}(T=\frac{t+t'}{2},\tau=t-t')=g(t,t')$, where $T$ ($\tau$) is the average (relative) time variable~\cite{Tsuji2008}. The Floquet representation of this correlation function is then obtained by the following transformation~\cite{Tsuji2008}}:
\begin{equation}
\begin{split}
\bm{g}_{mn}(E)=&\frac{1}{\period}\int_0^\period \mathrm{d}T\;\ee^{\ii(m-n)\Omega t}\\ &\times \int_{\mathbb{R}}\mathrm{d}\tau\;\ee^{\ii(E+\frac{m+n}{2}\hbar\Omega)\frac{\tau}{\hbar}}\ \bar{g}(T,\tau). \label{eq:FloquetRepresentation}
\end{split}
\end{equation}
The Floquet representation is especially convenient when manipulating convolution integrals, because it turns them into (infinite) matrix product~\cite{Tsuji2008}: $[\int_{\mathbb{R}}\mathrm{d}\tilde{t}\; f(t,\tilde{t})g(\tilde{t},t')]_{mn}(E) =  \sum_{k\in\mathbb{Z}}\bm{f}_{mk}(E)\bm{g}_{kn}(E)$. The following convention for Floquet representation of Green's functions will be used.
If the spatial components of a Green's function in the Floquet representation need to be specified, we write $\bm{g}_{(i,j;m,n)}(E)$, with  the first two indices of the subscript $(i,j;m,n)$ corresponding to lattice site indices $i,j$, and the last two indices $m,n$ corresponding to Floquet indices as in Eq.~(\ref{eq:FloquetRepresentation}).

\subsection{Transport quantities}
If a periodic steady state is attained, the correlation functions in the Wigner representation are also periodic with respect to the average time $T$, i.e., $\bar{g}(T,\tau)=\bar{g}(T+\period,\tau)$. Then { the Floquet decomposition is made possible and} it can be shown that the DC component of the current is given by~\cite{Kohler2005,Cuevas2010}:
\begin{equation}
\begin{split}
I^L_{\mathrm{DC}} = e\int_\mathbb{R}\frac{\mathrm{d}E}{2\pi\hbar}\sum_{k\in\mathbb{Z}}& \Big\{\TrSite\left[
\bm{G}^r_{k0}\bm{\Gamma}^L_{00}\bm{G}^a_{0k}\bm{\Gamma}^R_{kk} \right] \fermi^L \\ &- \TrSite\left[
\bm{G}^r_{k0}\bm{\Gamma}^R_{00}\bm{G}^a_{0k}\bm{\Gamma}^L_{kk} \right] \fermi^R
	\Big\} , \label{eq:KLH}
\end{split}
\end{equation}
where $\bm{G}^{r(a)}_{k0}(E)$ is the Floquet representation of retarded (advanced) Green's functions $G^{r(a)}(t,t')$, defined by:
\begin{equation}
\begin{split}
G^{r}(t,t') &= - \frac{\ii}{\hbar} \theta \left( t- t'\right) \langle\{c(t),c^\dag(t')\}\rangle, \\
G^{a}(t,t') &=  \frac{\ii}{\hbar} \theta \left( t'- t\right) \langle\{c(t),c^\dag(t')\}\rangle,
\end{split}
\end{equation}
and $f^{L(R)}$ is the Fermi function of the left~(right) lead. $\Gamma^{L(R)}=\ii\{\Sigma^r_{L(R)}-[\Sigma^r_{L(R)}]^\dagger\}$ is the {level width function}
of the left~(right) lead, with $\bm{\Gamma}^{L(R)}_{kk}$ being the corresponding matrix in the Floquet representation.  In general, all these functions depend explicitly on energy $E$.

\subsubsection{DC Conductance}
We now define the \emph{transmission coefficient} at an energy $E$ from the left to right lead as $T_{LR}(E)=\sum_{k\in\mathbb{Z}} \TrSite\left[\bm{G}^r_{k0}\bm{\Gamma}^L_{00}\bm{G}^a_{0k}\bm{\Gamma}^R_{kk}\right]$ [and analogously for $T_{RL}(E)$ with $L\leftrightarrow R$]. When a bias voltage $V$ is applied between the two reservoirs, the chemical potentials of the leads become $\mu_{L/R}=\mu_0\pm eV/2$. At zero temperature and to first order in $V$, one can linearize the Fermi function as: $f_{L/R}(E)=\theta(\mu_0\pm eV/2 -E)\approx \theta(\mu_0-E)\mp \delta(\mu_0-E) eV/2$. Under this condition, Eq.~\eqref{eq:KLH} simplifies to:
\begin{equation}
\begin{split}
I^L_{\mathrm{DC}} &= e\int_{-\infty}^{\mu_0}\frac{\mathrm{d}E}{2\pi\hbar}\left[T_{LR}(E)-T_{RL}(E)\right] \\ &+ \frac{e^2}{h}\frac{T_{LR}(\mu_0)+T_{RL}(\mu_0)}{2}V\label{eq:ILDC}.
\end{split}
\end{equation}
When there is an inversion symmetry between left and right leads (which is the case in all our numerical calculations), we have $T_{LR}(E)=T_{RL}(E)$~\cite{Foatorres2014}. Under this condition, the DC conductance ${\rm d}I^L_{\mathrm{DC}}/{\rm d}V$ \emph{at a given energy $E$} is simply given by $e^2/h$ times the transmission coefficient $T(E)=T_{LR}(E)=T_{RL}(E)$, which reads
\begin{equation}
T(E) = \sum_{k\in\mathbb{Z}} \TrSite\left[\bm{G}^r_{k0}(E)\bm{\Gamma}^L_{00}(E)\bm{G}^a_{0k}(E)\bm{\Gamma}^R_{kk}(E)\right]. \label{eq:Transmission}
\end{equation}

\subsubsection{\label{subsubsec:DCprofile}Local DC Profile}
The Heisenberg equation for the particle number operator $n_i=c_i^{\dagger}c_i$ at lattice site $i$ is given by~\cite{footnote1}:
\begin{equation}
\begin{split}
& \langle\dot{n}_i(t)\rangle = J_{i+\hatx,i}(t)G^<_{i,i+\hatx}(t,t)+J_{i-\hatx,i}(t)G^<_{i,i-\hatx}(t,t) \\ &+J_{i+\haty,i}(t)G^<_{i,i+\haty}(t,t)+J_{i-\haty,i}(t)G^<_{i,i-\haty}(t,t) + \cc,
\end{split}
\end{equation}
which is similar to the case without a driving field~\cite{Cresti2003,Waintal2008,Lewenkopf2013}, apart from the explicit time dependences of the hopping terms. We may then define the local current from site $i+\hatx$ to site $i$ as: $I_{i\leftarrow i+\hatx}(t) = J_{i+\hatx,i}(t)G^<_{i,i+\hatx}(t,t) + \cc=-I_{i+\hatx\leftarrow i}$, {and analogously for all other neighbors of $i$. From this one defines the time-averaged local current vector:
\begin{equation}
\begin{split}
\vec{I}_i &= \frac{1}{\period}\int_0^\period \mathrm{d}t\Big\{ \hatx[I_{i\leftarrow i-\hatx}(t)+I_{i+\hatx \leftarrow i}(t)]\\ &+\haty [I_{i\leftarrow i-\haty}(t)+I_{i+\haty \leftarrow i}(t)]\Big\},
\end{split}
\end{equation}
which represents both the magnitude and direction of the local current at site $i$}. Numerically, the time-averaged local current, e.g., $\frac{1}{\period}\int_0^\period\mathrm{d}t\;I_{i\leftarrow i-\hatx}(t)$, is given in the Floquet representation by:
\begin{equation}
\begin{split}
\frac{1}{\period}\int_0^\period\mathrm{d}t&\;I_{i\leftarrow i-\hatx}(t) \\ &= \sum_{m\in\mathbb{Z}}\int_{\mathbb{R}}\frac{\mathrm{d}E}{2\pi\hbar}\tilde{J}_{i-\hatx,i}(m)\bm{G}^<_{(i,i-\hatx;0,m)}(E),
\end{split}
\end{equation}
where $\tilde{J}_{i-\hatx,i}(m)=\frac{1}{\period}\int_0^\period\mathrm{d}t\,\ee^{\ii m\Omega t}J_{i-\hatx,i}(t)$ is the $m$th Fourier component of the $\period$-periodic function $J_{i-\hatx,i}(t)$. Furthermore, at a particular energy $E$, the \emph{energy-resolved} time-averaged local current is defined as: $I_{i\leftarrow i-\hatx}(E)=\sum_{m\in\mathbb{Z}}\tilde{J}_{i-\hatx,i}(m)\bm{G}^<_{(i,i-\hatx;0,m)}(E)$. Correspondingly, the energy-resolved local DC component of current vector at site $i$ reads:
\begin{widetext}
\begin{equation}
\begin{split}
\vec{I}_i(E) = \sum_{m\in\mathbb{Z}} &\hatx\big[ \tilde{J}_{i-\hatx,i}(m)\bm{G}^<_{(i,i-\hatx;0,m)}(E)+ \tilde{J}_{i,i+\hatx}(m)\bm{G}^<_{(i+\hatx,i;0,m)}(E)  \big]\\
+ &\haty\big[\tilde{J}_{i-\haty,i}(m)\bm{G}^<_{(i,i-\haty;0,m)}(E) + \tilde{J}_{i,i+\haty}(m)\bm{G}^<_{(i+\haty,i;0,m)}(E)	\big] +\cc
\end{split}\label{eq:localcurrent}
\end{equation}
\end{widetext}
The DC profile is constructed by plotting this quantity at every site $i$ of the central system.

\subsubsection{Time-averaged LDOS}
The time-averaged LDOS discribes the spatial distribution of states, thus allowing us to recognize whether there are edge states at a certain energy $E$. For static systems, the LDOS is defined by $\mathrm{LDOS}_i(E)=-\frac{1}{\pi}\mathrm{Im}G^r_{ii}(E)$. To extend this notion to Floquet systems, we need to decompose the dynamics into two parts, corresponding to the average time $T$ and relative time $\tau$ in the Wigner representation of Green's functions~\cite{Mikami2016}. When a periodic steady state is achieved, it is natural to average the Green's function over a driving period, yielding the generalized expression of time-averaged LDOS: $\textrm{T-LDOS}_i(E) = -\frac{1}{\pi} \mathrm{Im}\frac{1}{\period}\int_0^{\period}\mathrm{d}T\;\bar{G}^r_{ii}(T,E)$.
In the Floquet representation, it has the following expression:
\begin{equation}
\textrm{T-LDOS}_i(E)=-\frac{1}{\pi}\mathrm{Im}\bm{G}^r_{(i,i;0,0)}(E),
\label{eq:TLDOS}
\end{equation}
which is used in our computational study.

\subsection{\label{subsec:RFGF}Computational approach to Floquet-Green's function}
Thus far, we have expressed physical quantities in terms of Green's functions and self-energies. We also have the following relations to help simplify algebraic manipulations: (i) causality, i.e., $[\bm{G}^r_{(i,j;m,n)}]^\dag(E)=\bm{G}^a_{j,i;n,m}(E)$, and (ii) fluctuation-dissipation relations for leads, i.e., $\bm{\Sigma}^<_{L/R}=-[\bm{\Sigma}^r_{L/R}-(\bm{\Sigma}^r_{L/R})^\dag]\fermi^{L/R}$ \cite{Haug2008,Wang2008}. As shown in Eqs.~\eqref{eq:Transmission},~\eqref{eq:localcurrent} and~\eqref{eq:TLDOS}, for the calculations of physical quantities, it all boils down to two Green's functions: the retarded and lesser. Hence, we give here their governing equations, which can be calculated with the center Hamiltonian and lead self-energies as input.

The retarded Green's function satisfies the Dyson equation, which in the Floquet representation is given by~\cite{Kitagawa2011}:
\begin{equation}
\begin{split}
\sum_{k\in\mathbb{Z}}\Big[&(E+m\hbar\Omega)\delta_{mk} -H_{m-k} \\ &-(\bm{\Sigma}^r_L+\bm{\Sigma}^r_R)_{mk} \Big] \bm{G}^r_{kn}(E) = \delta_{mn}. \label{eq:FloquetDyson}
\end{split}
\end{equation}
The solution of this equation is found by truncating over Fourier components, followed by a matrix inversion. In practice, if we need to deal with large system sizes and large number of harmonics, direct inversion is not feasible, since its complexity scales like $(N_x N_y \nF)^3$, where $N_x$~($N_y$) is the number of lattice sites along $x$~($y$) direction, and $\nF$ is the truncated number of harmonics in frequency space. To surmount this difficulty, we adapt the recursive Green's function method~\cite{Lewenkopf2013,Ryndyk2015} to Floquet systems (see App.~\ref{app:RFGF}). The lesser Green's function is related to the retarded Green's function through the Keldysh equation~\cite{Haug2008,Wang2008,Jishi2014}. In the Floquet representation, it reads:
\begin{equation}
{\bm G}^<_{mn}(E) = \sum_{p,q\in\mathbb{Z}}{\bm G}^r_{mp}(E)({\bm \Sigma}^<_L+{\bm \Sigma}^<_R)_{pq}(E){\bm G}^a_{qn}(E).
\end{equation}

\section{\label{sec:model}Model}

\begin{figure}[H]
	\centering
	\def\svgwidth{0.5\columnwidth}
	\begingroup%
  \makeatletter%
  \providecommand\color[2][]{%
    \errmessage{(Inkscape) Color is used for the text in Inkscape, but the package 'color.sty' is not loaded}%
    \renewcommand\color[2][]{}%
  }%
  \providecommand\transparent[1]{%
    \errmessage{(Inkscape) Transparency is used (non-zero) for the text in Inkscape, but the package 'transparent.sty' is not loaded}%
    \renewcommand\transparent[1]{}%
  }%
  \providecommand\rotatebox[2]{#2}%
  \ifx\svgwidth\undefined%
    \setlength{\unitlength}{441.83893568bp}%
    \ifx\svgscale\undefined%
      \relax%
    \else%
      \setlength{\unitlength}{\unitlength * \real{\svgscale}}%
    \fi%
  \else%
    \setlength{\unitlength}{\svgwidth}%
  \fi%
  \global\let\svgwidth\undefined%
  \global\let\svgscale\undefined%
  \makeatother%
  \begin{picture}(1,0.55456154)%
    \put(0,0){\includegraphics[width=\unitlength,page=1]{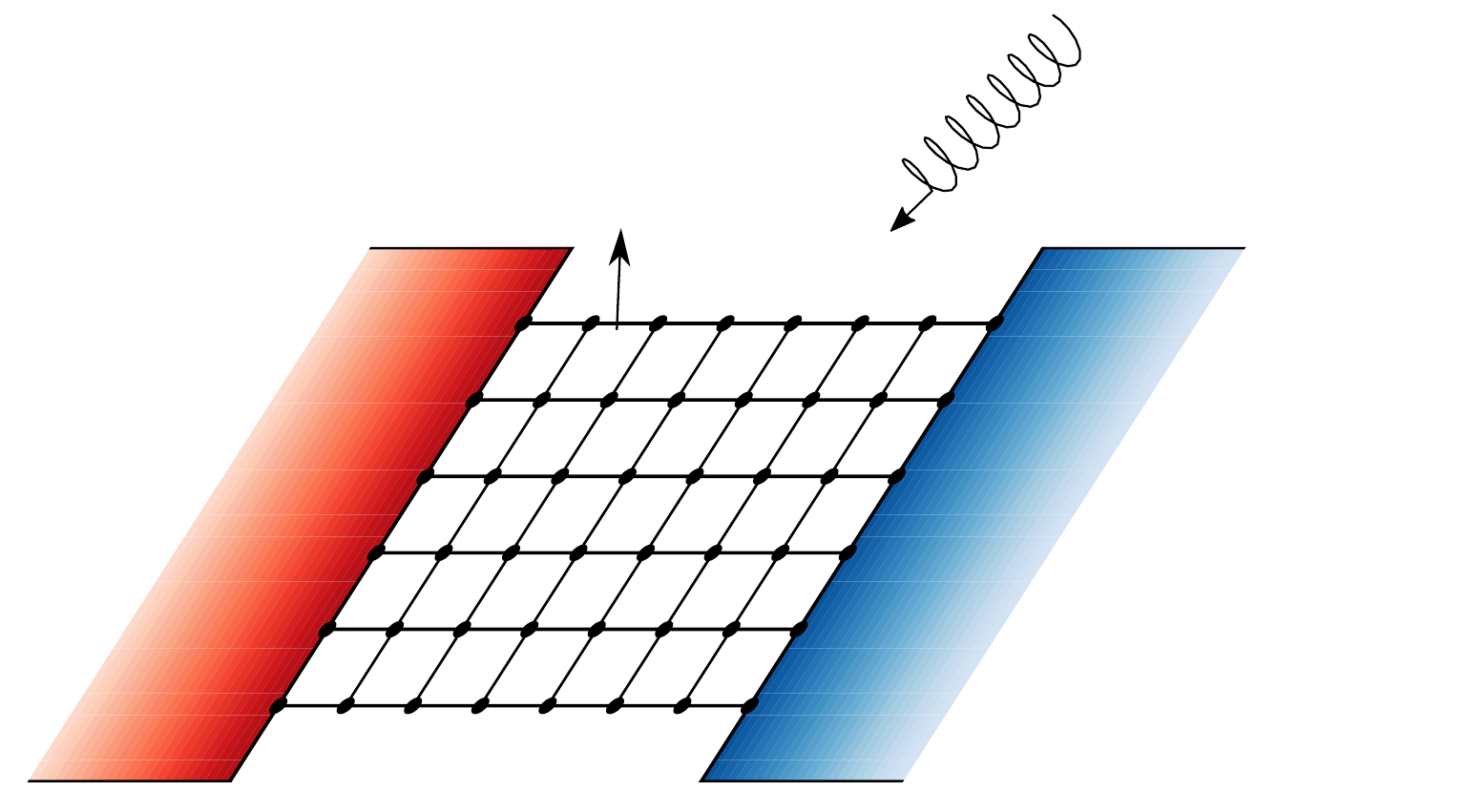}}%
    \put(0.41028281,0.43828048){\color[rgb]{0,0,0}\makebox(0,0)[lt]{\begin{minipage}{0.21550681\unitlength}\raggedright $\vec{B}$\end{minipage}}}%
    \put(0.1600211,0.22580974){\color[rgb]{0,0,0}\makebox(0,0)[lt]{\begin{minipage}{0.21550679\unitlength}\raggedright $\mu_L$\end{minipage}}}%
    \put(0.663394,0.22580924){\color[rgb]{0,0,0}\makebox(0,0)[lt]{\begin{minipage}{0.21550679\unitlength}\raggedright $\mu_R$\end{minipage}}}%
    \put(0.66386565,0.4316183){\color[rgb]{0,0,0}\makebox(0,0)[lt]{\begin{minipage}{0.41634649\unitlength}\raggedright $J_y\cos(\Omega t)$\end{minipage}}}%
    \put(0,0){\includegraphics[width=\unitlength,page=2]{schematicSystemEvenSmallerWithAxes.pdf}}%
    \put(0.80897614,0.04141596){\color[rgb]{0,0,0}\makebox(0,0)[lt]{\begin{minipage}{0.10319834\unitlength}\raggedright $x$\end{minipage}}}%
    \put(0.76104657,0.13143502){\color[rgb]{0,0,0}\makebox(0,0)[lt]{\begin{minipage}{0.10319833\unitlength}\raggedright $y$\end{minipage}}}%
  \end{picture}%
\endgroup%
	\caption{(color online) Schematic of the system studied in this work. The left~(right) lead has chemical potential $\mu_L$~$(\mu_R)$. The central system is a square tight-binding lattice subject to a perpendicular magnetic field $\vec{B}$ and a harmonic driving field with driving amplitude $J_{y}\cos({\Omega}t)$.}
	\label{Fig:model}
\end{figure}
We consider a two-terminal device as illustrated in Fig.~\ref{Fig:model}. The central system is a quantum well heterostructure described by the harmonically driven Hofstadter model (HDHM). This model has been shown to host many interesting Floquet topological phases, e.g. with nearly flat bands, very large Chern numbers, and counter-propagating edge states~\cite{Zhou2014,Zhao2014}. In this work, we put this model in a two-terminal transport setup: the central system is coupled to two metallic leads, which are kept separately in thermal equilibrium with chemical potentials $\mu_{L}$ and $\mu_{R}$. The total Hamiltonian $\mathcal{H}(t)$ describing the central system, leads and their couplings is given by:
\begin{equation}
\mathcal{H}(t) =  H_L + V_{L} + H(t) + V_{R} + H_R,
\end{equation}
where
\begin{equation}
\begin{split}
H(t) = \sum_{i}& \{J_x c^\dag_{i}c_{i+\hatx} \\ &+ J_y\left[s+\cos(\Omega t)\right]\ee^{\ii 2\pi\alpha x_i}c^\dagger_{i}
c_{i+\haty}+\Hc	\} \label{Hamiltonian}
\end{split}
\end{equation}
is the Hamiltonian of HDHM~\cite{Zhou2014,Zhao2014} at the center, with a magnetic flux $\alpha=p/q\in\mathbb{Q}$ per unit area. $q\in\mathbb{N}$ is the number of sublattices in each magnetic unit cell, and $\Omega$ is the frequency of the driving field. We abbreviate the lattice coordinates as $i\equiv (x_i,y_i)$. The leads are modeled as two semi-infinite (along $x$) square lattices, with nearest-neighbor hopping amplitudes $t_{x/y}$ along $x/y$ directions and a uniform onsite potential $v$:
\begin{equation}
H_{L/R} = \sum_i \left[\left( t_x a^\dag_i a_{i+\hatx} + t_y a^\dag_i a_{i+\haty} +\Hc \right)+va^\dag_i a_i\right]. \label{eq:lead}
\end{equation}
In our transport calculations, we take the tunneling amplitude between the leads and central system along $x$-direction to be the same as $t_x$. Under this choice, the Hamiltonian describing the coupling between the left (right) lead and the central system is give by:
\begin{equation}
V_{L(R)} = t_x\sum_{\ell,i:\left[x_\ell=-1(+1),x_i=0\right]} \left(a_\ell^\dag c_i + \Hc\right).
\end{equation}
Along $y$-direction, we take open boundary condition for the central system in all our transport calculations below.

\subsection{\label{subsec:Hofstadter}Floquet spectrum}
Taking periodic boundary conditions along $x/y$-direction and open boundary condition for the other direction, we can Fourier transform the coordinate $x/y$ in Eq.~(\ref{Hamiltonian}) into a quasimomentum $k_x/k_y$. The resulting Hamiltonian is given by either of the following:
\begin{equation}
\begin{split}
H_{k_y}&(t) = \sum_{x=0}^{N_x-1}\{J_x\big(c^\dag_{x,k_y}c_{x+1,k_y}+\Hc	\big)\\ & +2J_y\left[s+\cos(\Omega t)\right]\cos(2\pi\alpha x+k_y)c^\dag_{x,k_y}c_{x,k_y}	\}, \\
H_{k_x}&(t) = \sum_{y=0}^{N_y-1}\bigg\{ J_x\Big(c^\dag_{q-1,y}c_{0,y}\ee^{\ii k_x q}+ \sum_{j=0}^{q-2}c^\dag_{j,y}c_{j+1,y}	\Big) \\&+ J_y\left[s+\cos(\Omega t)\right]\sum_{j=0}^{q-1}\left(c^\dag_{j,y}c_{j,y+1}\ee^{\ii 2\pi\alpha j}+\Hc\right) \bigg\}.
\end{split}
\end{equation}
These are just $1$D descendants of the HDHM, also called the driven Harper model~\cite{Kolovsky2012,Zhou2014}. One can then obtain the Floquet spectrum with respect to $k_x$ or $k_y$ by solving the Floquet-Schr\"{o}dinger equation~\cite{Shirley1965,Sambe1973} using $H_{k_x}(t)$ or $H_{k_y}(t)$. The nontrivial topology of a Floquet band is then signaled by the presence of gapless edge states. Furthermore, {as the quasienergy is defined modulo the driving frequency, it is possible for gapless edge modes to wind around the Floquet-Brillouin zone, a scenario absent in} static systems~\cite{Rudner2013,Zhou2014,Zhao2014,Derek2014,GomezLeon_counter}. 
Note also that, one may intuitively prefer to inspect the edge states with open boundary condition along $y$, so as to plot the dispersion relation of the edge states as a function of $k_x$ to understand the transport along the $x$ direction.  However, in terms of understanding the bulk-edge correspondence and locating the edge states in connection with the Floquet spectrum gap (exceptions to be discussed below), choosing $H_{k_y}(t)$ or $H_{k_x}(t)$ for the Floquet spectrum to digest the transport results is just a matter of taste.

\section{\label{sec:results}Computational Results}
Using the theoretical and computational methods discussed in Sec.~\ref{sec:theory}, we now investigate the transport of Floquet edge states in the two-terminal device introduced in Sec.~\ref{sec:model}. We first look into the DC conductance of our model system in the pristine limit, where we  observe a conductance quantization by the Floquet sum rule~\cite{Kundu2013,Farrell2015}. Next, we introduce defects and disorder into our system, and present a detailed comparison of the robustness among three different types of gapless Floquet edge states. Finally, we consider a practical situation in which the leads have a finite bandwidth, and discuss its impact on Floquet edge state transport. In all our calculations presented, we take $J_x=1$ as the unit of energy.

\subsection{\label{subsec:large} Edge-state transport under harmonic driving}
\begin{figure}
\captionsetup[subfloat]{captionskip=-4pt}
\centering
\subfloat[\label{fig:TransmissinCleana}]{
\includegraphics[width=0.45\columnwidth]{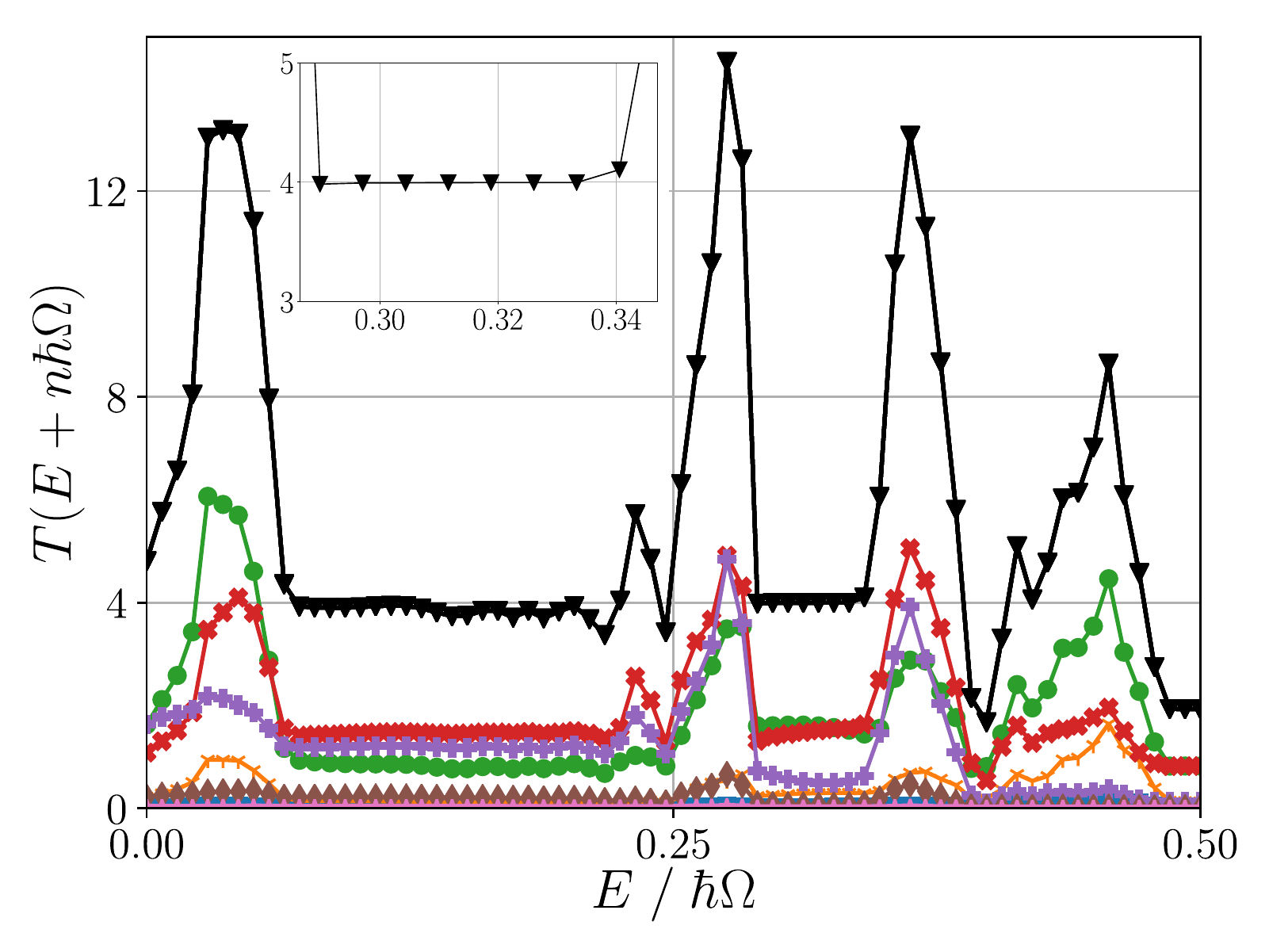}
}
\subfloat[\label{fig:TransmissinCleanb}]{
\includegraphics[width=0.45\columnwidth]{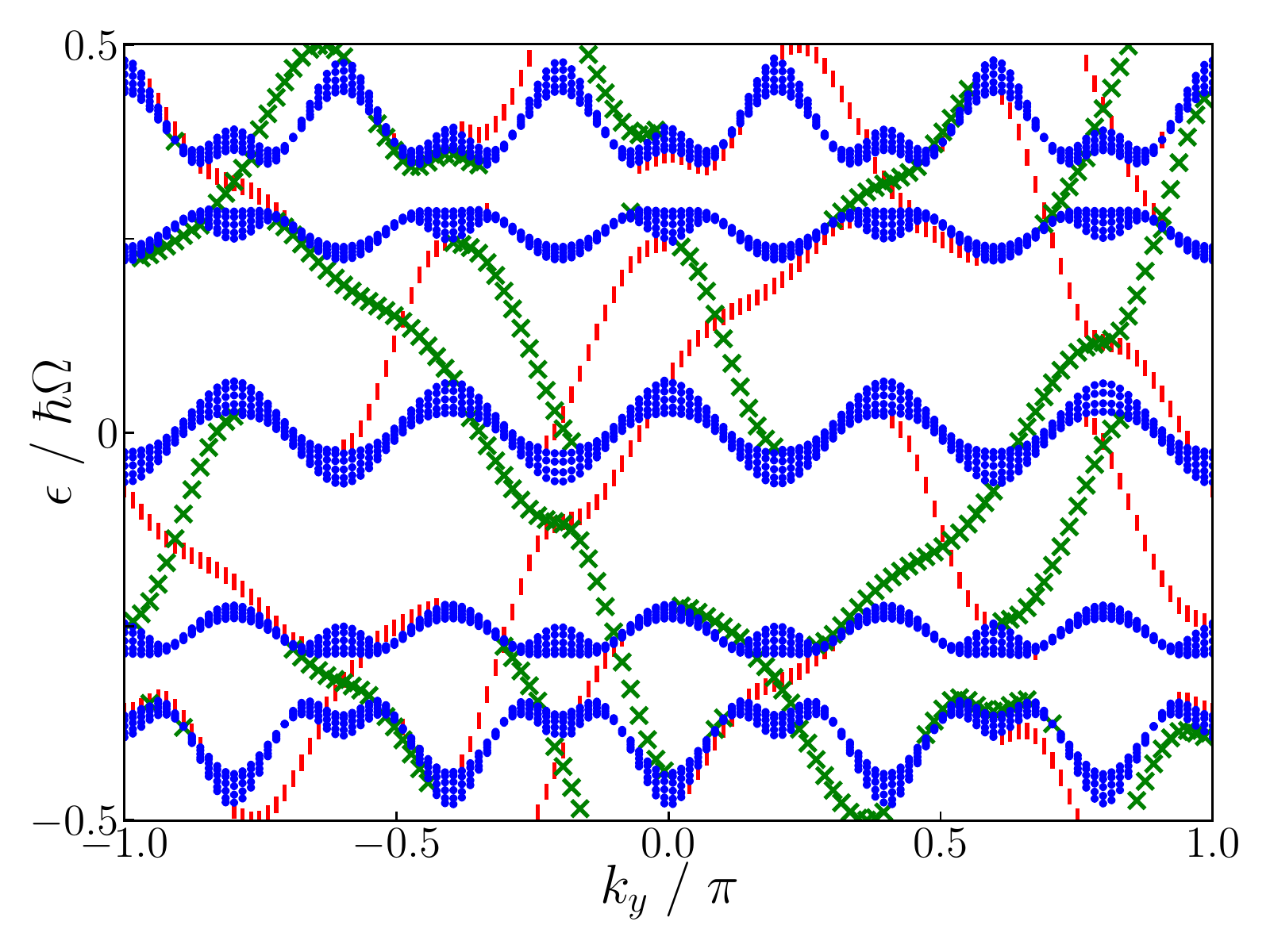}
}
\\[-1ex]
\centering
\subfloat[\label{fig:TransmissinCleanc}]{
\includegraphics[width=0.45\columnwidth]{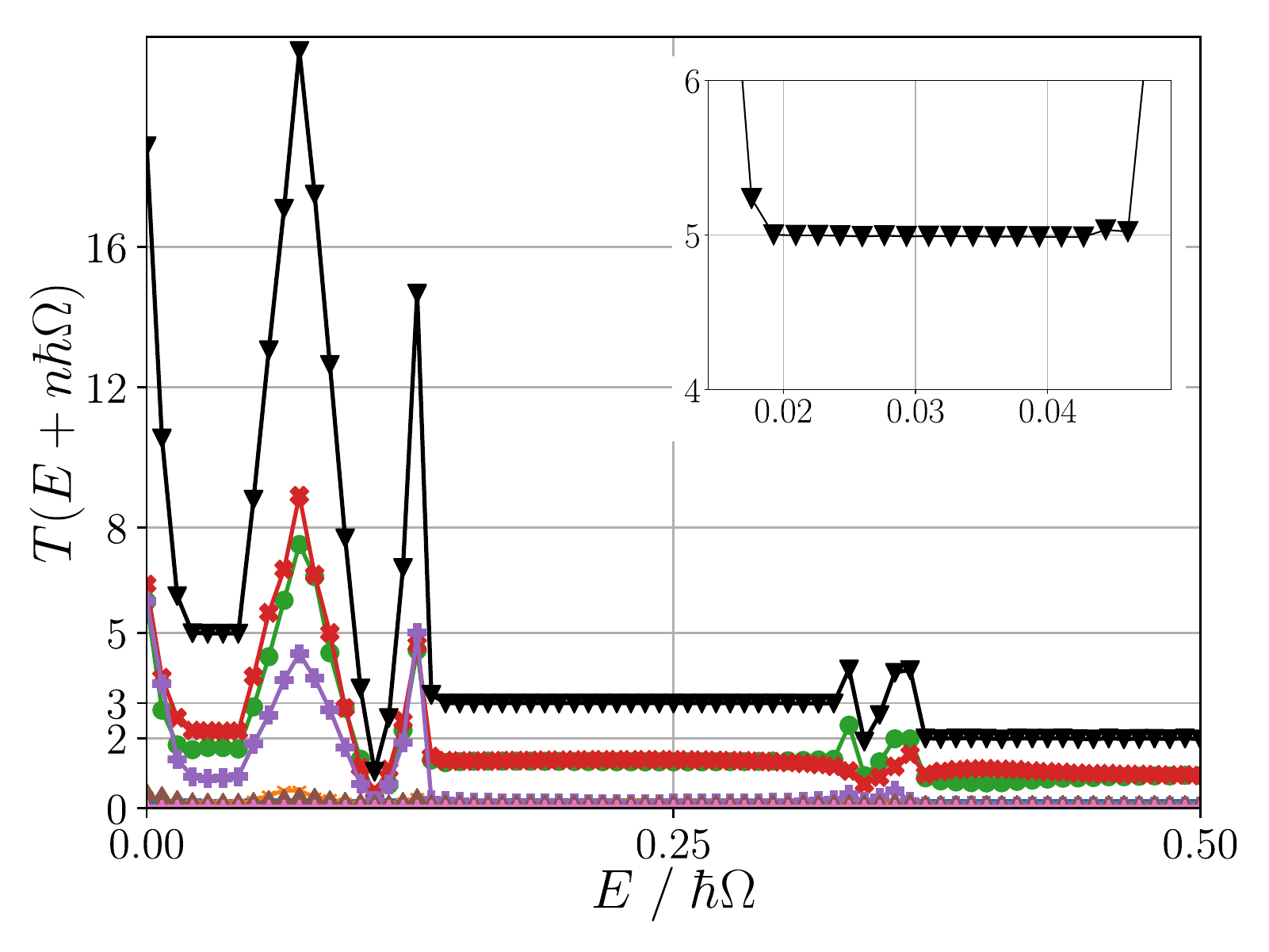}
}
\subfloat[\label{fig:TransmissinCleand}]{
\includegraphics[width=0.45\columnwidth]{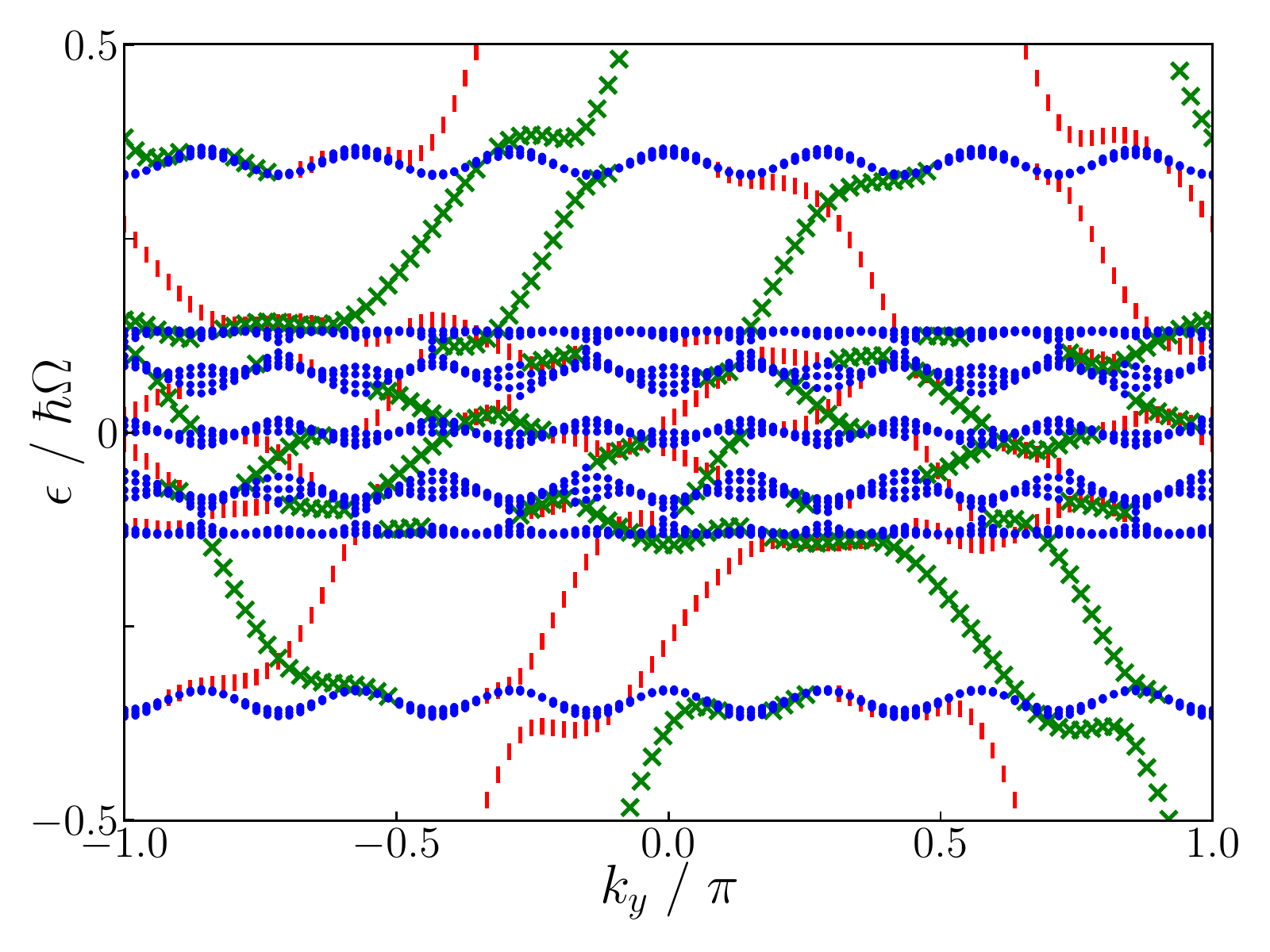}
}
\\[-1ex]
\centering
\subfloat[\label{fig:TransmissinCleane}]{
\includegraphics[width=0.45\columnwidth]{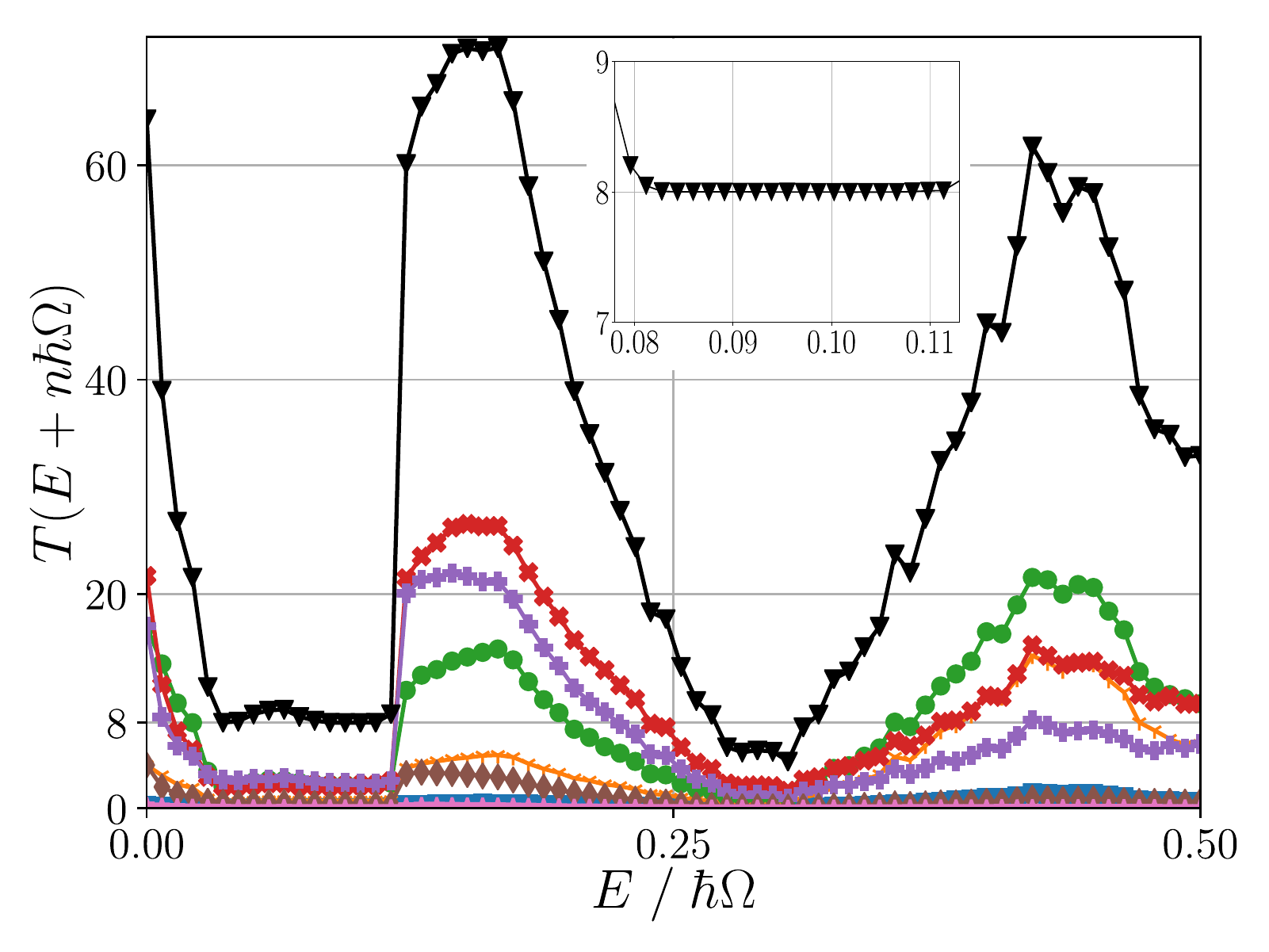}
}
\subfloat[\label{fig:TransmissinCleanf}]{
\includegraphics[width=0.45\columnwidth]{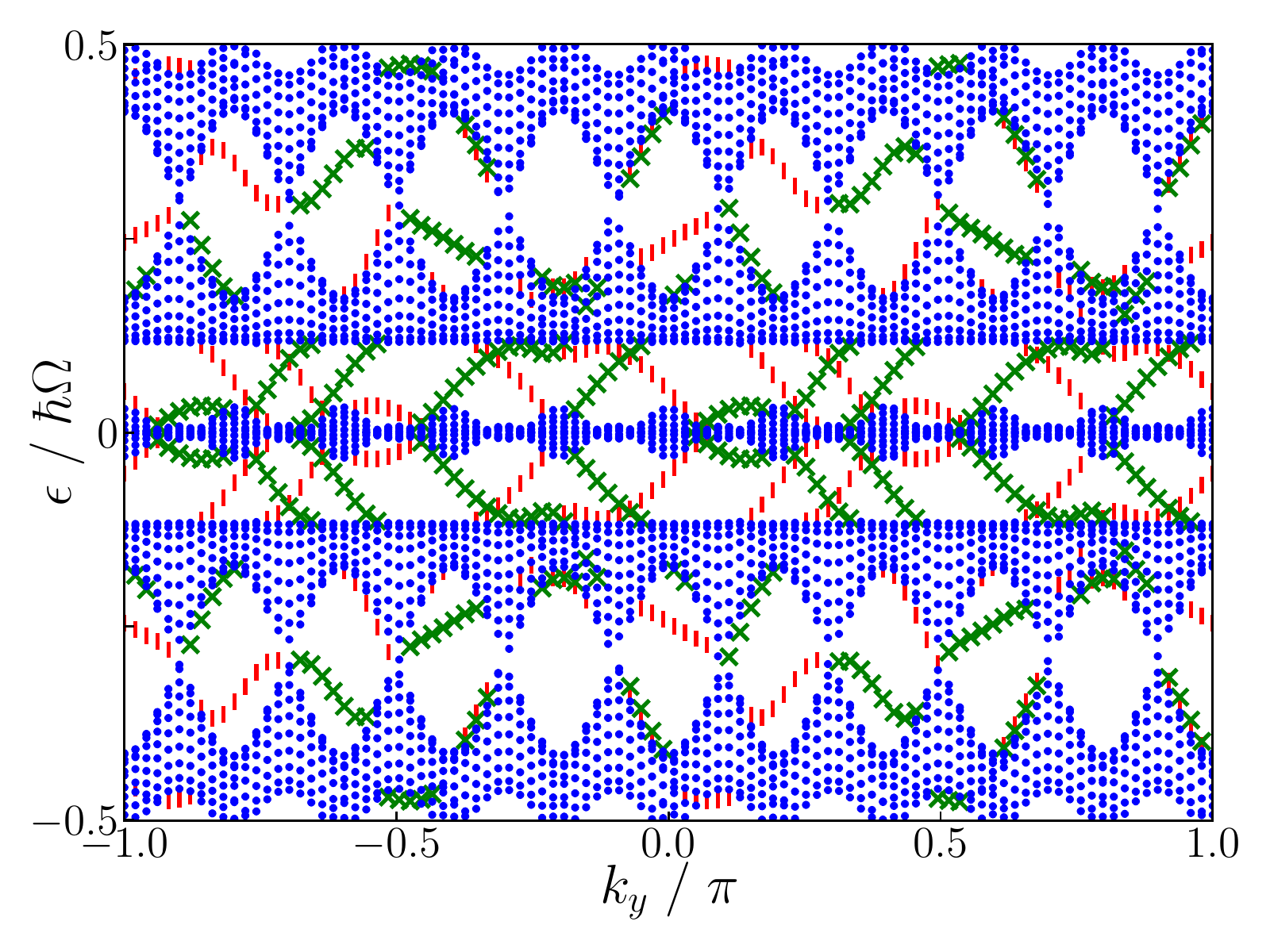}
}
\\[-1ex]
\subfloat{
\includegraphics[width=0.9\columnwidth]{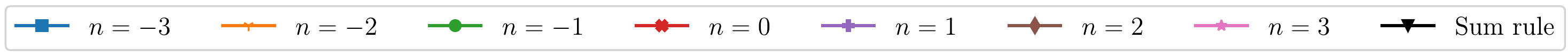}
}
\caption{(color online) Two-terminal DC conductance of the HDHM at zero temperature. Left panels: transmission $T(E+n\hbar\Omega)$ vs. energy $E$ of an incoming electron at different Floquet indices $n$. Results are shown for $E$ in the upper half of the quasienergy Brillouin zone. Curves with black triangles are obtained after applying the Floquet sum rule, with insets zoomed in to the plateaux of interest. Right panels: Floquet spectrum of the same model with PBC (OBC) along $y$($x$)-direction. Red $``|"$s and green crosses denote states localized at left and right edges of the sample. System parameters are $J_y=1.6,\alpha=1/5,\Omega=\pi,s=1,N_x=N_y=100$ for panels (a) and (b),  $J_y=1.3,\alpha=2/7,\Omega=\pi,s=1,N_x=N_y=80$ for panels (c) and (d), and $J_y=1.3,\alpha=1/5,\Omega=\pi/2,s=0,N_x=N_y=100$ for panels (e) and (f).}
\label{fig:TransmissinClean}
\end{figure}

Following the discussions of Sec.~\ref{sec:theory}, the DC conductance at low bias in zero temperature is given by Eq.~(\ref{eq:Transmission}) in units of $e^2/h$. Using the HDHM (Eq.~\eqref{Hamiltonian}) as our Hamiltonian, we now present results of DC conductance for different sets of system parameters (Fig.~\ref{fig:TransmissinClean}). The curves (except black triangles) in the left panels of Fig.~\ref{fig:TransmissinCleana} represent $T(E+n\hbar\Omega)$, i.e. the DC conductance in units of $e^2/h$ \emph{at energy} $E+n\hbar\Omega$, with different values of $n$. In other words, $T(E+n\hbar\Omega)$ is the contribution of the $n$th Floquet sideband to the DC conductance \emph{at a given quasienergy} $\mu=E\in[-\hbar\Omega/2,\hbar\Omega/2]$.

We take the wide-band approximation (WBA) for the self-energy of the leads, which is valid when the lead has a much larger bandwidth compared to that of the central system~\cite{Haug2008,Jauho1994,Lewenkopf2017,Velicky2010,Stefanucci2004}. Under this approximation, the self energies due to the leads are independent of energy $E$. More explicitly, these self-energy functions satisfy $\Sigma^L_{ij}\propto -\ii \delta_{x_i,0}\delta_{i,j} \Gamma^L/2$ and $\Sigma^{R}_{ij}\propto -\ii \delta_{x_i,N_x-1}\delta_{i,j} \Gamma^R/2$, where $\Gamma^{L/R}>0$ are constants taken to be the same as $J_x$. In Sec.~\ref{subsec:finiteBW}, we shall discuss the consequences if this assumption is lifted. Also, in all of our calculations, we have made sure that the number of Floquet sidebands $\nF$ is large enough. Specifically, $\nF=13$ is a truncation dimension that guarantees convergence in all our computational examples.

For the parameters used to compute Fig.~\ref{fig:TransmissinCleana}, the system admits five Floquet bulk bands, as shown in Fig.~\ref{fig:TransmissinCleanb}. In the first gap above the middle band, there are four Floquet edge states, with a net gap chirality of $2$.  Within this quasienergy gap, the transmissions of each Floquet sideband (curves other than black triangles) form well-defined plateaux. The fact that these plateaux are not quantized can be understood as follows \cite{Farrell2015,Farrell2016}. Each Floquet sideband $n$, corresponding to an incoming state at \emph{energy} $E+n\hbar\Omega$, has a non-unity overlap with the Floquet edge state at \emph{quasienergy} $\epsilon=E$ hosted by the central system, leading to a transmission smaller than the expected number of edge states. This understanding \cite{Farrell2015,Farrell2016} provides an intuitive motivation to the so-called Floquet sum rule~\cite{Kundu2013}, a transport feature unique to Floquet topological phases that is not yet well recognized.

More concretely, the Floquet sum rule dictates that, if one observes gapless edge modes in the Floquet spectrum, then quantized transmission \emph{at a quasienergy} $\epsilon_0=E_0\in[-\hbar\Omega/2,\hbar\Omega/2]$ can be recovered by summing over the transmissions \emph{at energies} $E+n\hbar\Omega$ due to different Floquet sidebands:
\begin{equation}
T(\epsilon_0) = \sum_{n\in\mathbb{Z}}T(E_0+n\hbar\Omega). \label{eq:sumrule}
\end{equation}
As shown in Fig.~\ref{fig:TransmissinCleana}, we found $T(\epsilon_0)=4,4,2$ for $\epsilon_0$ in the first, second and third quasienergy band gaps above the middle Floquet band, an observation fully consistent with the Floquet gap features (such as the number of edge states) shown in Fig.~\ref{fig:TransmissinCleanb}.  In addition to several studies regarding the Floquet sum rule in Floquet quantum spin Hall insulators~\cite{Farrell2015,Farrell2016} and Floquet Chern insulators~\cite{Fruchart2016}, here we confirm it in  Floquet quantum Hall insulators. Furthermore, in the first and third quasienergy gaps above $\epsilon=0$ in Fig.~\ref{fig:TransmissinCleanb}, we observe counter-propagating gapless edge modes at the same edge~\cite{Zhou2014,Zhao2014,Derek2014}, which are unique to Floquet topological phases \cite{footnote2}. The corresponding DC conductance calculations (Fig.~\ref{fig:TransmissinCleana}) confirm that these counter-propagating edge modes also adhere to the Floquet sum rule.

%


We note that experimental observation of the above-confirmed Floquet sum rule is not so straightforward. One needs to prepare different initial states at chemical potentials $E+n\hbar\Omega$ to execute different experimental runs in order to measure the contributions from all Floquet sidebands to the transport at quasienergy $\epsilon=E$. Nevertheless, it would be of great interest to experimentally confirm the results in Figs.~\ref{fig:TransmissinCleanc} and \ref{fig:TransmissinCleane}.  There, one sees that at suitable system parameters, the DC conductances of Floquet edge states (after applying the Floquet sum rule) can reach $5e^2/h$ and even $8e^2/h$, which surpass the largest edge state conductance ever realized experimentally in undriven systems.  By contrast, for transport situations where the chemical potential is fixed (not allowed to adjust), the Floquet edge state transport is unlikely to be substantially larger than static systems, as indicated by the individual sidebands (curves other than black triangles) in Figs.~\ref{fig:TransmissinCleana},~\ref{fig:TransmissinCleanc} and~\ref{fig:TransmissinCleane}. The robustness of such Floquet edge state transport will be studied in the next subsection.

\subsection{\label{subsec:robust}Topological protection of Floquet edge currents}
One advantage of topological edge state transport is its robustness against {structural imperfections. While this has been demonstrated by the conductance plateaux in disordered quantum Hall samples~\cite{Klitzing1980}, the robustness of Floquet topological systems is yet to be confirmed experimentally.}

{Here we numerically study the response of gapless Floquet edge modes to disorder and defects. Thanks to the richness of the harmonically-driven Hofstadter model, we are able to realize: (i) co-propagating, (ii) counter-propagating, (iii) symmetry-restricted gapless edge states, and probe their robustness with the tools we developed.}

\begin{figure}[H]
\subfloat[]{
\includegraphics[width=.25\textwidth]{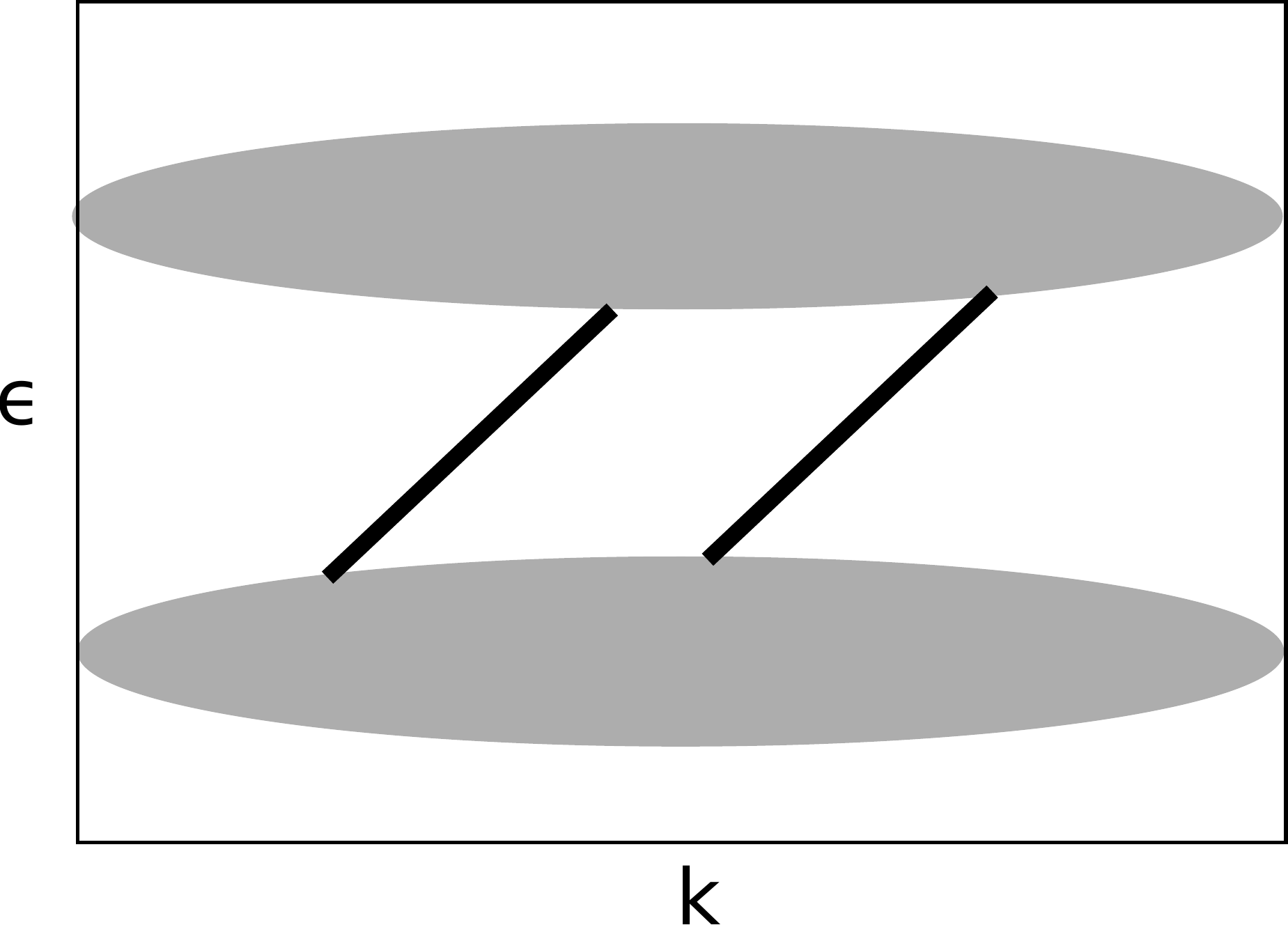}
}
\subfloat[]{
\includegraphics[width=.25\textwidth]{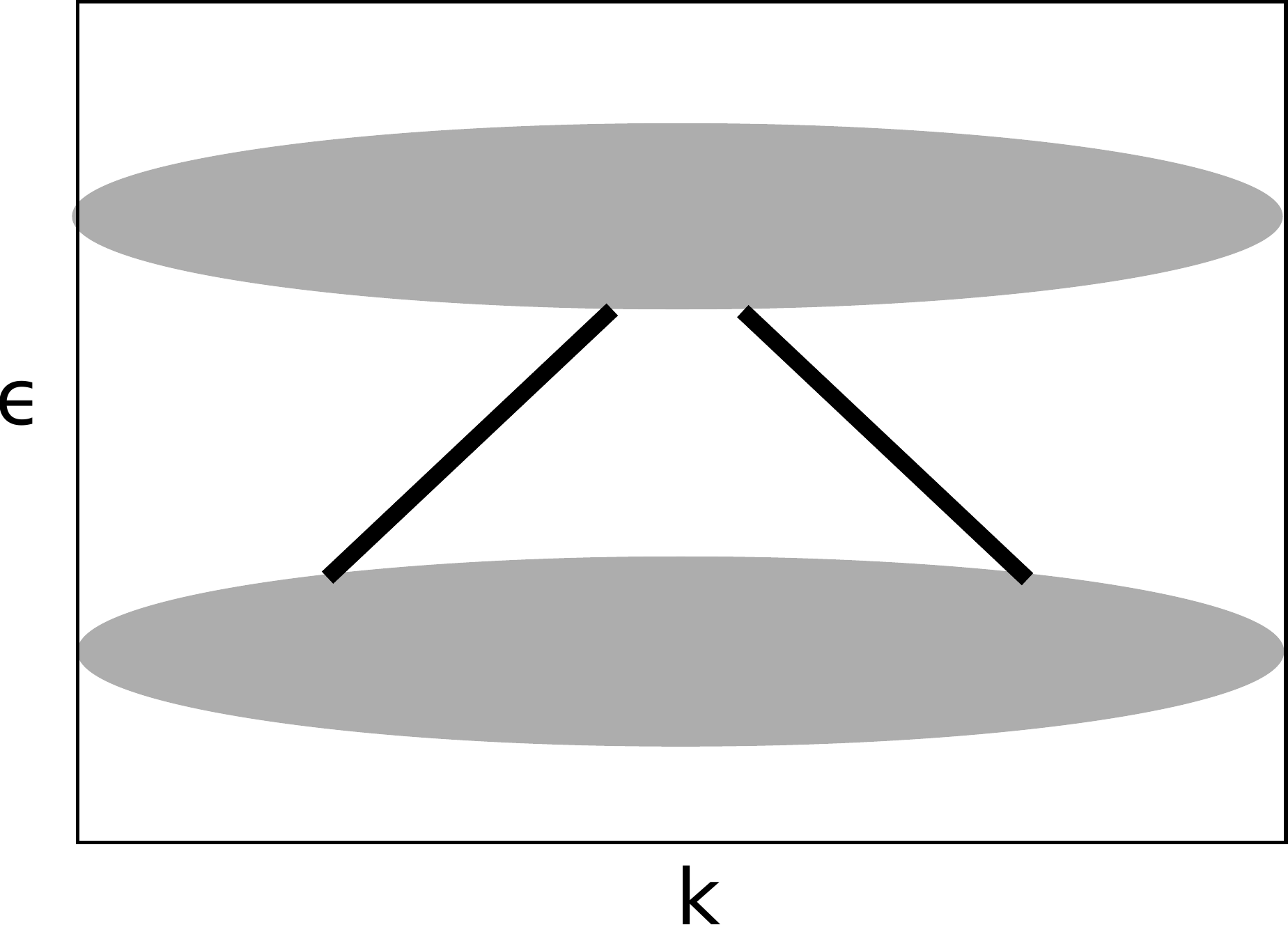}
}
\subfloat[]{
\includegraphics[width=.5\textwidth]{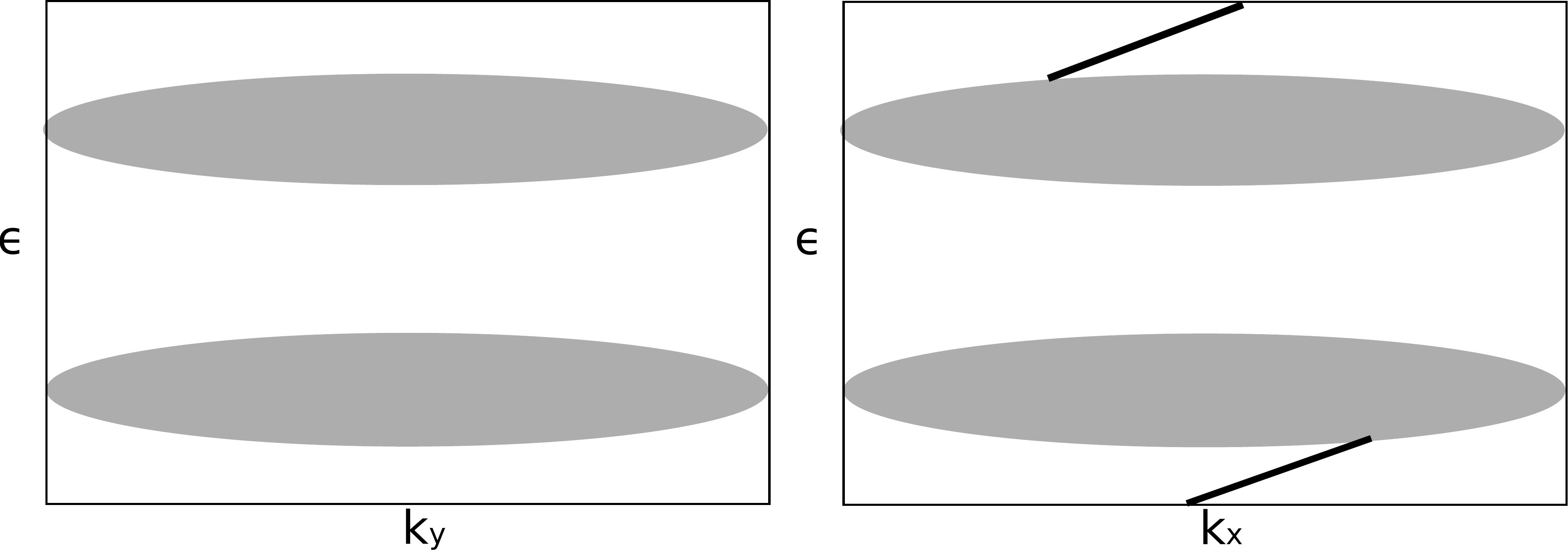}
}
\caption{ Schematic of three different types of edge states. The open boundary condition is taken along only one direction. Shaded regiones represent bulk bands. Lines represent co-propagating edge bands in panel (a), counter-propagating edge bands in panel (b), and symmetry-restricted gapless edge bands in panel (c). In panel (c), only when the open (periodic) boundary condition is taken along $y$-direction ($x$-direction) does one observe the edge states.} \label{fig:schematicEdge}
\end{figure}

{Before presenting the numerical results, we briefly describe each type of edge modes. The first kind, co-propagating Floquet edge states, are chiral in the following sense: under a stripe geometry, each edge only hosts states dispersing in a fixed direction. They have a modified bulk-edge correspondence to the nontrivial topology of the bulk bands \cite{Titum2015} as there may exist anomalous winding edge states in Floquet systems. The second type, counter-propagating edge modes, are as the name suggests: under a stripe geometry, each edge hosts states that are moving forward and backward. If the chirality (number of left movers minus number of right movers) is zero, unlike the spin-locked helical edge states in quantum spin Hall effect \cite{Zhang2010}, one generally does not expect protection of these counter-propagating edge modes against disorder. The third kind, dubbed symmetry-restricted edge modes, are gapless but only exist when the stripe is taken in a particular way. They may be co- or counter-propagating, but in our model the ones we found are counter-propagating. A schematic illustrating the Floquet spectra in presence of these edge modes is given in Fig.~\ref{fig:schematicEdge}.}

{
Since these three types of edge modes are not mutually exclusive (i.e. two of them may simultaneously exist at different quasienergies for the same system parameters), we organize the following discussions by the type of imperfections (disorder or defect). Our computational studies are summarized in Table~\ref{table:stability}.}

\subsubsection{Response to on-site disorder}
We introduce disorder to the HDHM by assigning to each site a random on-site potential. The resulting Hamiltonian of the disordered central system is given by:
\begin{equation}
{H}_{\textrm{dis}}(t) = {H}(t) + \sum_i \xi_i c_i^\dag c_i.
\end{equation}
The term $\sum_i \xi_i c_i^\dag c_i$ changes the potential on each site $i$ by $\xi_i$, where ${\xi_i}$ is a random number uniformly distributed in $[-W,W]$ {, where $W$ is the disorder strength.}

Consider first the case in which the Floquet spectrum of the HDHM without disorder is shown in Fig.~\ref{subfig:3edge5bandClean}, where there are three co-propagating chiral edge states in the first gap above the middle Floquet band, and a pair of counter-propagating edge states winding around quasienergies $\epsilon=\pm\hbar\Omega/2$. Such counter-propagating edge states are unique to Floquet systems~\cite{Lababidi2014,Derek2014}. To investigate their robustness to disorder, we calculate the transmission coefficients and T-LDOS at different quasienergies.

\begin{figure}
\subfloat[\label{subfig:DOSpristine0}]{
\includegraphics[width=0.25\columnwidth]{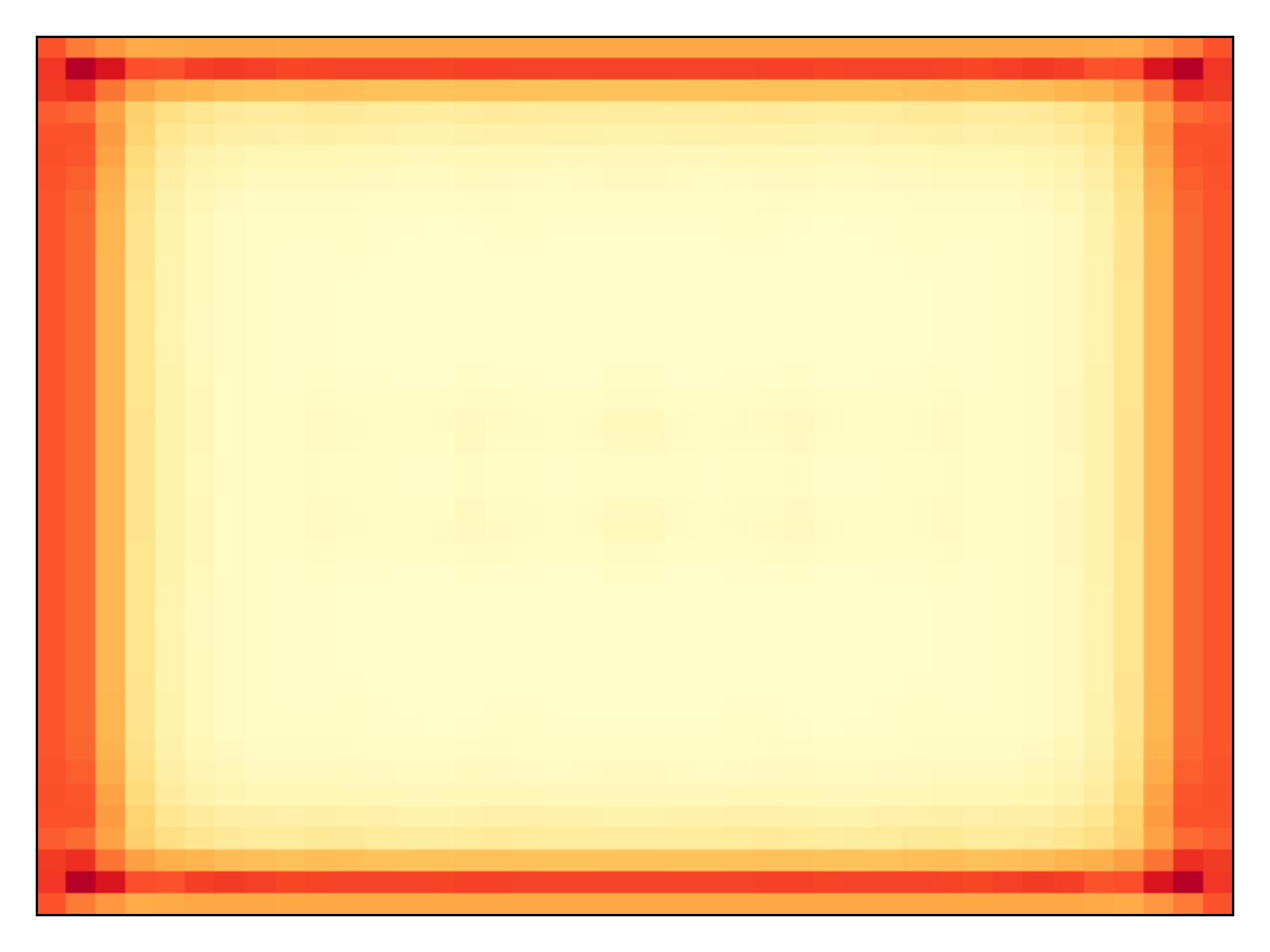}
}
\subfloat[\label{subfig:DOSdirty0}]{
\includegraphics[width=0.25\columnwidth]{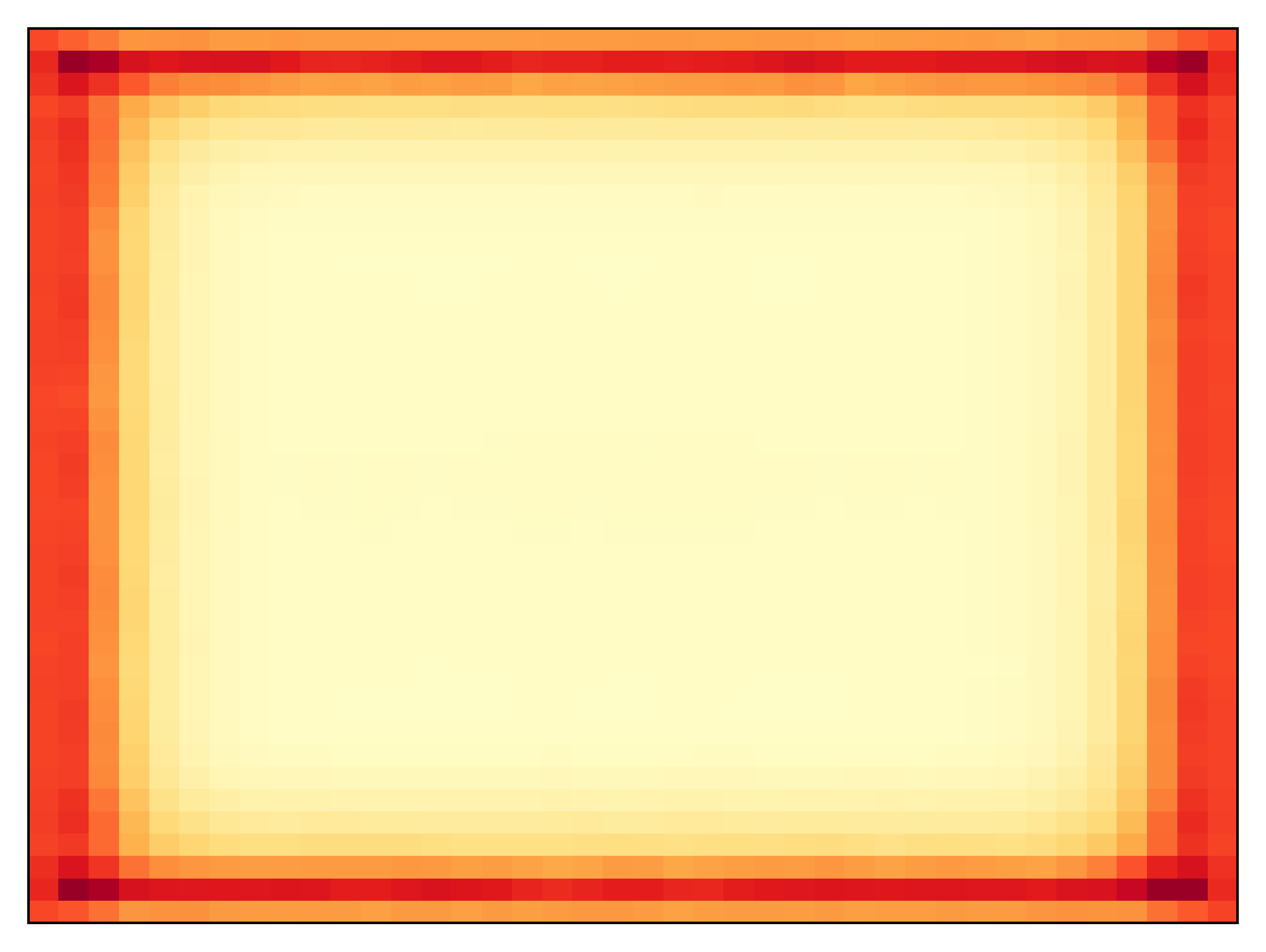}
}
\subfloat[\label{subfig:DOSpristine1}]{
\includegraphics[width=0.25\columnwidth]{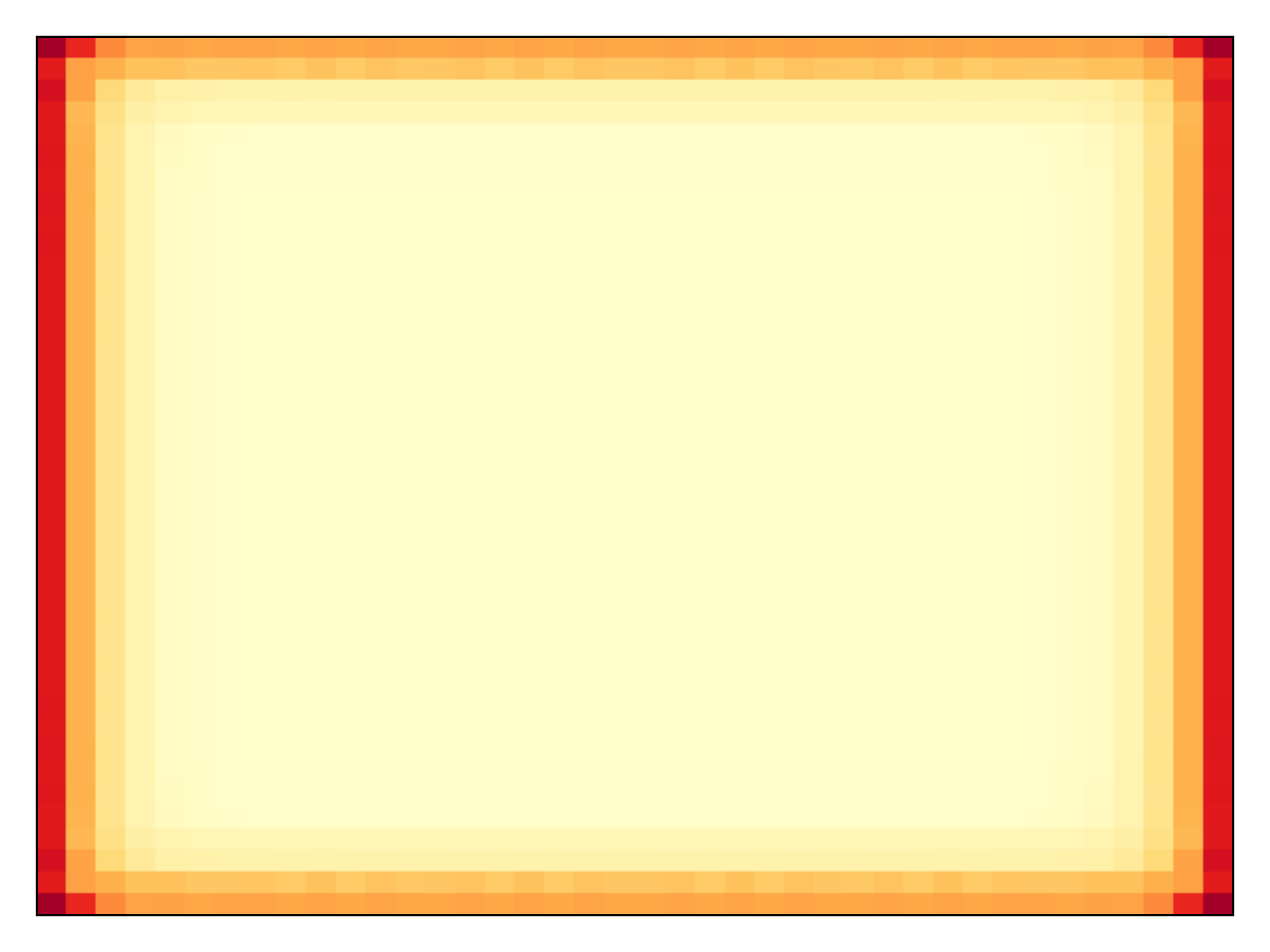}
}
\subfloat[\label{subfig:DOSdirty1}]{
\includegraphics[width=0.25\columnwidth]{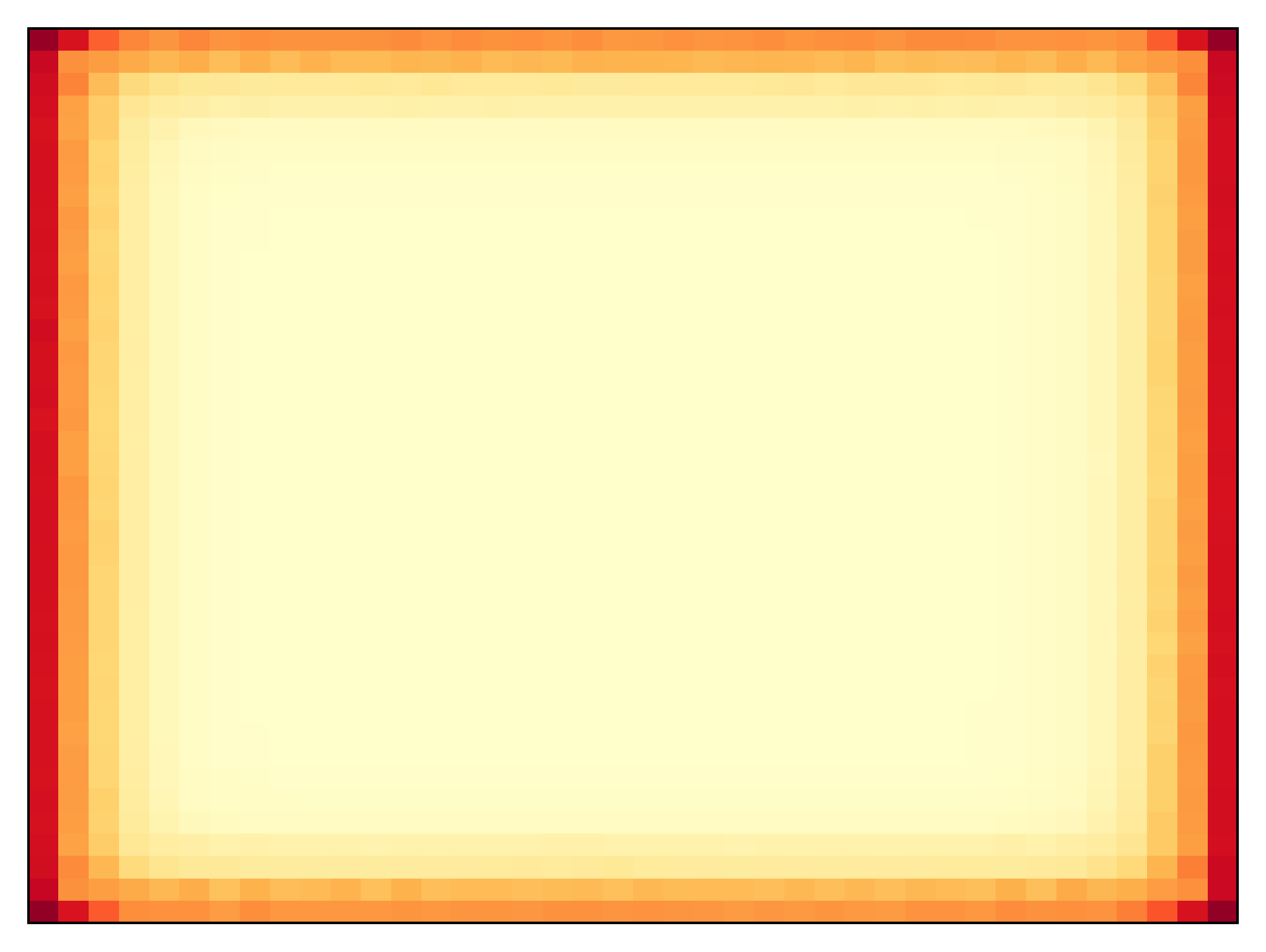}
}
\\
\centering
\subfloat{
\includegraphics[width=0.35\columnwidth]{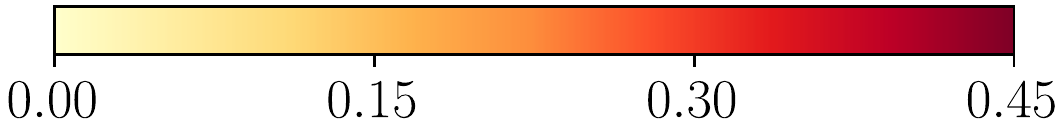}
}
\hspace{3cm}
\subfloat{
\includegraphics[width=0.35\columnwidth]{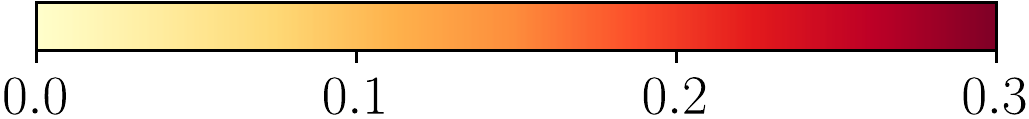}
}
\caption{(color online) Time-averaged LDOS at quasienergy $\epsilon=0.25$ for panels (a), (b), and $\epsilon=1.3$ for panels (c), (d), after summing over the contributions from Floquet sidebands as in the Floquet sum rule. {Panels (a) and (c)} show results for pristine central systems, {whereas panels (b) and (d)} show results for disordered central systems, each averaged over {$200$} different disorder realizations. The disorder is implemented via onsite random potentials, which penetrates all four edges by a depth of three layers each. The disorder strength is $W=0.7$. The system size is $N_x=N_y=40$. Other system parameters are $J_y=1.5,\alpha=1/5,\Omega=\pi$ and $s=0.7$.}
\label{fig:DOSdisord}
\end{figure}
\begin{figure}
\subfloat[\label{subfig:3edge5bandClean}]{
\includegraphics[height=5cm,width=0.5\columnwidth]{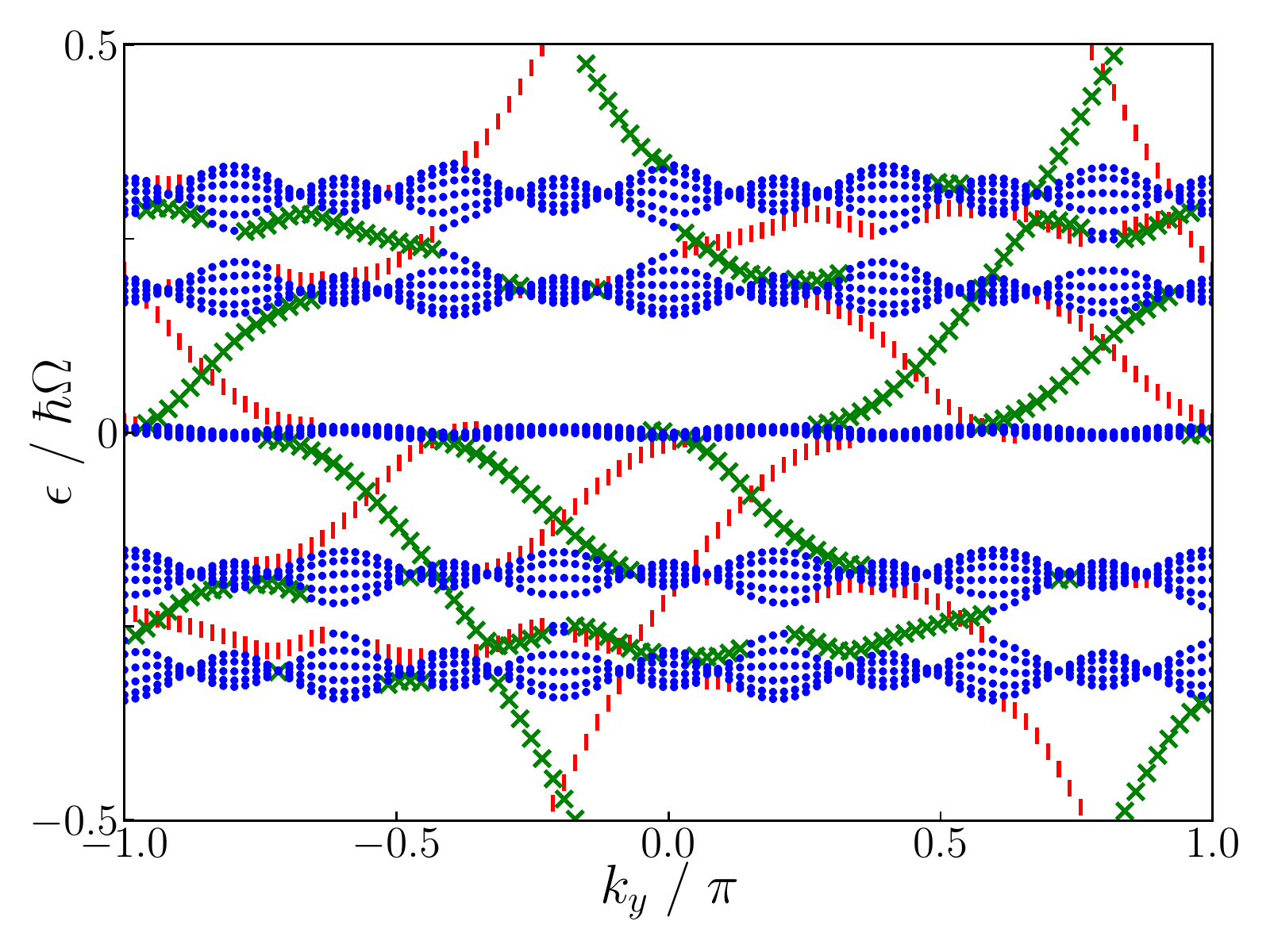}
}
\subfloat[\label{subfig:3edge5bandDisord}]{
\includegraphics[height=5cm,width=0.5\columnwidth]{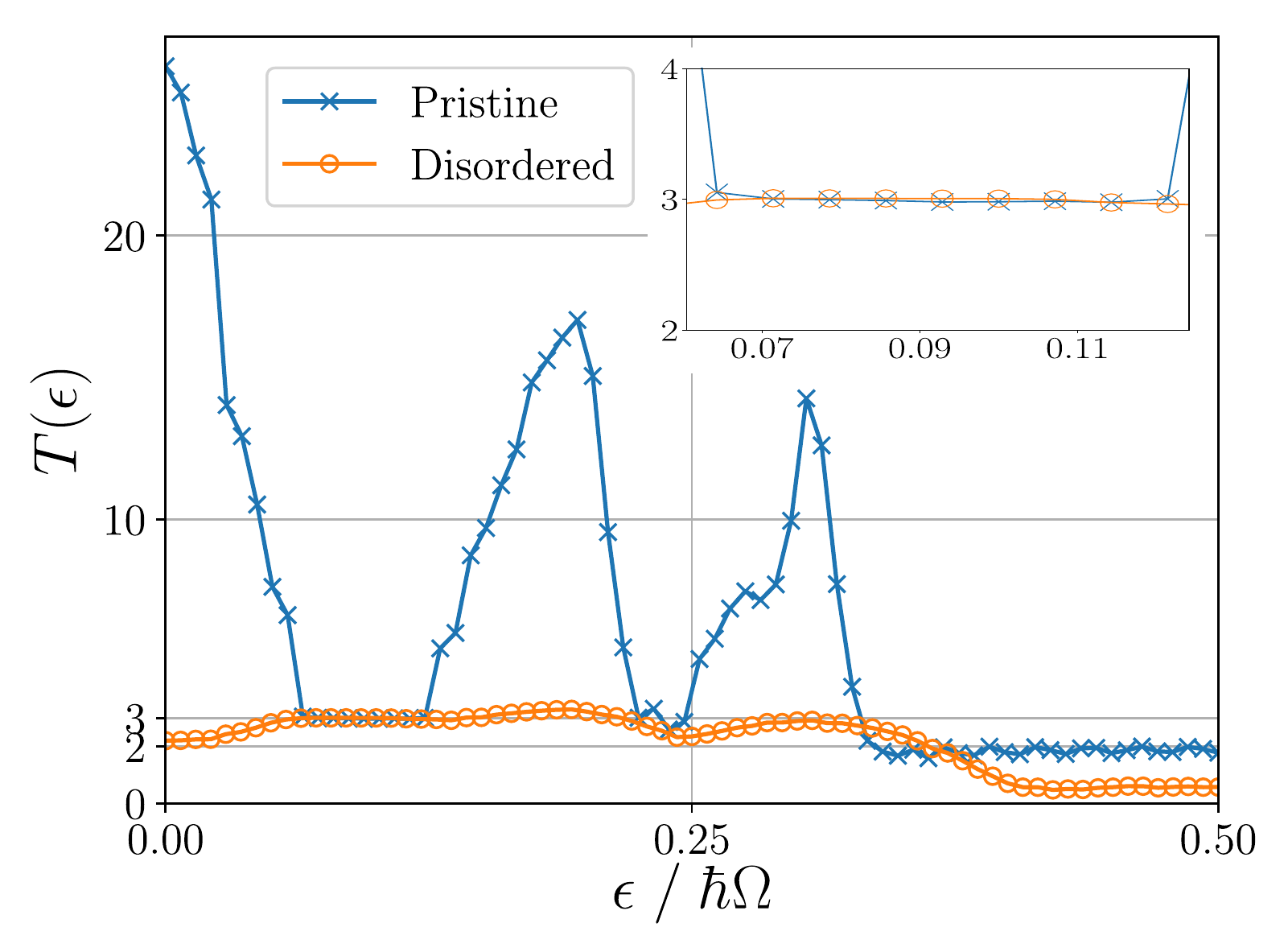}
}
\caption{(color online) Left panel: Floquet spectrum of the HDHM with PBC (OBC) along $y$($x$)-direction. Red $``|"$s and green crosses denote states localized at left and right edges of the sample. Right panel: transmission coefficient $T(\epsilon)$ (after applying the Floquet sum rule) at different quasienergies $\epsilon$ in the upper half of the Floquet quasienergy Brillouin zone for both pristine and disordered samples. The disorder strength {$W=1$}. The results for disordered sample are averaged over {$100$} different disorder realizations. Other system parameters are the same as those shown in the caption of Fig.~\ref{fig:DOSdisord}.}
\end{figure}
\begin{figure}
\subfloat[\label{subfig:copropaBandstruc}]{
\includegraphics[height=5cm,width=0.25\columnwidth]{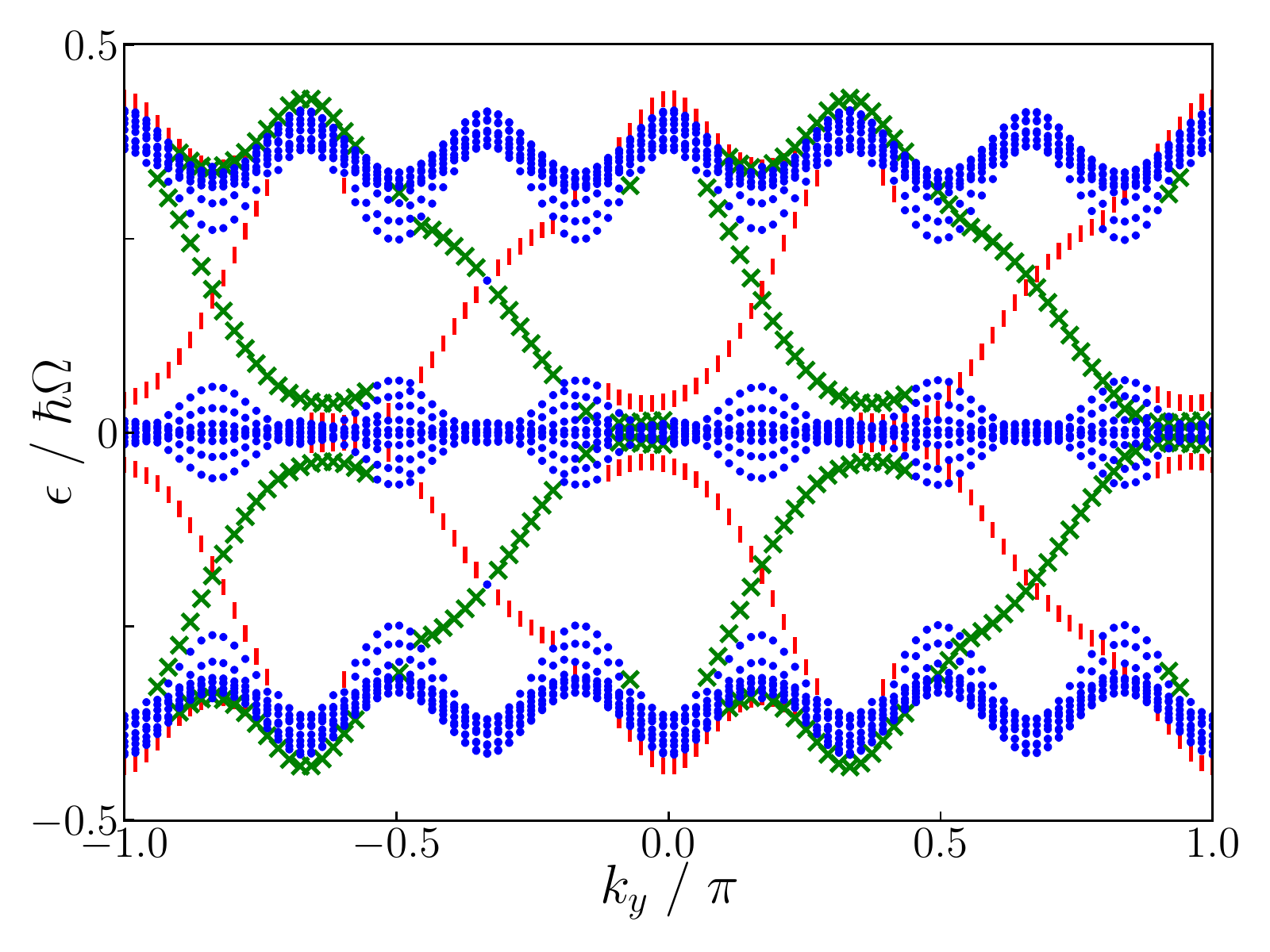}
}
\subfloat[\label{subfig:copropaBandstruc2}]{
	\includegraphics[height=5cm,width=0.25\columnwidth]{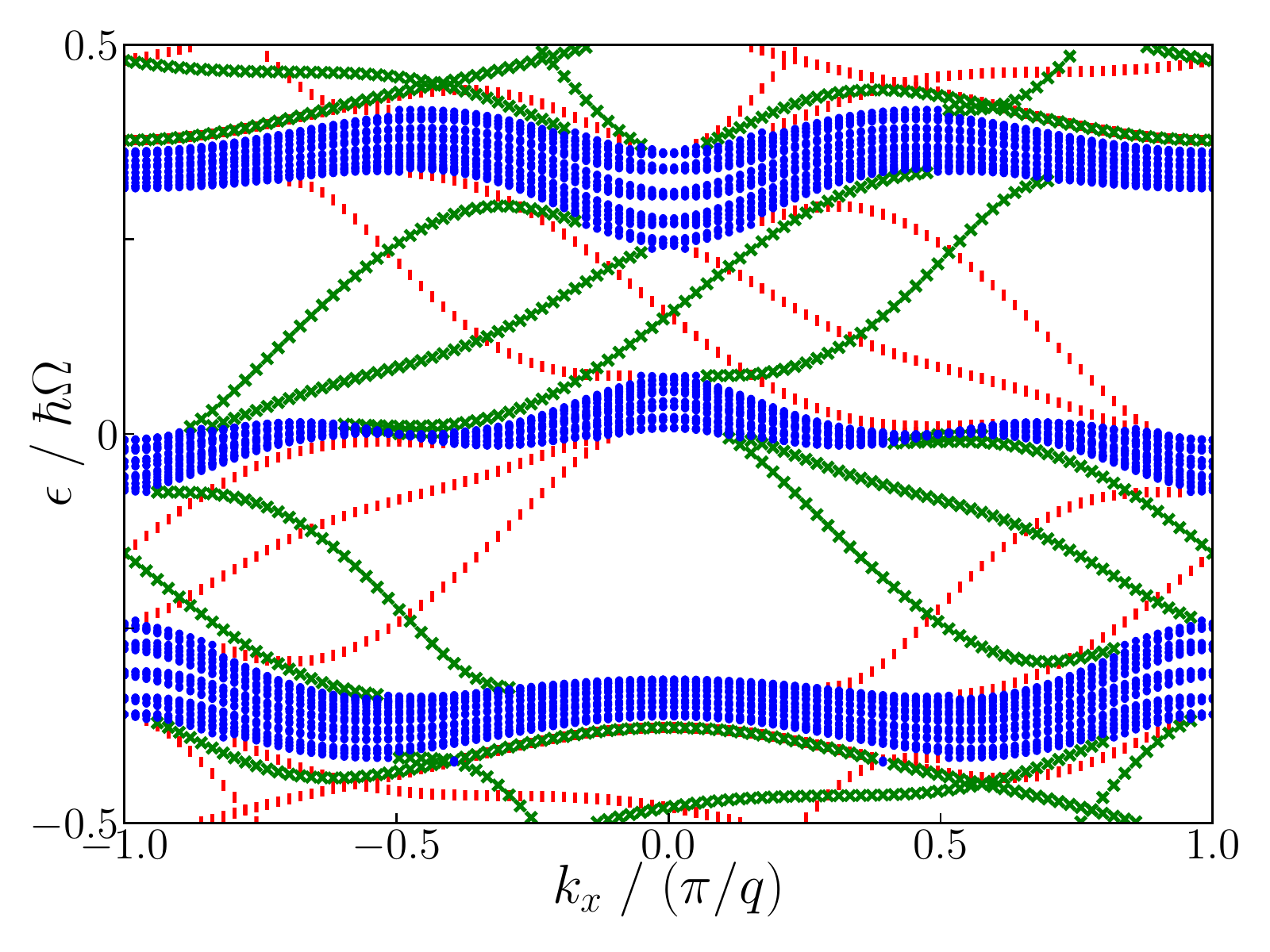}
}
\subfloat[\label{subfig:copropaTransmission}]{
\includegraphics[height=5cm,width=0.5\columnwidth]{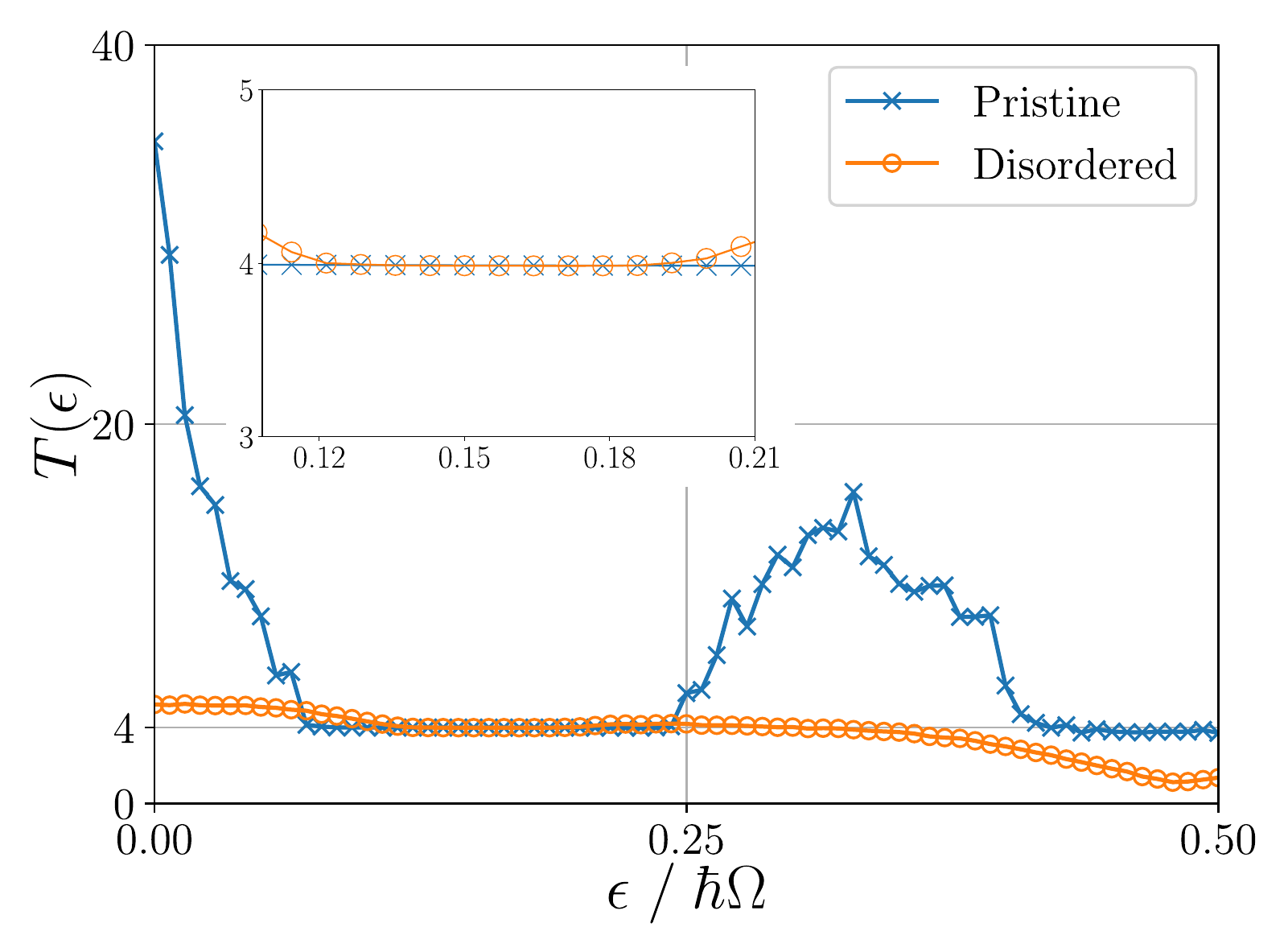}
}
\caption{ (color online) Panel (a): Floquet spectrum of the HDHM with PBC (OBC) along $y$($x$)-direction. Red $``|"$s and green crosses denote states localized at left and right edges of the sample. Panel (b): Floquet spectrum of the HDHM with the same parameters, but with PBC (OBC) along $x$($y$)-direction. Panel (c): transmission $T(\epsilon)$ (after applying the Floquet sum rule) at different quasienergies $\epsilon$ in the upper half of the Floquet quasienergy Brillouin zone for both pristine and disordered samples. The disorder strength $W=1$. The results for disordered sample are averaged over {$100$} different disorder realizations. The other system parameters are $J_y=1.25,\alpha=1/3,\Omega=\pi/2,s=0$ and $N_x=N_y=45$.}
\label{fig:copropa}
\end{figure}

From the T-LDOS for a pristine lattice depicted in Figs.~\ref{subfig:DOSpristine0} and \ref{subfig:DOSpristine1}, we see that the chiral modes at $\epsilon=0.25$ and the counter-propagating modes at $\epsilon=1.3$ are indeed localized at the edges. Upon the introduction of disorder---even at a strength comparable to the size of the spectral gap---the T-LDOS still retain their edge features. At first glance, this seems to suggest that both types of edge states are favorably robust against disorder.  However, if we inspect the DC conductance, where, as shown in Fig.~\ref{subfig:3edge5bandDisord} for $\epsilon/(\hbar\Omega) \sim 0.4-0.5$, the conductance plateau due to counter-propagating edge states is appreciably pushed down by disorder from being quantized at $2$. Thus from the perspective of conductance quantization, counter-propagating Floquet edge states are actually not robust. Indeed, the net winding number at quasienergies $\epsilon=\pm\hbar\Omega/2$ is zero. Hence the corresponding edge states are topologically trivial and are generally not expected to resist disorder. This is in contrast to the anomalous Floquet edge modes with non-vanishing winding numbers studied in Ref.~\cite{Titum2016}, which are robust against disorder. From Fig.~\ref{subfig:3edge5bandDisord} one can draw two more observations. First, the plateau due to co-propagating chiral edge states remains intact under disorder, akin to the static quantum Hall insulators. Of topological origin, this robustness is related to the Chern numbers of the Floquet bands which cannot change without closing the spectral gaps. Second, the very high conductance peaks are suppressed when disorder is introduced into the pristine sample. This is evidence that these peaks originate from Floquet bulk states and as such they are not topologically protected. This remarkable response of bulk states to disorder should allow one to draw a line between conductance contributions from bulk and from topological edge states.

As a side note, we mention another type of Floquet edge states restricted by the symmetry of the system. One such example is presented in Figs.~\ref{subfig:copropaBandstruc} and~\ref{subfig:copropaBandstruc2} for a 3-band case, where counter-propagating edge modes appear in quasienergy band gaps around $\pm\hbar\Omega/2$, only if the periodic boundary condition is taken along the $x$-direction.  This is similar to graphene, where zero-energy edge states appear only in ribbons with zigzag edges~\cite{Fujita1996}. To study the robustness of transport contributed by such kind of edge states, we again calculate the DC conductance in the presence of disorder. As shown in Fig.~\ref{subfig:copropaTransmission}, there are two plateaux corresponding to the same quantized conductance $4e^2/h$. The one closer to the central Floquet band (due to co-propagating Floquet edge modes) is manifestly more robust against disorder than the one further away (due to symmetry-restricted Floquet edge modes).

For completeness, we also studied the transport property of topologically trivial gapped Floquet edge states, and found that they are indeed not robust to disorder and defects. More details are presented in App.~\ref{app:gapped}.

\subsubsection{Response to sample defects}
{A hallmark of quantum Hall insulators is the existence of chiral edge modes, which can move past defects located at sample boundaries and maintain their propagation directions without being scattered backward.} In photonic and phononic analog of Floquet topological insulators, the robustness of edge states against sample defects have been demonstrated in Refs.~\cite{Rechtsman2013,Fleury2016,Bandres2016}. Here we propose the local DC profile (Sec.~\ref{subsubsec:DCprofile}) as a tool to study the response of edge states to defects in Floquet quantum Hall insulators.  We introduce a defect to the system by removing all terms coupled to the defect from the system Hamiltonian. For example, the Hamiltonian of the HDHM with a single defect at site $d\equiv(x_d,y_d)$ is given by:
\begin{equation}
\begin{split}
H_{\mathrm{def}}&(t) = H(t) -  \big\{J_x c^\dag_d c_{d\pm \hatx} \\ & + J_y \left[s+\cos(\Omega t)\right] \ee^{\pm\ii 2\pi\alpha x_d }c^\dag_d c_{d\pm \haty} + \Hc \big\}.
\end{split}
\end{equation}

\begin{figure}
\subfloat[\label{fig:bandstructcounterpropa}]{
\includegraphics[height=5cm,width=.5\columnwidth]{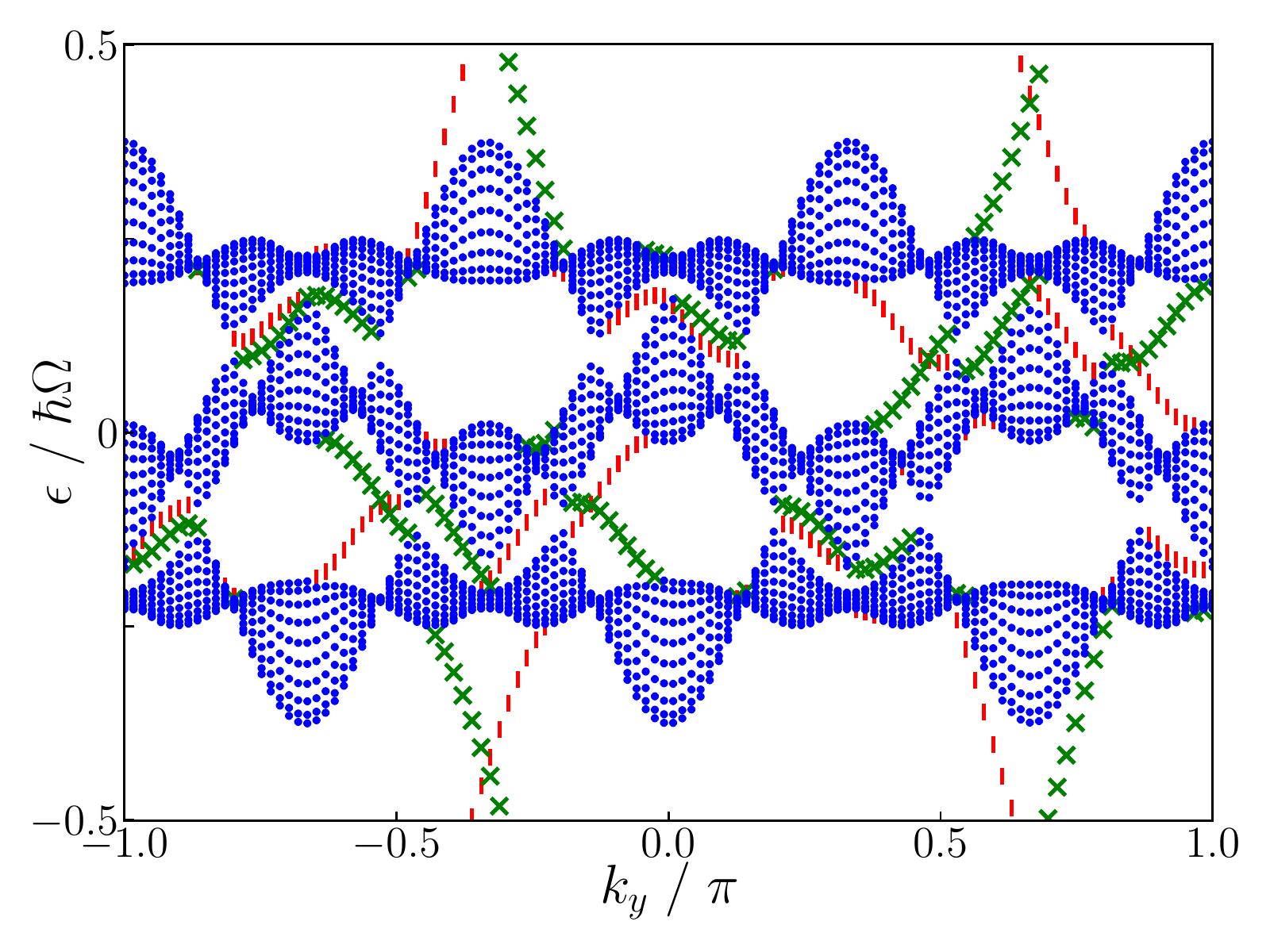}
}
\subfloat[\label{fig:transmDefectCounterpropa}]{
\includegraphics[height=5cm,width=.5\columnwidth]{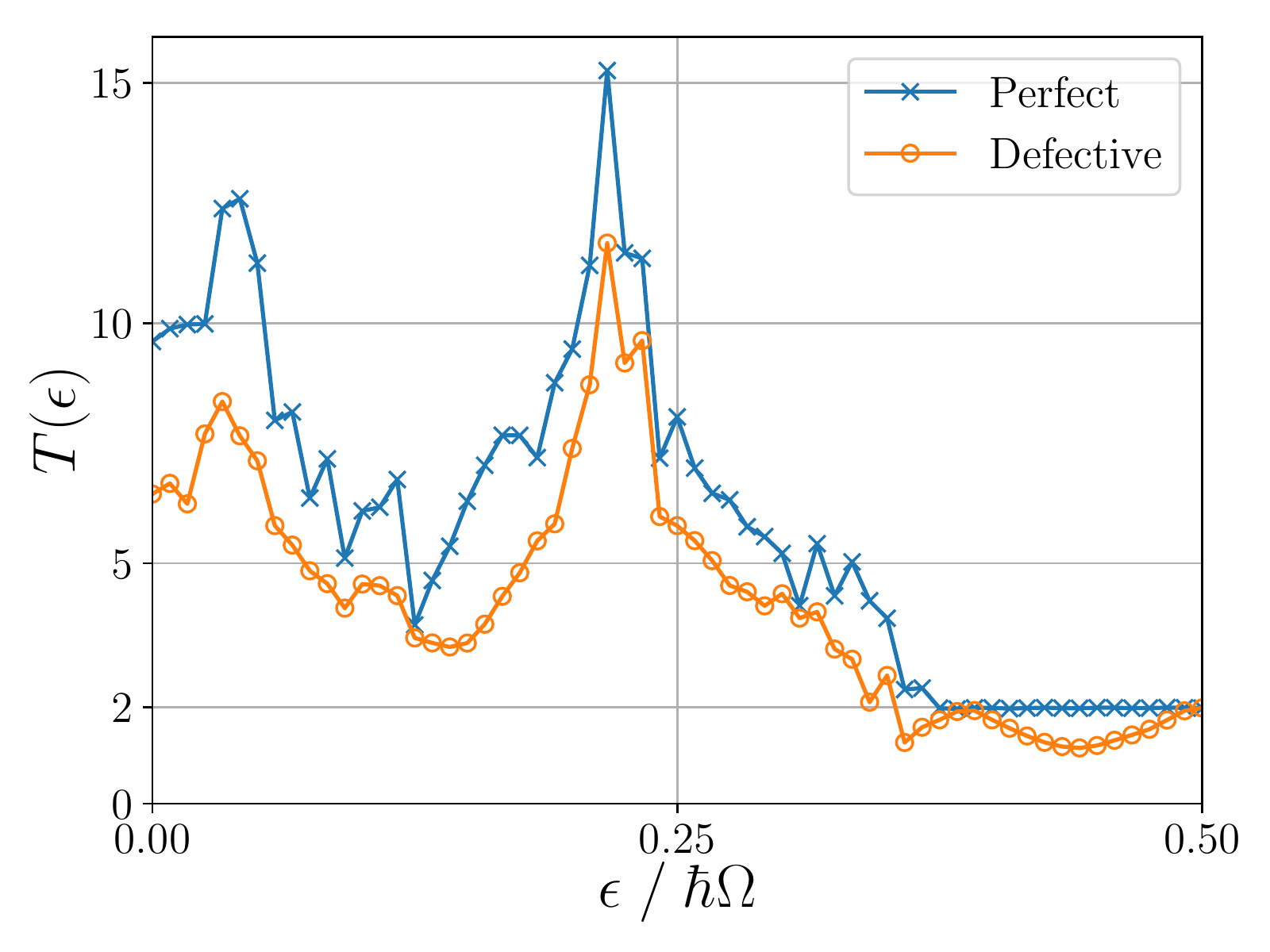}
}
\caption{(color online) Panel (a): Floquet spectrum of the HDHM with PBC (OBC) along $y$($x$)-direction. Red $``|"$s and green crosses denote states localized at left and right edges of the sample. Panel (b): transmission coefficient vs. quasienergy with the Floquet sum rule applied. The curve with crosses (circles) corresponds to the perfect (defective) sample shown in left (right) panels of Fig.~\ref{fig:DCprofileCounterpropa}. Other system parameters are chosen as $J_x=1,~J_y=1.6,~\alpha=1/3,~\Omega=\pi,~s=1$.}
\label{fig:bandstructtranscounterpropa}
\end{figure}

\begin{figure}
\subfloat[\label{subfig:perfectHarmonic-1}$n=-1$]{
\includegraphics[height=5cm,width=0.5\columnwidth]{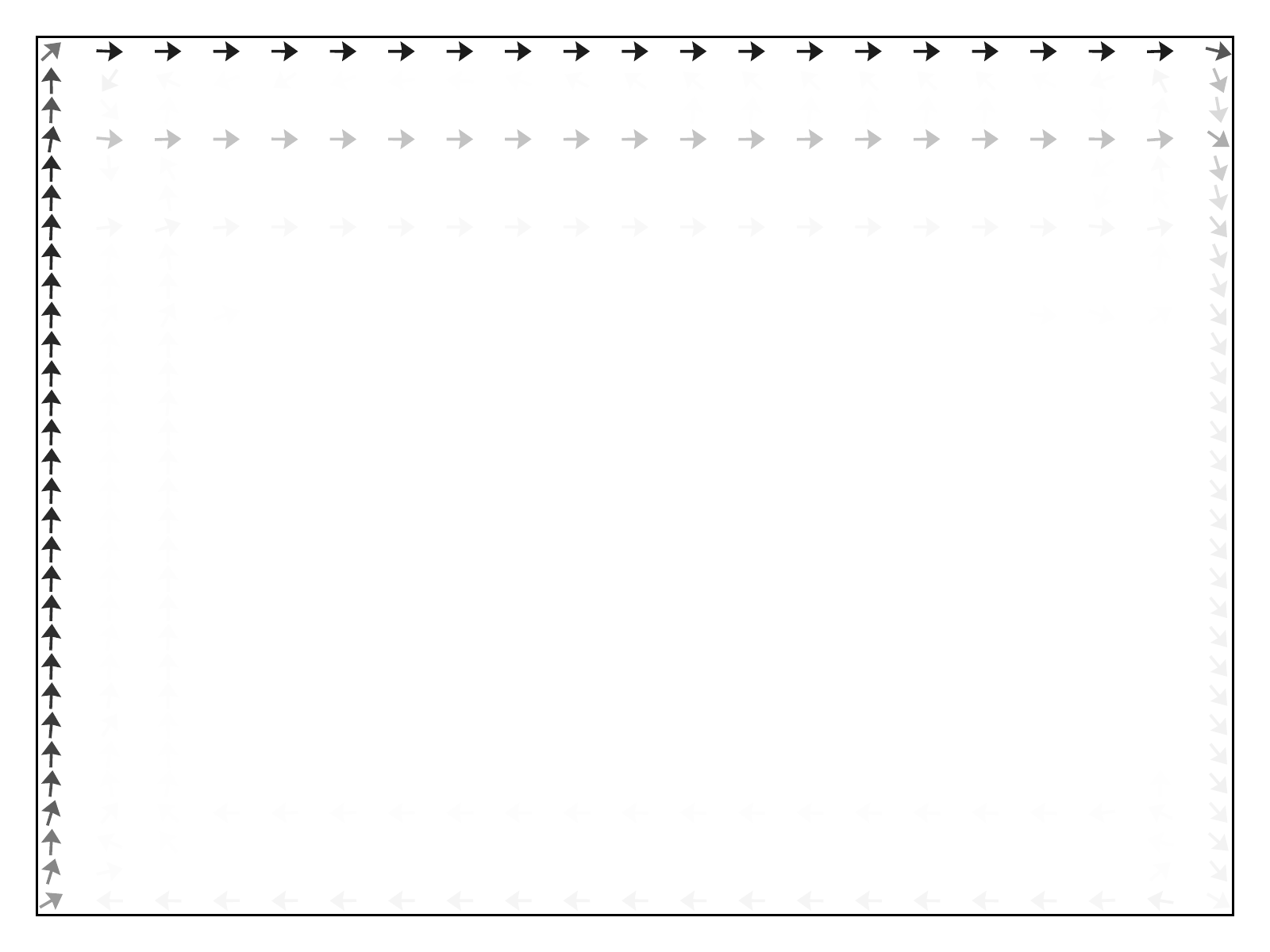}
}
\subfloat[\label{subfig:defectHarmonic-1}$n=-1$]{
\includegraphics[height=5cm,width=0.5\columnwidth]{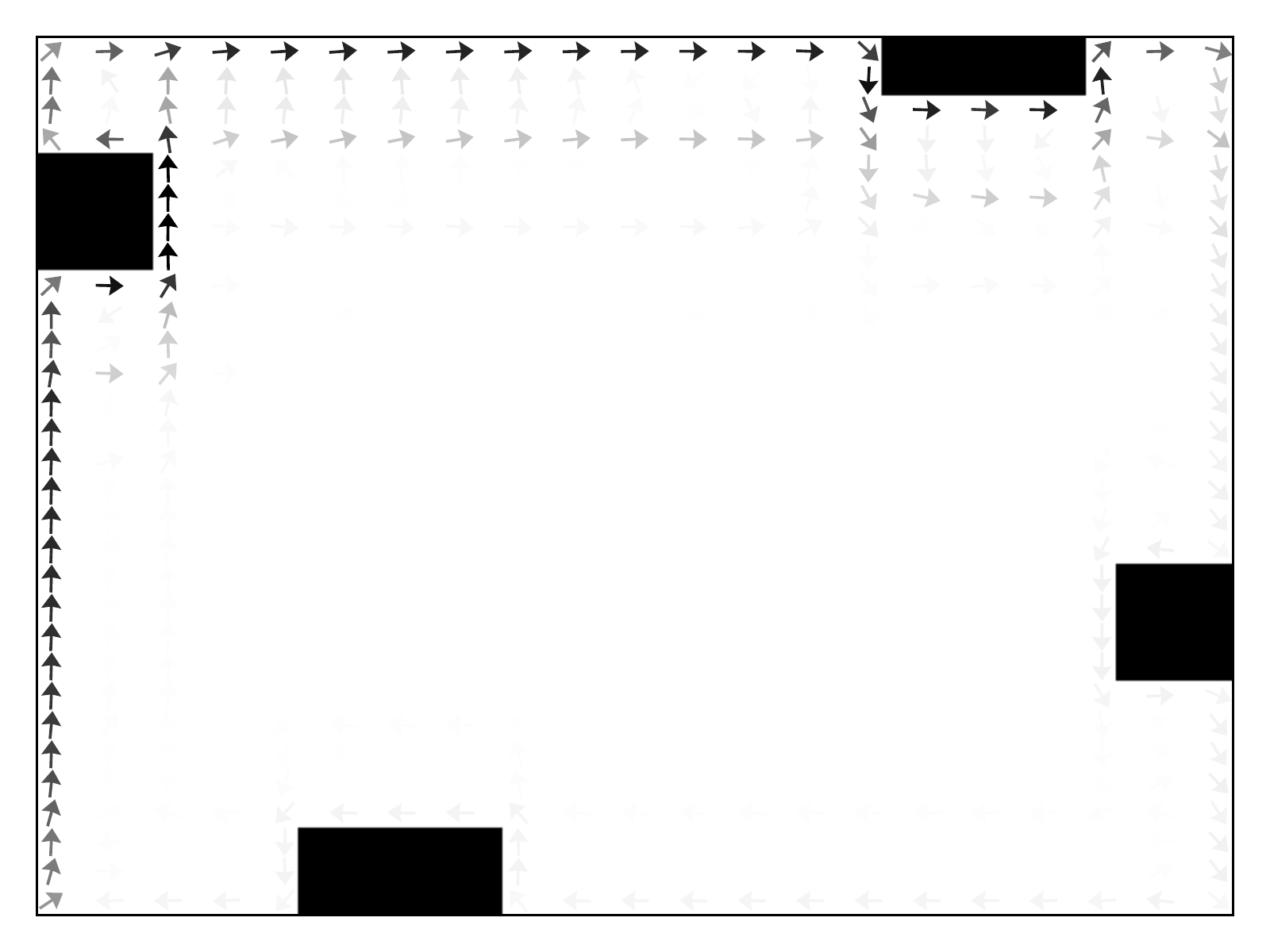}
}
\\
\subfloat[\label{subfig:perfectHarmonic0}$n=0$]{
\includegraphics[height=5cm,width=0.5\columnwidth]{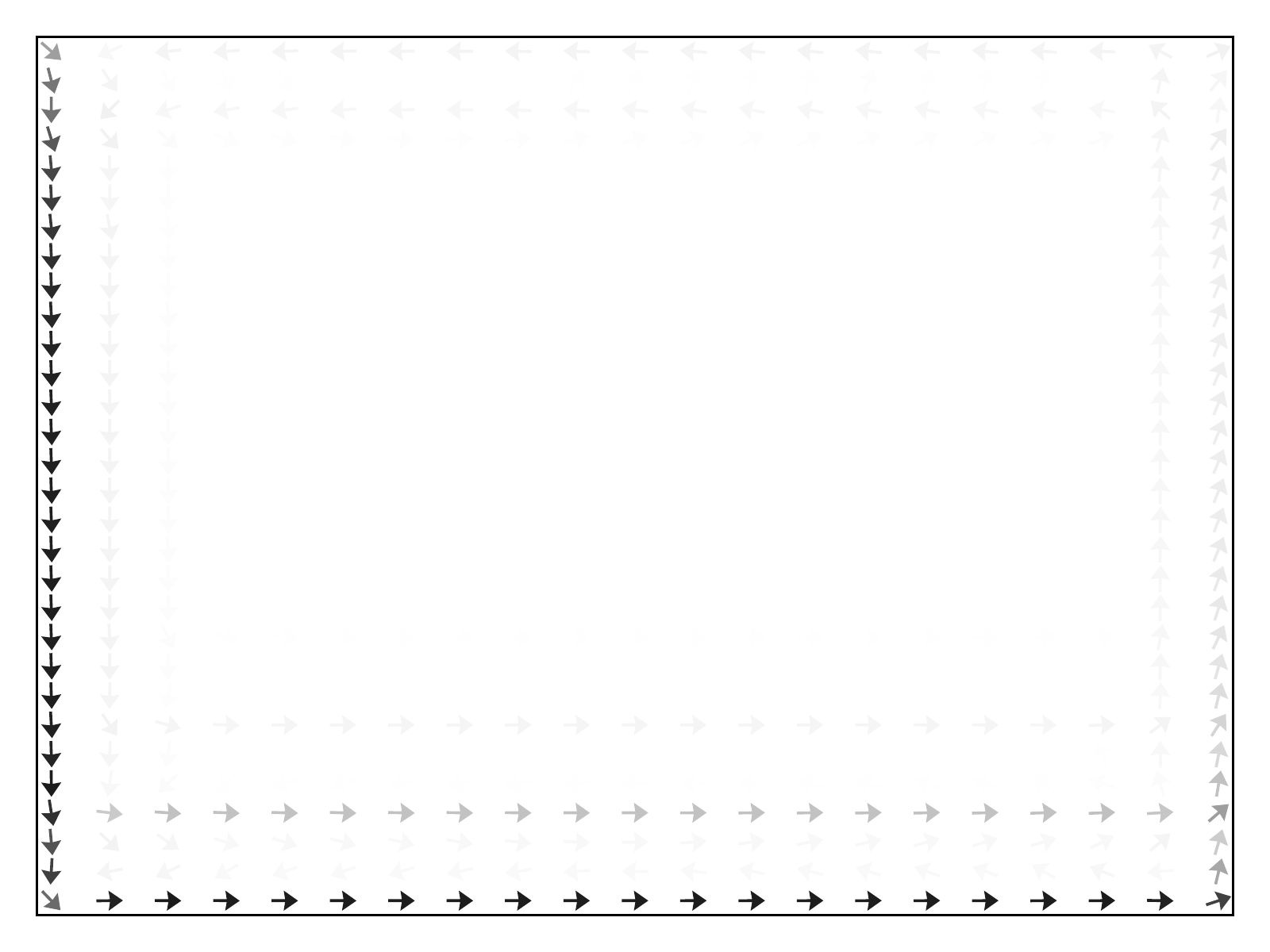}
}
\subfloat[\label{subfig:defectHarmonic0}$n=0$]{
\includegraphics[height=5cm,width=0.5\columnwidth]{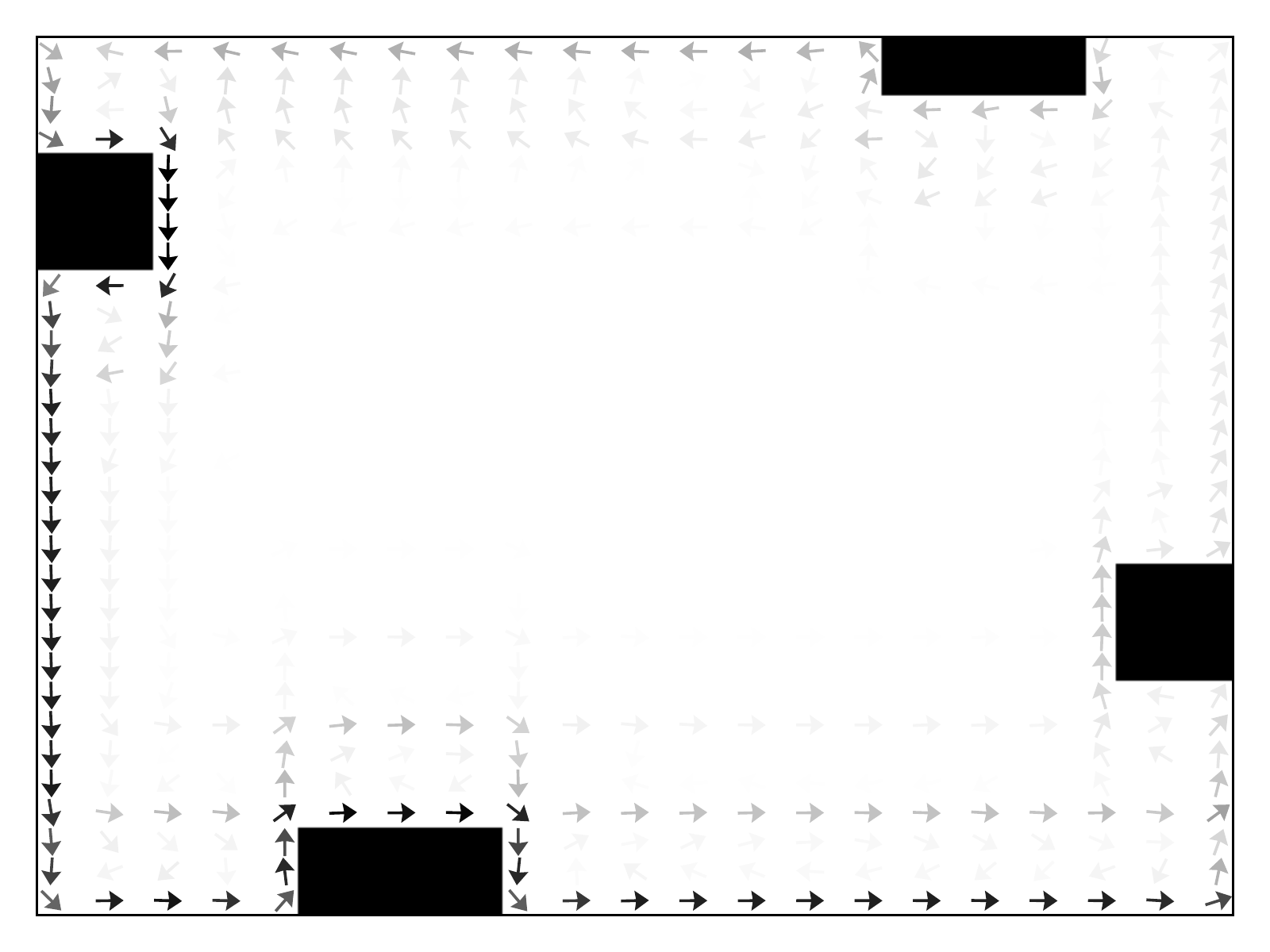}
}
\\
\subfloat[\label{subfig:perfectSumRule}Floquet sum rule applied]{
\includegraphics[height=5cm,width=0.5\columnwidth]{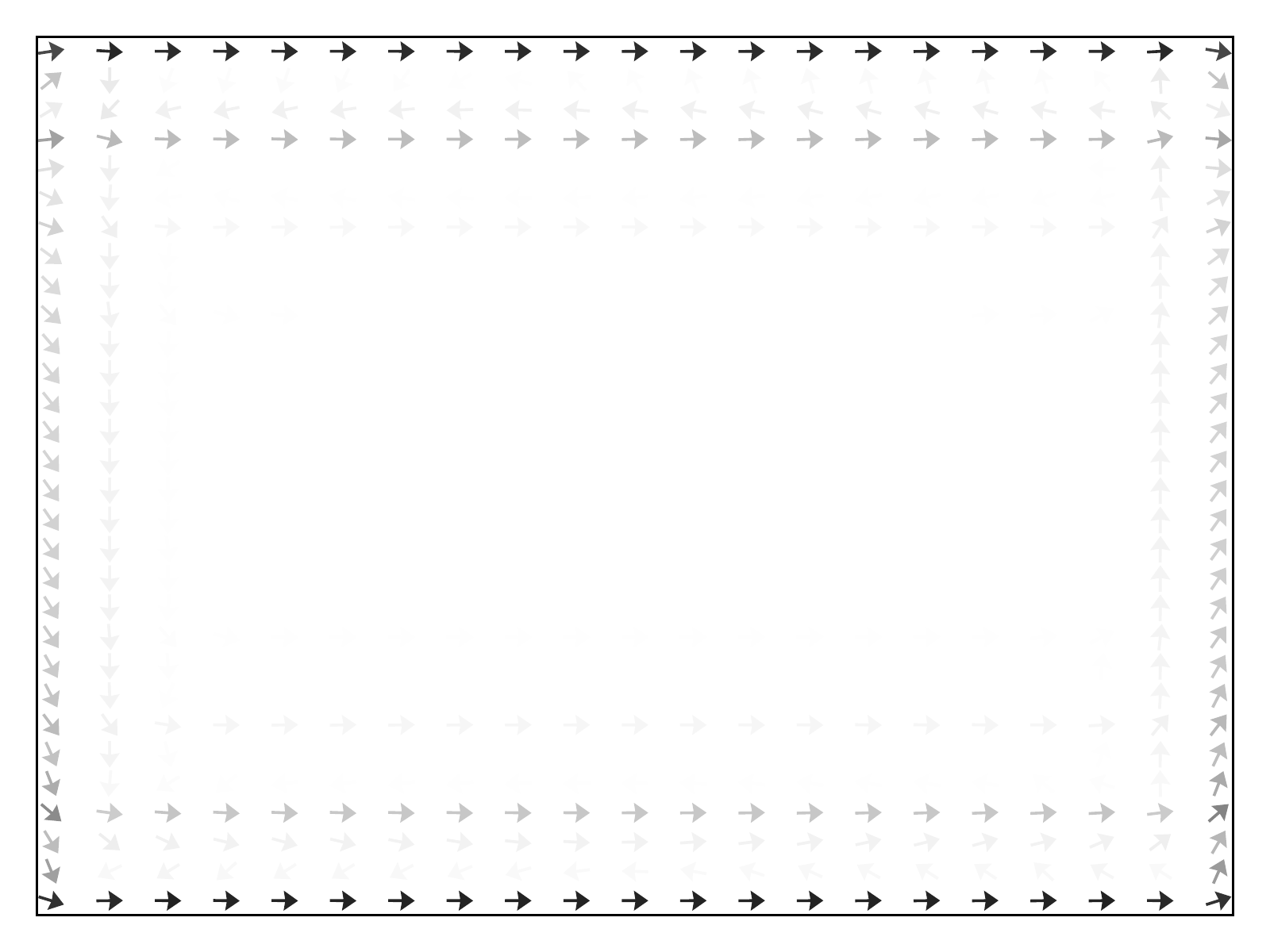}
}
\subfloat[\label{subfig:defectSumRule}Floquet sum rule applied]{
\includegraphics[height=5cm,width=0.5\columnwidth]{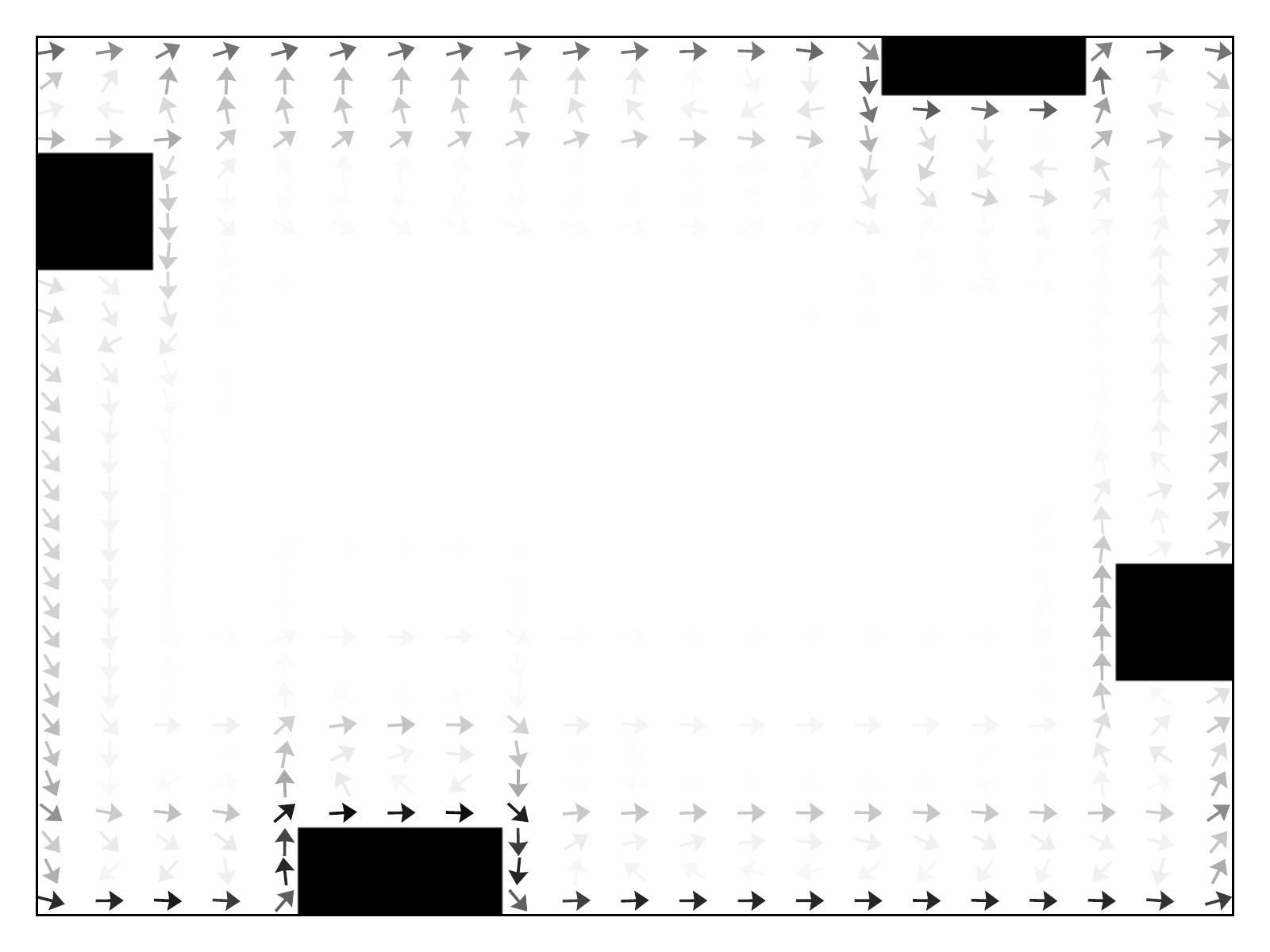}
}
\\
\centering
\subfloat{
	\includegraphics[width=0.35\columnwidth]{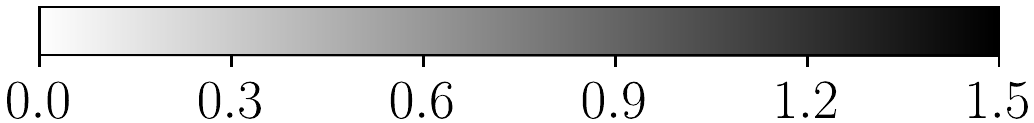}
}
\caption{Local DC profile for perfect (left panels) and defective (right panels) samples. Arrows point along the direction of local currents, whose magnitude is indicated by the color intensity as given in the colorbar (in units of $J_x/\hbar$). The system parameters are $E=1.5,\mu_L=1.51,\mu_R=1.49,N_x=41,N_y=30,J_y=1.6,\alpha=1/3,\Omega=\pi$ and $s=1$.}
 \label{fig:DCprofileCounterpropa}
\end{figure}
The local DC profile is then determined as follows: at each site of the lattice, evaluate Eq.~(\ref{eq:localcurrent}) and then represent the result as an arrow. Following the rationale behind the Floquet sum rule, this will be done for not just a single energy and chemical potential at $(E,\mu_L,\mu_R)$, but also for those that are at integer multiples of the driving frequency away, i.e., with chemical potentials at $(E+n\hbar\Omega,\mu_L+n\hbar\Omega,\mu_R+n\hbar\Omega)$.

We present the DC profile at energy $E=1.5+n\hbar\Omega$ in Fig.~\ref{fig:DCprofileCounterpropa}, with the same system parameters as in Fig.~\ref{fig:bandstructtranscounterpropa}. The chemical potentials of the left and right leads are set at $\mu_L=1.51+n\hbar\Omega$ and $\mu_R=1.49+n\hbar\Omega$, both within the Floquet band gap. We introduce small blocks of defects to our sample as shown in Fig.~\ref{fig:DCprofileCounterpropa}.  In Figs.~\ref{subfig:perfectHarmonic-1} and \ref{subfig:perfectHarmonic0}, we see that the local currents due to harmonics $n=-1$ and $n=0$ flow mainly along the top and bottom boundaries of the sample, respectively. Contributions from other Floquet sidebands are negligible and therefore not shown. After applying the Floquet sum rule, we obtain the DC profile as shown in Fig.~\ref{subfig:perfectSumRule}, which shows two edge channels propagating from the left to the right leads along the sample boundaries.

We then turn to defective samples. In Fig.~\ref{subfig:defectHarmonic-1}, the local current associated with harmonic $n=-1$ bypasses the defects and maintains its propagation direction along the upper edge. Similarly, along the lower edge, the local current due to harmonic $n=0$ is immune to backscattering (Fig.~\ref{subfig:defectHarmonic0}). Remarkably, along the upper boundary, the harmonic $n=0$ contributes a non-vanishing current that flows against the bias. This backward-moving channel clearly indicates the sensitivity of counter-propagating edge modes to sample defects.  Fig.~\ref{subfig:defectSumRule} presents the overall DC profile after summing over contributions from all Floquet sidebands at quasienergy $\epsilon=1.5$. Here a weaker edge current signal compared to that of Fig.~\ref{subfig:perfectSumRule} is observed, suggesting a deviation from conductance quantization due to defects. This is confirmed by the string of circles hanging off the plateau of crosses in the right end of the transmission characteristics shown in Fig.~\ref{fig:transmDefectCounterpropa}. For completeness, we also checked the case with co-propagating Floquet chiral edge modes, and find that they are robust against defects (Fig.~\ref{subfig:transmGappedCopropa} and~\ref{subfig:copropaDefectpattern} in App.~\ref{app:gapped}).

In summary, we have confirmed that counter-propagating Floquet edge states are not robust against defects, as expected from a vanishing winding number across the Floquet-Brillouin zone. Also, as a computational tool, the local DC profile carries more information than the DC conductance and T-LDOS, in that it is able to distinguish the chirality of different harmonics, which all correspond to the same quasienergy.

\begin{table}
\centering
\begin{tabular}{c|c|c|c|c|c|c|}
\cline{2-7}
& \multicolumn{2}{c|}{DC conductance} & \multicolumn{2}{c|}{Time-averaged LDOS} & \multicolumn{2}{c|}{Local DC profile}  \\
\cline{2-7}
& Plateau & Stable & Edge pattern & Persists under disorder & Flows on edge & Bypasses defect \\ \hline
\multicolumn{1}{|c|}{Gapped} & $\times$ & $\times$ & \checkmark & $\times$ & \checkmark & $\times$ \\ \hline
\multicolumn{1}{|c|}{Symmetry-restricted gapless} & \checkmark & $\times$ & \checkmark & $\times$ & \checkmark & $\times$    \\ \hline
\multicolumn{1}{|c|}{Counter-propagating gapless} & \checkmark & $\times$ & \checkmark & \checkmark & \checkmark & \checkmark \\ \hline
\multicolumn{1}{|c|}{Co-propagating gapless} & \checkmark & \checkmark & \checkmark & \checkmark & \checkmark & \checkmark \\ \hline
\end{tabular}
\caption{Four types of edge states and their response towards disorder and defect. The check mark (cross mark) means robust (not robust) to the corresponding perturbation.}
\label{table:stability}
\end{table}
In Table \ref{table:stability}, we summarize the robustness of Floquet edge states studied in this subsection. For completeness, the robustness of topologically trivial gapped edge states is also listed here, with more details spelled out in App.~\ref{app:gapped}. The co-propagating chiral edge states in Floquet quantum Hall insulators are the most robust against local perturbations, and therefore potentially most useful in realizing high-precision electronic transport devices.

\subsection{\label{subsec:finiteBW}Beyond the wide-band approximation}
Up to now, our calculations were done under the WBA~\cite{Haug2008,Jauho1994,Lewenkopf2017,Velicky2010,Stefanucci2004}, {where the leads are assumed to be able to accommodate, uniformly, all energy states of the central system}. However, in Floquet topological systems, the transport \emph{at a quasienergy} requires contributions from energies spanning several Floquet-Brillouin zones. Thus, it is all the more appealing to model the leads as having only finite bandwidths.
\begin{figure}
\subfloat[\label{subfig:bandstrucFiniteBW}]{
\includegraphics[height=5cm,width=0.5\columnwidth]{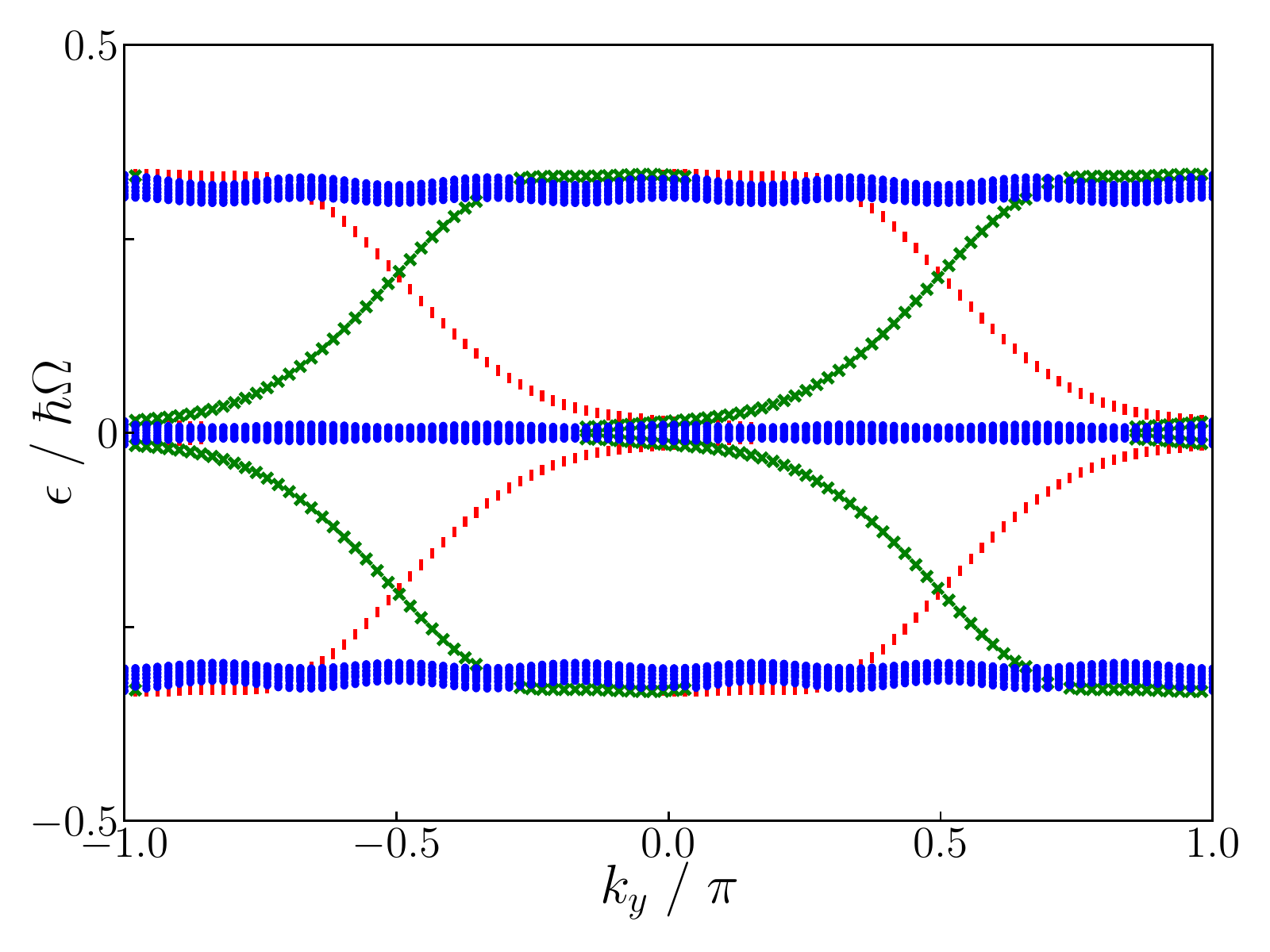}
}
\subfloat[\label{subfig:transmFiniteBW}]{
\includegraphics[height=5cm,width=0.5\columnwidth]{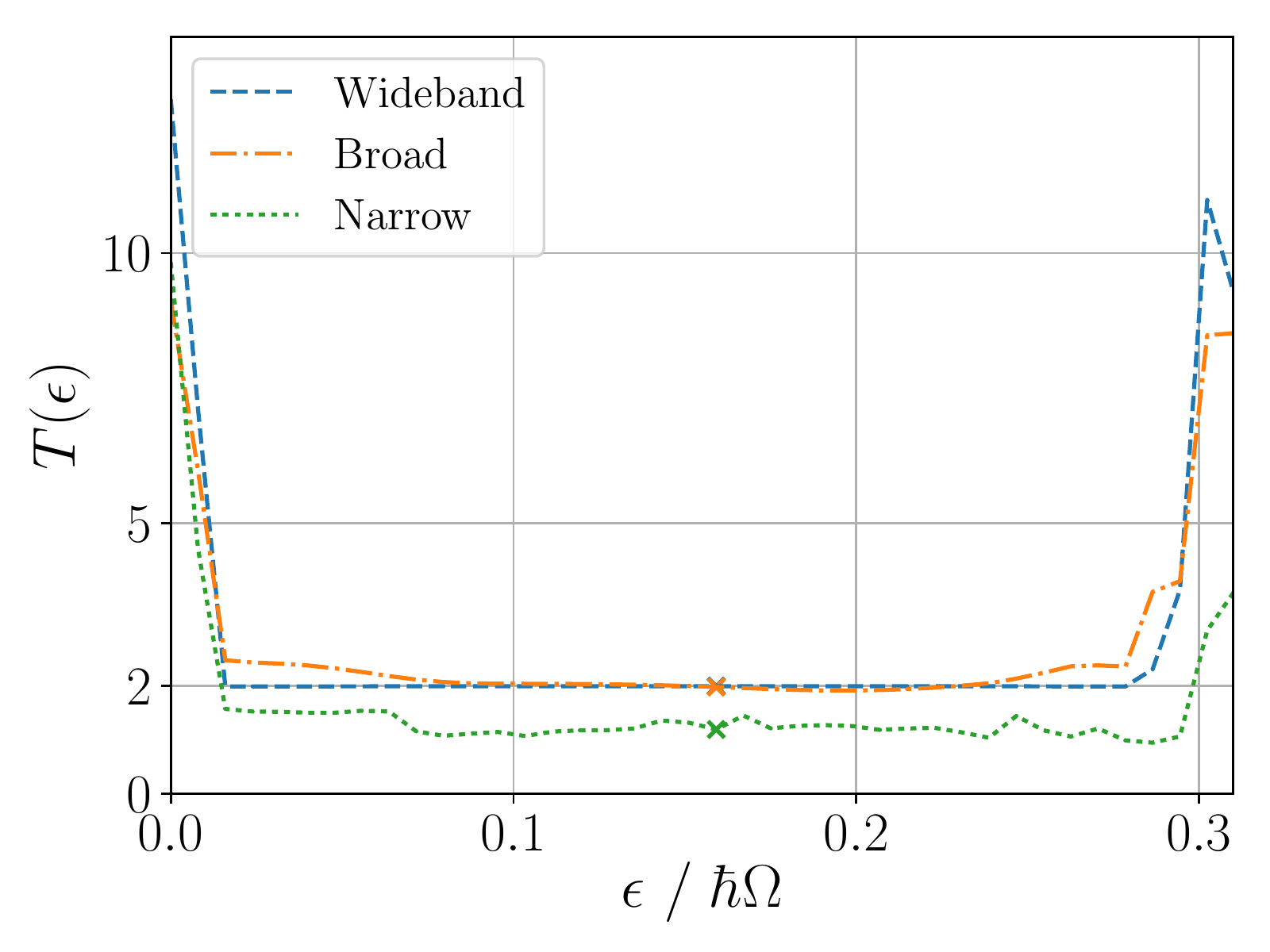}
}
\caption{(color online) Panel (a): Floquet spectrum of the HDHM with PBC (OBC) along $y$($x$)-direction. Panel (b): DC component of the transmission coefficient $T(\epsilon)$ at different quasienergies (with the Floquet sum rule applied) for leads with three types of bandwidth. The broad (narrow) bandwidth corresponds to $t_x=t_y=3$ ($t_x=t_y=0.5)$. The crosses mark the points at which we analyze the contribution of each side-peak to $T(\epsilon)$ in Figs.~\ref{fig:largeBW} and \ref{fig:narrowBW}. The systems parameters are taken as $J_y=1.6,\alpha=1/3,\Omega=\pi,s=0$ and $N_x=N_y=30$.}
\label{fig:WidebandBoardNarrow}
\end{figure}
For this purpose, it suffices to focus on a parameter regime as given in Fig.~\ref{subfig:bandstrucFiniteBW}. We model the leads as square lattices with nearest-neighbor hoppings described by Eq.~\eqref{eq:lead} and compute the DC transmission. For broadband leads (dash-dotted line in Fig.~\ref{subfig:transmFiniteBW}), upon applying the Floquet sum rule, the conductance quantization remains intact for quasienergies that are close to the center of the gap. On the contrary, for the case of a narrow-band leads (dotted line in Fig.~\ref{subfig:transmFiniteBW}), one no longer observes conductance quantization. These two results should be compared with the ideal case under the WBA, which corresponds to the use of leads with infinitely-broad band (dashed line in Fig.~\ref{subfig:transmFiniteBW}).
\begin{figure}[H]
\subfloat[\label{subfig:largeBW}]{
\def\svgwidth{0.75\columnwidth}
\begingroup%
  \makeatletter%
  \providecommand\color[2][]{%
    \errmessage{(Inkscape) Color is used for the text in Inkscape, but the package 'color.sty' is not loaded}%
    \renewcommand\color[2][]{}%
  }%
  \providecommand\transparent[1]{%
    \errmessage{(Inkscape) Transparency is used (non-zero) for the text in Inkscape, but the package 'transparent.sty' is not loaded}%
    \renewcommand\transparent[1]{}%
  }%
  \providecommand\rotatebox[2]{#2}%
  \ifx\svgwidth\undefined%
    \setlength{\unitlength}{458.03759781bp}%
    \ifx\svgscale\undefined%
      \relax%
    \else%
      \setlength{\unitlength}{\unitlength * \real{\svgscale}}%
    \fi%
  \else%
    \setlength{\unitlength}{\svgwidth}%
  \fi%
  \global\let\svgwidth\undefined%
  \global\let\svgscale\undefined%
  \makeatother%
  \begin{picture}(1,0.5108454)%
    \put(0,0){\includegraphics[width=\unitlength,page=1]{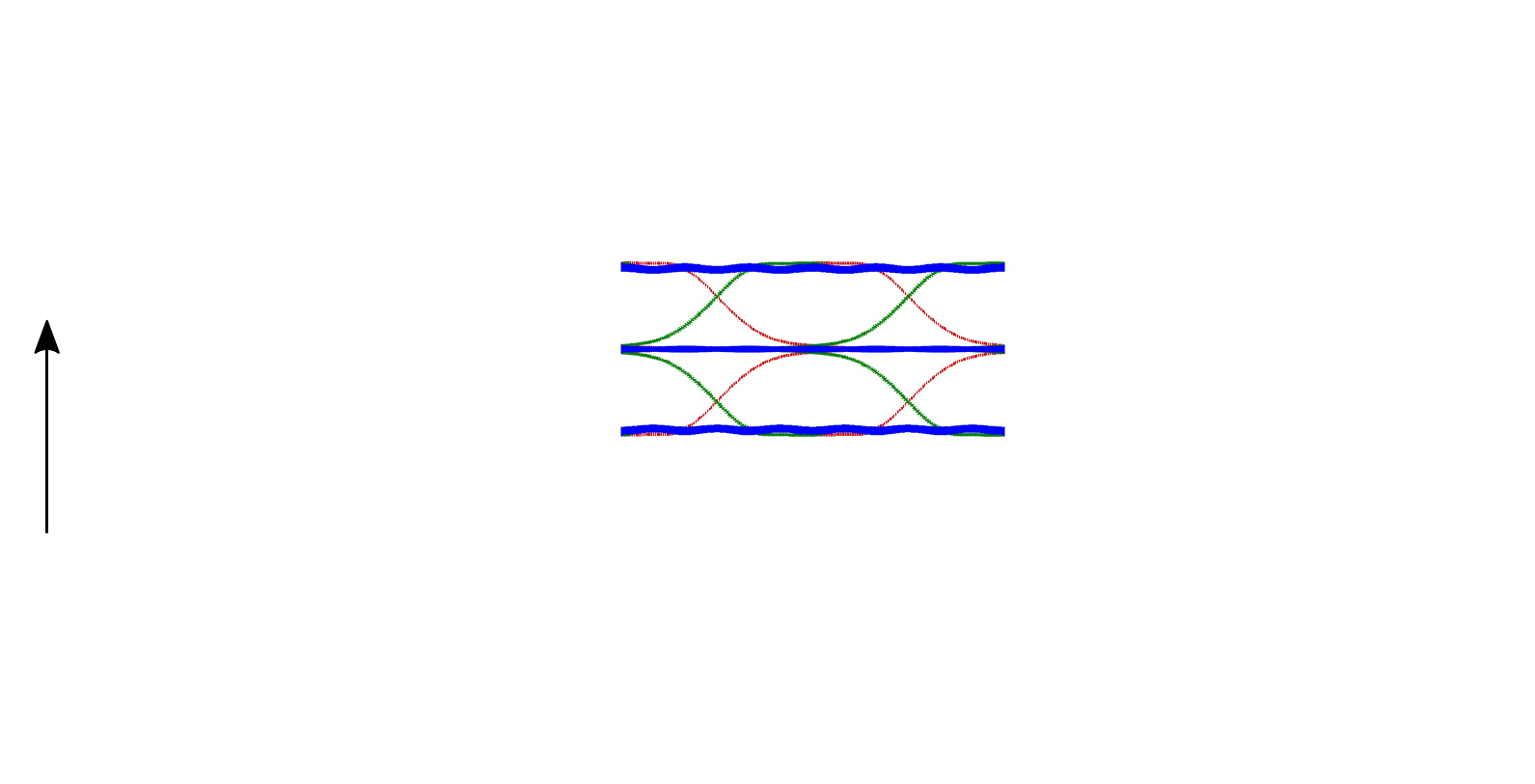}}%
    \put(1.13374113,0.37552678){\color[rgb]{0,0,0}\makebox(0,0)[lt]{\begin{minipage}{0.10978508\unitlength}\raggedright \end{minipage}}}%
    \put(-0.00315782,0.34183325){\color[rgb]{0,0,0}\makebox(0,0)[lt]{\begin{minipage}{0.31131668\unitlength}\raggedright $E/\hbar$\end{minipage}}}%
    \put(0,0){\includegraphics[width=\unitlength,page=2]{largeBWsmallFile.pdf}}%
  \end{picture}%
\endgroup%
}
\subfloat[\label{subfig:barlargeBW}]{
\includegraphics[angle=90,width=0.25\columnwidth]{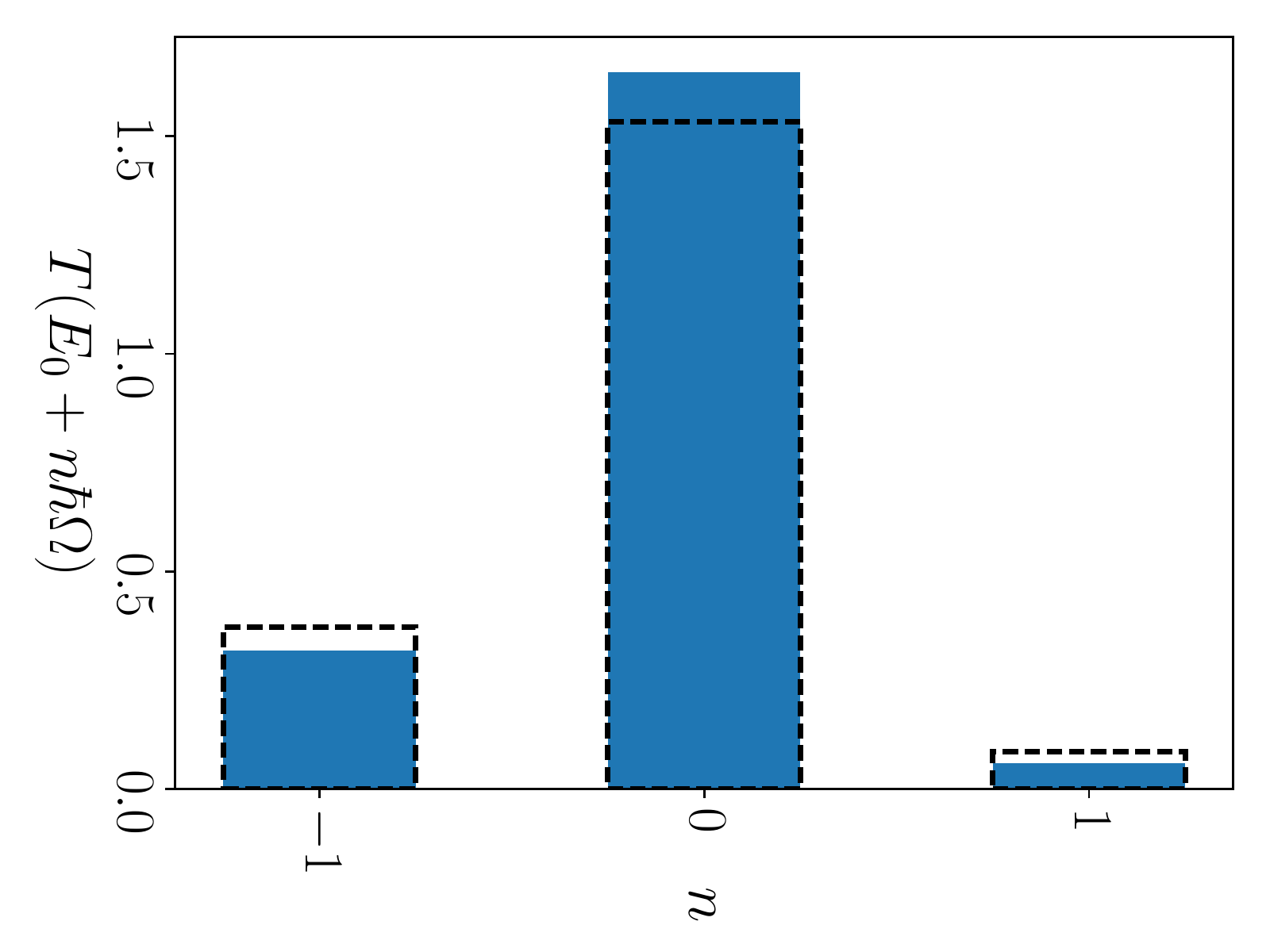}
}
\caption{(color online) When attached to leads with a finite, yet broad enough bandwidth $(t_x=t_y=3)$, quantized transmission is observed after applying the Floquet sum rule. In (a), the color indicates the local density of states of the leads, calculated over $0,\pm1$ Floquet zones, with energy as the vertical and $y$-direction as the horizontal axis. The HDHM has a quasienergy spectrum which repeats itself over integer multiples of the driving frequency $\Omega$. The red dashed line at $E_0=0.5$ refers to the energy of the incoming electron. In (b), the side-peak contributions to the quantized DC conductance are shown. Dashed rectangles indicate the case of wideband limit.} \label{fig:largeBW}
\end{figure}
To understand these observations, let us select a quasienergy in the middle of the gap, $\epsilon=E_0=1.5$ (crossed markers in Fig.~\ref{subfig:transmFiniteBW}). Consider now the three Floquet sidebands that contribute to a quantized conductance in the WBA (dashed rectangles in Figs.~\ref{subfig:barlargeBW} and~\ref{subfig:barnarrowBW}). When the WBA is relaxed, the leads must have non-vanishing density of states in these Floquet zones, in order for the corresponding sidebands to participate in the transport at $\epsilon=1.5$. This is the case for the broadband leads (Fig.~\ref{subfig:largeBW}), whose DOS spreads uniformly throughout the first three $(0,\pm1)$ Floquet zones. Hence, undeterred by a slight redistribution of the sideband weights (colored rectangles in Fig.~\ref{subfig:barlargeBW}), the DC conductance remains quantized as shown by the crossed marker in Fig.~\ref{subfig:transmFiniteBW}. Next, consider the case of narrow-band leads. Since their density of states is non-vanishing only at the $n=0$ Floquet zone (Fig.~\ref{subfig:narrowBW}), no electrons at energies $E_0\pm\hbar\Omega$ can contribute to the transport at quasienergy $\epsilon=E_0$. Indeed, as shown in Fig.~\ref{subfig:barnarrowBW}, there is only one single colored rectangle at the center, which is, worse still, diminished compared to the case of WBA (dashed rectangle). Therefore, even the Floquet sum rule cannot salvage its conductance quantization (green cross in Fig.~\ref{subfig:transmFiniteBW}), because the lead bandwidths are simply too narrow for anything other than $E=E_0$ to contribute to the transport.

{The above observation calls attention to the leads when studying the transport of Floquet topological systems. Let us consider the case in which the central system is governed by a certain energy scale $J$, with the bandwidths of the leads characterized by $\Gamma$, such that $\Gamma \gg J$. In static systems, one can then simplify the problem using the WBA. But if the Floquet edge modes of interest are scattered throughout $\mathcal{N}$ dominant sidebands, such that $\mathcal{N}\Omega \gg \Gamma$, then the leads have states only within $\Gamma$, which, by the assumption $\mathcal{N}\Omega \gg \Gamma$, do not span some of the Floquet zones. It would then be inappropriate, in this case, to invoke the WBA right at the beginning.}

In brief, unless sufficiently broad, the bandwidth of electronic leads may result in non-quantized DC edge conductances in Floquet  Hall insulators. This suggests that any attempt at detecting quantized DC conductance (upon applying the Floquet sum rule) must involve electronic leads with bands wide enough to cover all contributing Floquet sidebands.

\begin{figure}
\subfloat[\label{subfig:narrowBW}]{
\def\svgwidth{0.75\columnwidth}
\begingroup%
  \makeatletter%
  \providecommand\color[2][]{%
    \errmessage{(Inkscape) Color is used for the text in Inkscape, but the package 'color.sty' is not loaded}%
    \renewcommand\color[2][]{}%
  }%
  \providecommand\transparent[1]{%
    \errmessage{(Inkscape) Transparency is used (non-zero) for the text in Inkscape, but the package 'transparent.sty' is not loaded}%
    \renewcommand\transparent[1]{}%
  }%
  \providecommand\rotatebox[2]{#2}%
  \ifx\svgwidth\undefined%
    \setlength{\unitlength}{458.03759781bp}%
    \ifx\svgscale\undefined%
      \relax%
    \else%
      \setlength{\unitlength}{\unitlength * \real{\svgscale}}%
    \fi%
  \else%
    \setlength{\unitlength}{\svgwidth}%
  \fi%
  \global\let\svgwidth\undefined%
  \global\let\svgscale\undefined%
  \makeatother%
  \begin{picture}(1,0.51084558)%
    \put(0,0){\includegraphics[width=\unitlength,page=1]{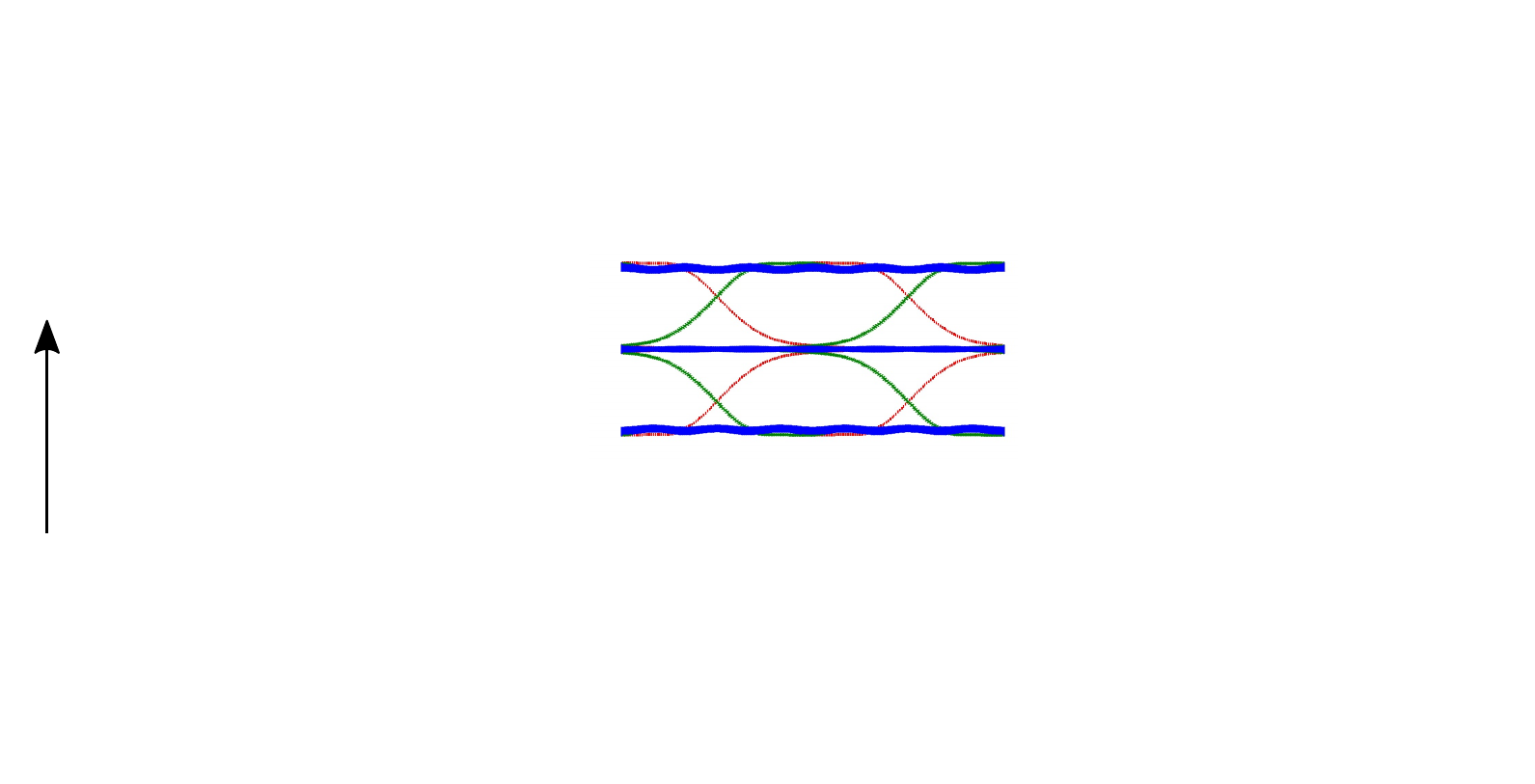}}%
    \put(1.13374113,0.37552676){\color[rgb]{0,0,0}\makebox(0,0)[lt]{\begin{minipage}{0.10978508\unitlength}\raggedright \end{minipage}}}%
    \put(-0.00315782,0.34183323){\color[rgb]{0,0,0}\makebox(0,0)[lt]{\begin{minipage}{0.31131668\unitlength}\raggedright $E/\hbar$\end{minipage}}}%
    \put(0,0){\includegraphics[width=\unitlength,page=2]{narrowBWsmallFile2.pdf}}%
  \end{picture}%
\endgroup%
}
\subfloat[\label{subfig:barnarrowBW}]{
\includegraphics[angle=90,width=0.25\columnwidth]{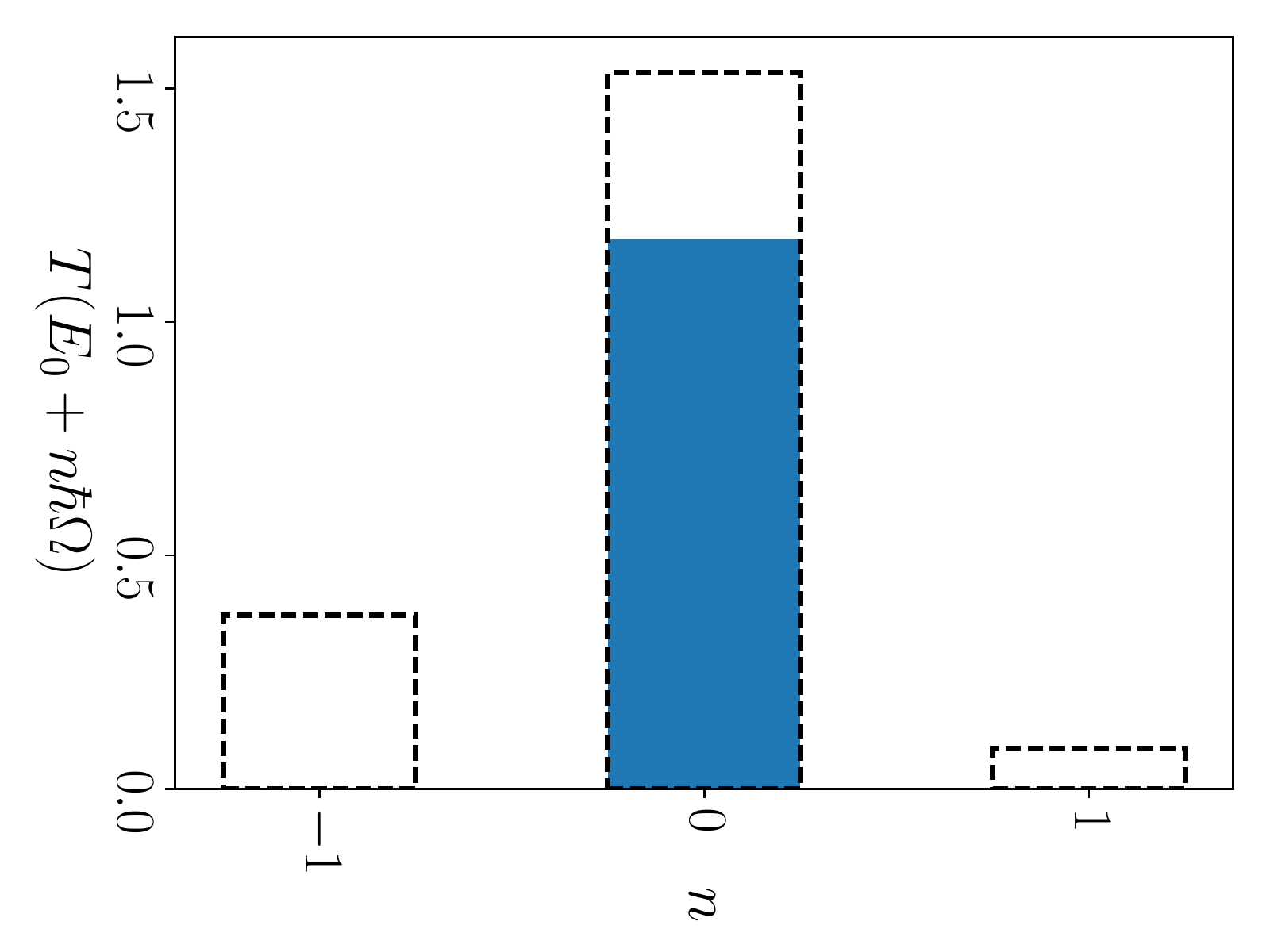}
}
\caption{(color online) For leads with narrow bandwidth $(t_x=t_y=0.5)$, there is no quantized transmission even after performing the Floquet sum rule. This figure is plotted the same way as Fig.~\ref{fig:largeBW}}.
\label{fig:narrowBW}
\end{figure}

\section{\label{sec:conclusion}Conclusion and outlook}

In this work, we have theoretically and computationally studied how
Floquet topological phases with many edge state channels may lead to robust transport in a two-terminal setup. Under the Keldysh-NEGF framework and using the recursive Floquet-Green's function method, we have investigated the two-terminal DC conductance, time-averaged LDOS and local DC profile of a harmonically-driven Hofstadter model in several parameter regimes.  Thanks to the Floquet sum rule, we are able to observe tunable and quantized Floquet edge state transport, with the largest DC conductance demonstrated in this work being $8e^2/h$.  We have also presented a detailed comparison among Floquet chiral, counter-propagating and symmetric-restricted edge states regarding their robustness to disorder and defects. Our results indicate that in terms of transport, Floquet chiral edge modes are the most resistant to sample imperfections. We have also shown that, even after invoking the Floquet sum rule, one cannot guarantee quantization of the DC conductance of Floquet edge modes, if the bandwidth of the lead is not wide enough.

Finally, we note that it remains unclear how to exploit the tunability of Floquet topological matter to \emph{directly} realize large, robust, and quantized DC conductance at a given chemical potential (namely, without using the Floquet sum rule).  Some results from Ref.~\cite{Cuevas2010} lead us to believe that this could be made possible by driving the leads appropriately, which may be an interesting future direction.

\begin{acknowledgments}
	H.H.~Y. thanks Patrick Haughian for fruitful discussions. This work is supported by the Singapore NRF grant No. NRF-NRFI2017-04 (WBS No. R-144-000-378-281) and by the Singapore Ministry of Education Academic Research Fund Tier I~(WBS No. R-144-000-353-112).
\end{acknowledgments}

\appendix

\section{\label{app:RFGF}Recursive Floquet-Green's function method}
The recursive Green's function method is a powerful method which speeds up calculations and reduces memory usage substantially. Here we follow Ref.~\cite{Lewenkopf2013} and extend the method to incorporate a periodic driving field. We first describe the simpler case for transmission, which involves a single, unidirectional renormalization of the left self energy. We then discuss the DC profile and the time-averaged LDOS, which involve $N_x$ bidirectional renormalization of both left and right self energies.

\subsection{DC conductance}
Recall the expression for DC conductance:
\begin{equation}
T(E) = \sum_{k\in\mathbb{Z}}\TrSite\Big[\bm{G}^r_{k0}\bm{\Gamma}^L_{00}\bm{G}^a_{0k}\bm{\Gamma}^R_{kk}\Big].
\end{equation}
This has a form that is almost identical to the static Caroli formula \cite{Caroli1971,Khomyakov2005}, apart from the additional summation over Floquet indices. Thus consider the Floquet-Dyson equation \eqref{eq:FloquetDyson}, which we write here as \cite{Kitagawa2011}:
\begin{equation}
[E\bm{I}+\hbar\bm{\Omega}-\bm{H}-\{\bm{\Sigma}^r_L(E)+
\bm{\Sigma}^r_R(E)\} ]\bm{G}^r(E) = \bm{I}
\end{equation}
Let $n_\mathrm{F}=2n_{\mathrm{H}}+1$ be the truncation dimension (or Floquet dimension), where $n_{\mathrm{H}}$ is the number of harmonics. We write:
\begin{widetext}
\begin{equation}
\begin{split}
&\bm{\Omega} = \begin{bmatrix}
{-\nH \Omega}  &       &  & & \\
 			  & \ddots &       & & \\
              &       &   {0}    & & \\
              &        &       & \ddots &   \\
              &        &        &   &    {\nH \Omega}
\end{bmatrix},
\bm{\Sigma}^r_{L/R}(E) =  \begin{bmatrix}
\Sigma^r_{L/R}(E-\nH\Omega)  &       &  & & \\
 			  & \ddots &       & & \\
              &       &   \Sigma^r_{L/R}(E)    & & \\
              &        &       & \ddots &   \\
              &        &        &   &    \Sigma^r_{L/R}(E+\nH\Omega)
\end{bmatrix},
\\
&\bm{H} = \begin{bmatrix}
H_{0}  & H_{1}  &  		   & 		 & \\
H_{-1} & \ddots &   H_1    &		 & \\
       &  H_{-1}&   H_0    & \ddots  & \\
       &        & \ddots   & \ddots  & H_1  \\
       &        &          & H_{-1}  & H_0
\end{bmatrix},
\bm{G}^r(E) =\left[
\begin{array}{*{5}c}
\cline{3-3}
G^r_{-\nH,-\nH}(E)  & \cdots  &  \multicolumn{1}{|c|}{G^r_{-\nH,0}(E)}		   & 	\cdots	 & G^r_{-\nH,\nH}(E)\\
\vdots & \ddots &   \multicolumn{1}{|c|}{\vdots}    &   \cdots		 & \vdots \\
G^r_{0,-\nH}(E)       &  \cdots&  \multicolumn{1}{|c|}{G^r_{0,0}(E)}    & \cdots  & G^r_{0,\nH}(E)\\
   \vdots    &  \cdots     &  \multicolumn{1}{|c|}{\vdots}  & \ddots  & \vdots  \\
  G^r_{\nH,-\nH}(E)     &   \cdots     &  \multicolumn{1}{|c|}{G^r_{\nH,0}(E)}        & \cdots & G^r_{\nH,\nH}(E) \\
 \cline{3-3}
\end{array} \right] .
\end{split} \label{eq:matrices}
\end{equation}
\end{widetext}
where we boxed the center column of $\bm{G}^r$ to indicate the part of the Floquet-Green's function that needs to be extracted in conductance calculation, and stress that in \eqref{eq:matrices}, each entry is itself a $N_xN_y\times N_xN_y$ matrix.

Thanks to the trace over space, the Floquet-Caroli formula for transmission \eqref{eq:Transmission} requires only the ``first-to-last'' correlation, $\bm{G}^r_{0,N_x-1}$. { Therefore we slice the lattice into slabs and couple them consecutively starting from the left.} This procedure is equivalent to renormalizing the left self-energy by coupling the system to it, layer by layer, until the last system slab is arrived.

\begin{figure}
\centering
\def\svgwidth{0.9\columnwidth}
\begingroup%
  \makeatletter%
  \providecommand\color[2][]{%
    \errmessage{(Inkscape) Color is used for the text in Inkscape, but the package 'color.sty' is not loaded}%
    \renewcommand\color[2][]{}%
  }%
  \providecommand\transparent[1]{%
    \errmessage{(Inkscape) Transparency is used (non-zero) for the text in Inkscape, but the package 'transparent.sty' is not loaded}%
    \renewcommand\transparent[1]{}%
  }%
  \providecommand\rotatebox[2]{#2}%
  \ifx\svgwidth\undefined%
    \setlength{\unitlength}{897.42335695bp}%
    \ifx\svgscale\undefined%
      \relax%
    \else%
      \setlength{\unitlength}{\unitlength * \real{\svgscale}}%
    \fi%
  \else%
    \setlength{\unitlength}{\svgwidth}%
  \fi%
  \global\let\svgwidth\undefined%
  \global\let\svgscale\undefined%
  \makeatother%
  \begin{picture}(1,0.32454157)%
    \put(0,0){\includegraphics[width=\unitlength,page=1]{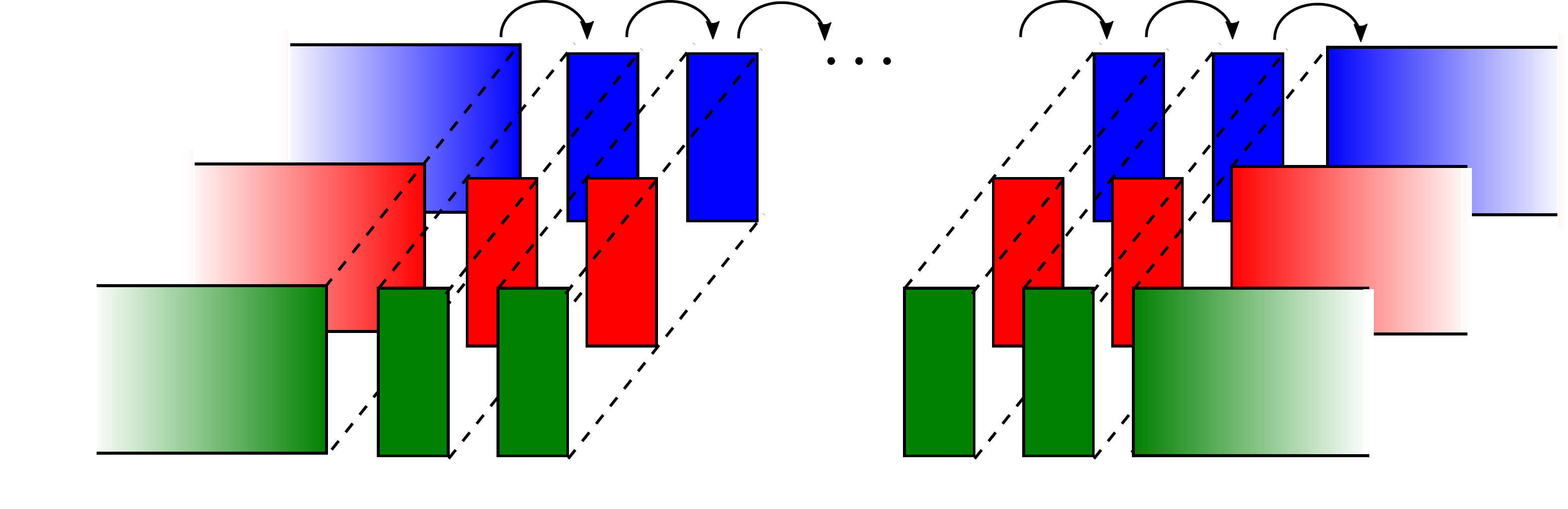}}%
    \put(0.05419861,0.19013926){\color[rgb]{0,0,0}\makebox(0,0)[lt]{\begin{minipage}{0.16045939\unitlength}\raggedright $n=0$\end{minipage}}}%
    \put(0.11084584,0.2760411){\color[rgb]{0,0,0}\makebox(0,0)[lt]{\begin{minipage}{0.16045939\unitlength}\raggedright $n=-1$\end{minipage}}}%
    \put(-0.00166579,0.09577327){\color[rgb]{0,0,0}\makebox(0,0)[lt]{\begin{minipage}{0.16045939\unitlength}\raggedright $n=1$\end{minipage}}}%
    \put(0.11182067,0.02472212){\color[rgb]{0,0,0}\makebox(0,0)[lt]{\begin{minipage}{0.35657642\unitlength}\raggedright $\bm{g^r_L}$\end{minipage}}}%
    \put(0.25114016,0.02293969){\color[rgb]{0,0,0}\makebox(0,0)[lt]{\begin{minipage}{0.35657642\unitlength}\raggedright $0$\end{minipage}}}%
    \put(0.32372893,0.02293969){\color[rgb]{0,0,0}\makebox(0,0)[lt]{\begin{minipage}{0.35657642\unitlength}\raggedright $1$\end{minipage}}}%
    \put(0.55728645,0.02472239){\color[rgb]{0,0,0}\makebox(0,0)[lt]{\begin{minipage}{0.35657642\unitlength}\raggedright $N_x-2$\end{minipage}}}%
    \put(0.63573326,0.02472239){\color[rgb]{0,0,0}\makebox(0,0)[lt]{\begin{minipage}{0.35657642\unitlength}\raggedright $N_x-1$\end{minipage}}}%
    \put(0.77658103,0.02472212){\color[rgb]{0,0,0}\makebox(0,0)[lt]{\begin{minipage}{0.35657642\unitlength}\raggedright $\bm{g^r_R}$\end{minipage}}}%
    \put(0,0){\includegraphics[width=\unitlength,page=2]{RGF_Singlesweep.pdf}}%
  \end{picture}%
\endgroup%
\caption{(color online)Schematic of the recursive Floquet-Green's function method for DC transmission. ${\bm{\mathcal{G}}}^r_{x,x'}$ is the ``slab-wise'' Floquet-Green's function which is represented by dashed cuboids, enclosing rectangles corresponding to different harmonics $n$. Black arrows indicate the left-to-right sweep.}
\label{RGF_Singlesweep}
\end{figure}

Below we denote by cursive letters the ``slab-wise'' Floquet-Green's function: ${\bm{\mathcal{G}}}^r_{x_1,x_2}$. They are indexed by a pair of $x$-coordinates $(x_1,x_2)$, and, for each pair, they are matrices of size $N_y\nF \times N_y\nF$. The same goes for lowercase ${\bm{g}}^r_{xx}$, which are the slab-wise decoupled Floquet-Green's function at site $x$. Schematically, this process is demonstrated in Fig.~\ref{RGF_Singlesweep}.
\noindent We need one more ingredient: the interslab coupling matrix $\bm{\mathcal{V}}$. It is defined by considering the Dyson equation when we couple two slabs \cite{Wang2008}:
\begin{widetext}
\begin{equation}
\begin{bmatrix}
\bm{\mathcal{G}}_{x,x} & \bm{\mathcal{G}}_{x,x+1} \\
\bm{\mathcal{G}}_{x+1,x} & \bm{\mathcal{G}}_{x+1,x+1}
\end{bmatrix} =
\begin{bmatrix}
\bm{\mathrm{g}}_{x,x} & \bm{0} \\
\bm{0} & \bm{\mathrm{g}}_{x+1,x+1}
\end{bmatrix} + \begin{bmatrix}
\bm{\mathrm{g}}_{x,x} & \bm{0} \\
\bm{0} & \bm{\mathrm{g}}_{x+1,x+1}
\end{bmatrix}
\begin{bmatrix}
\bm{0} & \bm{\mathcal{V}} \\
\bm{\mathcal{V}}^\dag  & \bm{0}
\end{bmatrix}\begin{bmatrix}
\bm{\mathcal{G}}_{x,x} & \bm{\mathcal{G}}_{x,x+1} \\
\bm{\mathcal{G}}_{x+1,x} & \bm{\mathcal{G}}_{x+1,x+1}
\end{bmatrix}.\label{eq:DysonCoupling}
\end{equation}
\end{widetext}
It varies depending on the system (sublattice degree of freedom and how the lattice couples horizontally), as well as the drive protocol (whether $x$-hopping is driven). Let us specify to the case where $x$-hopping is undriven (if not, one most likely can gauge it away), and the slabs couple uniformly. In this case $\bm{\mathcal{V}}$ is simply proportional to a $N_y\nF \times N_y\nF$ identity matrix: $\bm{\mathcal{V}}_{(yy';mn)}=J_x\delta_{y,y'}\delta_{m,n}$. Finally, we remark that Eq.\eqref{eq:DysonCoupling} is valid for contour Green's functions. Therefore one can derive from it the retarded (for time-averaged LDOS) and lesser component (for DC profile) of Green's functions using the Langreth rules~\cite{Haug2008}.

\begin{algorithm}[H]
\caption{Unidirectional sweep for DC transmission}
\label{algo:lefttoright}
 \begin{algorithmic}
	\State $\bm{\mathcal{G}}^r_{xx}\gets[(\bm{\mathrm{g}}^r_{00})^{-1}-\bm{\mathcal{V}}\mathrm{g}^r_L\bm{\mathcal{V}}]^{-1}$ \Comment{Initialization using left surface Green's function}
	\State $\bm{\mathcal{G}}^r_{0x}\gets \bm{\mathcal{G}}^r_{xx}$
	\For {$x\in[1,N_x-2]$}
	    \State $\bm{\mathcal{G}}^r_{xx}\gets [(\bm{\mathrm{g}^r}_{xx})^{-1}-\bm{\mathcal{V}}\bm{\mathcal{G}}^r_{xx}\bm{\mathcal{V}}]^{-1}$
		\State $\bm{\mathcal{G}}^r_{0x}\gets \bm{\mathcal{G}}^r_{0x}\bm{\mathcal{V}}\bm{\mathcal{G}}^r_{xx}$
	\EndFor
	\State $\bm{\mathcal{G}}^r_{xx}\gets[(\bm{g}^r_{N_x-1,N_x-1})^{-1}-\bm{\mathcal{V}}\bm{g}^r_{R}\bm{\mathcal{V}}-\bm{\mathcal{V}}\bm{\mathcal{G}}^r_{xx}\bm{\mathcal{V}}]$\Comment{Wraps the right boundary condition}
	\State $\bm{\mathcal{G}}^r_{0,x} \gets \bm{\mathcal{G}}^r_{0,x}\bm{\mathcal{V}}\bm{\mathcal{G}}^r_{x,x}$
 \end{algorithmic}
\end{algorithm}
The pseudocode for the main loop of our calculation is given in Algorithm \ref{algo:lefttoright}. It has the same structure as the static one~\cite{Lewenkopf2013,Ryndyk2015}. The Floquet-Green's function thus obtained has its $x$-components resolved to $(0,N_x-1)$. Next, we extract all relevant Floquet components, $\bm{\mathcal{G}}^r_{0,N_x-1}[k,0]$, and calculate the transmission \emph{at energy $E$} by multiplying the relevant Floquet components and performing a trace. This process is summarized in Algorithm \ref{algo:DC}. Here a square bracket $[m,n]$ means extracting (as a $N_y\times N_y$ matrix) the $m,n$ Floquet component, from a $N_y\nF\times N_y\nF$ Floquet matrix. Finally, Floquet sum rule is achieved by looping Algorithm~\ref{algo:lefttoright} and then Algorithm~\ref{algo:DC} over $n$ for $T(E+n\hbar\Omega)$ and summing them.
\begin{algorithm}[H]
\caption{Extracting Floquet components to compute DC transmission at energy $E$}
\label{algo:DC}
\begin{algorithmic}
\State $T(E)\gets 0$
\State $\gamma^L\gets \ii(\bm{\Sigma}^r_L-(\bm{\Sigma}^r_L)^\dag)[0,0]$
\For {$k\in[-\nH,\nH]$}
	\State $g\gets (\bm{\mathcal{G}}^r_{0,N_x-1})[k,0]$
	\State $\gamma^R \gets \ii(\bm{\Sigma}^r_R-(\bm{\Sigma}^r_R)^\dag)[k,k]$
	\State $T(E)\gets T(E)+ \TrSite(g\gamma^Lg^\dag \gamma^R)$
\EndFor
\end{algorithmic}
\end{algorithm}
\begin{algorithm}[H]
\caption{Bidirectional sweep for DC profile (and time-averaged LDOS)}
\label{algo:localDCprofile}
\begin{algorithmic}[1]
\For {$x\in[0,N_x-1]$}
	\State $\bm{\mathcal{G}}^r_L\gets \bm{\mathrm{g}}^r_L$ \Comment{Initialize left retarded Green's function}
	\State $\bm{\mathcal{G}}^<_L\gets -(\bm{\mathcal{G}}^r_L- (\bm{\mathcal{G}}^r_L)^\dag )\fermi^L$ \Comment{Initialize left lesser Green's function}
	\For {$x_L\in[0,x)$} \Comment{Left sweep towards $x$}
	\State $\bm{\mathcal{G}}^r_{L} \gets [(\bm{\mathrm{g}}^r_{x_Lx_L})^{-1}-\bm{\mathcal{V}}\bm{\mathcal{G}}^r_{L}\bm{\mathcal{V}}]^{-1}$ \Comment{Renormalize the left retarded Green's function}
	\State $\bm{\mathcal{G}}^<_L \gets \bm{\mathcal{G}}^r_L\bm{\mathcal{V}} \bm{\mathcal{G}}^<_L \bm{\mathcal{V}}(\bm{\mathcal{G}}^r_L)^\dag$ \Comment{Renormalize the left lesser Green's function}
	\EndFor
	\State $\bm{\mathcal{G}}^r_R\gets \bm{\mathrm{g}}^r_R$ \Comment{Initialize right retarded Green's function}
	\State $\bm{\mathcal{G}}^<_R\gets -( \bm{\mathcal{G}}^r_R - (\bm{\mathcal{G}}^r_R)^\dag )\fermi^R $ \Comment{Initialize right lesser Green's function}
	\For {$x_R\in (x,N_x-1]$} \Comment{Right sweep towards $x$}
	\State $\bm{\mathcal{G}}^r_R\gets [(\bm{\mathrm{g}}^r_{x_Rx_R})^{-1}-\bm{\mathcal{V}}\bm{\mathcal{G}}^r_{R}\bm{\mathcal{V}}]^{-1} $ \Comment{Renormalize the right retarded Green's function}
	\State $\bm{\mathcal{G}}^<_R\gets \bm{\mathcal{G}}^r_R\bm{\mathcal{V}}\bm{\mathcal{G}}^<_R\bm{\mathcal{V}}(\bm{\mathcal{G}}^r_R)^\dag  $ \Comment{Renormalize the right lesser Green's function}
	\EndFor
	\State $\bm{\mathcal{G}}^r_{xx}\gets[(\bm{\mathrm{g}}^r_{xx})^{-1}-\bm{\mathcal{V}}(\bm{\mathcal{G}}^r_L+\bm{\mathcal{G}}^r_R)\bm{\mathcal{V}}]^{-1}$
	\Comment{Solve the retarded Green's function for slab $x$.}
	\State $\textrm{T-LDOS}_{(x,y)}(E)\gets -\frac{1}{\pi}\mathrm{Im}(\bm{\mathcal{G}}^r_{xx})_{yy}[0,0]$ \Comment{Diagonal component $yy$ of the $0,0$ Floquet block at slab $x$}
	
	\State $\bm{\mathcal{G}}^<_{xx}\gets\bm{\mathcal{G}}^r_{xx}\bm{\mathcal{V}}(\bm{\mathcal{G}}^<_{L}+\bm{\mathcal{G}}^<_{R} ) \bm{\mathcal{V}}(\bm{\mathcal{G}}^r_{xx})^\dag$ \Comment{Calculate the lesser Green's function for slab $x$.}
	
	\State $I_{x,y}^\uparrow\gets 2\mathrm{Re}\sum_{m=-\nF}^{\nF}\big\{\hat{J}_{y-1,y}(m)(\bm{\mathcal{G}}^<_{xx})_{y,y-1}[0,m]+\hat{J}_{y,y+1}(m)(\bm{\mathcal{G}}^<_{xx})_{y+1,y}[0,m] \big\}$
	
	\State $\bm{\mathcal{G}}^<_{x+1,x}\gets \bm{\mathcal{G}}^r_R\bm{\mathcal{V}}\bm{\mathcal{G}}^<_{xx} + \bm{\mathcal{G}}^<_R\bm{\mathcal{V}}(\bm{\mathcal{G}}^r_{xx})^\dag $ \Comment{Calculate the interslab $(x+1,x)$ lesser Green's function}
	\State $I_{x+1\leftarrow x,y}=2\mathrm{Re}\sum_{m=-\nF}^{\nF}\big\{\hat{J}_{x,x+1}(m)(\bm{\mathcal{G}}^<_{x+1,x})_{yy}[m,0]		\big\} $

	\State $\bm{\mathcal{G}}^<_{x-1,x} \gets \bm{\mathcal{G}}^r_L\bm{\mathcal{V}}\bm{\mathcal{G}}^<_{xx}+\bm{\mathcal{G}}^<_L\bm{\mathcal{V}}(\bm{\mathcal{G}}^r_{xx})^\dag$
	\Comment{Calculate the interslab $(x-1,x)$ lesser Green's function}
	\State $I_{x-1\leftarrow x,y}\gets 2\mathrm{Re}\sum_{m=-\nF}^{\nF}\big\{\hat{J}_{x,x-1}(m)(\bm{\mathcal{G}}^<_{x,x-1})_{yy}[0,m]	\big\}$
\EndFor
\end{algorithmic}
\end{algorithm}
\subsection{DC profile}
\begin{figure}
\centering
\def\svgwidth{0.9\columnwidth}
\begingroup%
  \makeatletter%
  \providecommand\color[2][]{%
    \errmessage{(Inkscape) Color is used for the text in Inkscape, but the package 'color.sty' is not loaded}%
    \renewcommand\color[2][]{}%
  }%
  \providecommand\transparent[1]{%
    \errmessage{(Inkscape) Transparency is used (non-zero) for the text in Inkscape, but the package 'transparent.sty' is not loaded}%
    \renewcommand\transparent[1]{}%
  }%
  \providecommand\rotatebox[2]{#2}%
  \ifx\svgwidth\undefined%
    \setlength{\unitlength}{990.22335695bp}%
    \ifx\svgscale\undefined%
      \relax%
    \else%
      \setlength{\unitlength}{\unitlength * \real{\svgscale}}%
    \fi%
  \else%
    \setlength{\unitlength}{\svgwidth}%
  \fi%
  \global\let\svgwidth\undefined%
  \global\let\svgscale\undefined%
  \makeatother%
  \begin{picture}(1,0.29412676)%
    \put(0,0){\includegraphics[width=\unitlength,page=1]{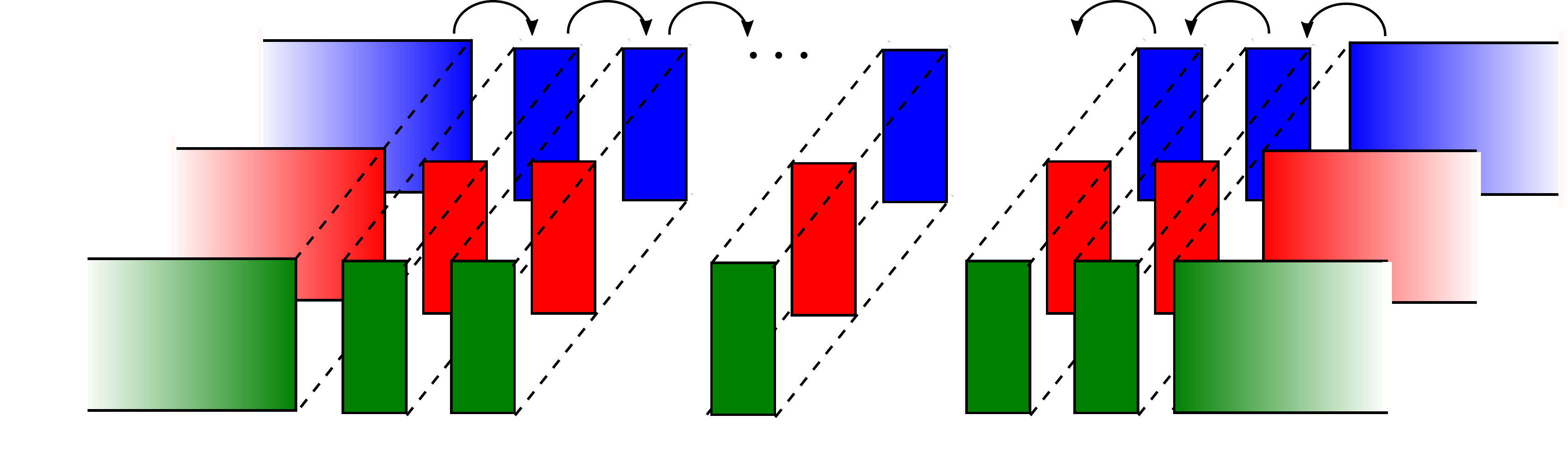}}%
    \put(0.10134126,0.02240526){\color[rgb]{0,0,0}\makebox(0,0)[lt]{\begin{minipage}{0.32315942\unitlength}\raggedright $\bm{g^r_L}$\end{minipage}}}%
    \put(0.21952527,0.02078987){\color[rgb]{0,0,0}\makebox(0,0)[lt]{\begin{minipage}{0.32315942\unitlength}\raggedright $0$\end{minipage}}}%
    \put(0.28531129,0.02078987){\color[rgb]{0,0,0}\makebox(0,0)[lt]{\begin{minipage}{0.32315942\unitlength}\raggedright $1$\end{minipage}}}%
    \put(0.59069691,0.0224055){\color[rgb]{0,0,0}\makebox(0,0)[lt]{\begin{minipage}{0.32315942\unitlength}\raggedright $N_x-2$\end{minipage}}}%
    \put(0.66825517,0.0224055){\color[rgb]{0,0,0}\makebox(0,0)[lt]{\begin{minipage}{0.32315942\unitlength}\raggedright $N_x-1$\end{minipage}}}%
    \put(0.78944003,0.02240526){\color[rgb]{0,0,0}\makebox(0,0)[lt]{\begin{minipage}{0.32315942\unitlength}\raggedright $\bm{g^r_R}$\end{minipage}}}%
    \put(0.04911932,0.17232012){\color[rgb]{0,0,0}\makebox(0,0)[lt]{\begin{minipage}{0.14542174\unitlength}\raggedright $n=0$\end{minipage}}}%
    \put(0.10045779,0.25017157){\color[rgb]{0,0,0}\makebox(0,0)[lt]{\begin{minipage}{0.14542174\unitlength}\raggedright $n=-1$\end{minipage}}}%
    \put(-0.00150968,0.08679776){\color[rgb]{0,0,0}\makebox(0,0)[lt]{\begin{minipage}{0.14542174\unitlength}\raggedright $n=1$\end{minipage}}}%
    \put(0.45496995,0.0224055){\color[rgb]{0,0,0}\makebox(0,0)[lt]{\begin{minipage}{0.32315942\unitlength}\raggedright $x$\end{minipage}}}%
    \put(0,0){\includegraphics[width=\unitlength,page=2]{RGF_Mixedsweep.pdf}}%
  \end{picture}%
\endgroup%
\caption{(color online) Schematic of the recursive Floquet-Green's function method for local quantities (DC profile, LDOS). For each $x$, one performs a bidirectional sweep (black arrows), which amounts to considering slab $x$ as the system, with all slabs to the left (including the left lead) acting as left self energy, and the same for right. Dashed lines enclose photon-dressed Green's function at different harmonics $n$.}
\label{RGF_Mixedsweep}
\end{figure}

For quantities that involve spatial distributions, ``local'' Green's functions of the like ${\bm G}^{r/<}_{x_i,x_i}$ are needed instead, and this applies to all system slabs $x_i\in[0,N_x-1]$. Thus the recursive method consists of $N_x$ sweeps, each sweep being bidirectional, as shown in Fig.~\ref{RGF_Mixedsweep}.

The strategy is similar as before: promote the Green's function to Floquet space, perform recursive calculations as in the static case~\cite{Lewenkopf2013}, then extract the relevant Floquet components. We first rewrite the local current at site $i=(x_i,y_i)$ given by Eq.~\eqref{eq:localcurrent} as:
\begin{widetext}
\begin{equation}
\begin{split}
\vec{I}_i &= \hatx I^\rightarrow_i + \haty I^\uparrow_i \\
&=  \hatx \left(  I_{i+\hatx\leftarrow i}-I_{i-\hatx\leftarrow i}  \right) + \haty I^\uparrow_i \\
&= 2\mathrm{Re}\sum_{m\in\mathbb{Z}}\Big\{ \hatx \Big[ \tilde{J}_{i,i+\hatx}(m)\bm{G}^<_{(i+\hatx,i;m,0)} -\tilde{J}_{i,i-\hatx}(m)\bm{G}^<_{(i-\hatx,i;0,m)}   \Big]	\\ &+\haty \Big[  \tilde{J}_{i-\haty,i}(m)\bm{G}^<_{(i,i-\haty;0,m)} + \tilde{J}_{i,i+\haty}(m)\bm{G}^<_{(i+\haty,i;0,m)} \Big]	\Big\}.
\end{split} \label{eq:localcurrent2}
\end{equation}
\end{widetext}
Thus, for the horizontal current, we need the interslab lesser Floquet-Green's function $\bm{\mathcal{G}}^<_{x+1,x}$, and for the vertical current, we need the local lesser function $\bm{\mathcal{G}}^<_{xx} $. From the Dyson equation \eqref{eq:DysonCoupling}, we apply the Langreth rule~\cite{Haug2008} for lesser projection to find:
\begin{equation}
\begin{split}
\bm{\mathcal{G}}^<_{x+1,x}&=\bm{\mathrm{g}}^r_{x+1,x+1}\bm{\mathcal{V}}\bm{\mathcal{G}}^<_{xx}+\bm{\mathrm{g}}^<_{x+1,x+1}\bm{\mathcal{V}}\bm{\mathcal{G}}^a_{xx},\\
\bm{\mathcal{G}}^<_{x-1,x}&=\bm{\mathrm{g}}^r_{x-1,x-1}\bm{\mathcal{V}}\bm{\mathcal{G}}^<_{xx}+\bm{\mathrm{g}}^<_{x-1,x-1}\bm{\mathcal{V}}\bm{\mathcal{G}}^a_{xx}.
\label{eq:horizontalCurrent}
\end{split}
\end{equation}
The task then is to calculate, for each slab $x$, the lesser and advanced Green's function $\bm{\mathcal{G}}^<_{xx},\bm{\mathcal{G}}^a_{xx}$. The latter is given by the Dyson equation, from which one can calculate the former using the Keldysh equation. This is implemented by Algorithm~\ref{algo:localDCprofile}. The core of the algorithm is very similar to that described in Ref.~\cite{Lewenkopf2013} (despite all the Green's functions being promoted to the Floquet version). A difference arises when we calculate the time-averaged LDOS (line 15), the vertical current (line 17) and the horizontal currents (lines 19 and 21) in Algorithm~\ref{algo:localDCprofile}.

\subsection{Time-averaged LDOS}
The calculation of the time-averaged retarded Floquet-Green's function is needed in the calculation of DC profile, hence we shall not repeat it here. If the DC profile is not needed, one simply removes all lesser-related steps in Algorithm~\ref{algo:localDCprofile}. We end this section by writing explicitly again what we mean by the time-averaged LDOS:
\begin{equation}
\begin{split}
& \textrm{T-LDOS}_{(x,y)}(E)\\
&=-\frac{1}{\pi}\mathrm{Im}\left[-\frac{\ii}{\hbar}\theta(t-t')\protect\langle c_{x,y}(t)c^\dag_{x,y}(t')\protect\rangle\right]_{00}(E),
\end{split}
\end{equation}
where $[\cdots]_{00}(E)$ means the Floquet representation Eq.~\eqref{eq:FloquetRepresentation}.

\section{Expressions of local DC of the HDHM}
The expressions of the time-averaged local current, given in Eq.~\eqref{eq:localcurrent} and~\eqref{eq:localcurrent2}, are rather general. As a concrete example, here we give the corresponding expressions in the harmonically-driven Hofstadter model.

\subsubsection{Time-averaged longitudinal current}
The $x$-hopping is undriven in our model, so we have the longitudinal DC (longitudinal with respect to the applied bias):
\begin{equation}
I^\rightarrow_i = 2J_x \mathrm{Re}\Big[\bm{G}^<_{(i,i-\hatx;0,0)}+\bm{G}^<_{(i+\hatx,i;0,0)}	\Big] .
\end{equation}
\subsubsection{Time-averaged transverse current}
With the $y$-hopping driven and modulated over $x$-direction, the transverse current (transverse with respect to the applied bias) needs a bit more work. Given the $y$-hopping term as $J_y\left[s+\cos(\Omega t)\right]\ee^{\ii 2\pi\alpha x_i} c^\dag_ic_{i+\haty}+\Hc$, we have the transverse direct current:
\begin{widetext}
\begin{equation}
\begin{split}
I^{\uparrow}_{i} &= J_y\Bigg\{ s\times 2 \mathrm{Re}\Big[\ee^{\ii 2\pi\alpha x_i}\big(\bm{G}^<_{(i+\haty,i;0,0)}+\bm{G}^<_{(i,i-\haty;0,0)}\big)\Big] \\
&+\frac{1}{2}\Bigg(\ee^{\ii 2\pi\alpha x_i}\Big[\bm{G}^<_{(i+\haty,i;1,0)}+\bm{G}^<_{(i+\haty,i;-1,0)}+\bm{G}^<_{(i,i-\haty;1,0)}+\bm{G}^<_{(i,i-\haty;-1,0)}	\Big]\\&-	\ee^{-\ii 2\pi\alpha x_i}\Big[\bm{G}^<_{(i,i+\haty;1,0)}+\bm{G}^<_{(i,i+\haty;-1,0)}+\bm{G}^<_{(i-\haty,i;1,0)}+\bm{G}^<_{(i-\haty,i;-1,0)}	\Big]		\Bigg)	\Bigg\}.
\end{split}
\end{equation}
\end{widetext}

\section{\label{app:gapped}Supplementary results on co-propagating and gapped edge states}
We consider the parameters $J_y=1.6,\alpha=1/3,\Omega=\pi,$ and $s=0$. The corresponding phase has two pairs of co-propagating edge states in the first gap and gapped edge states in the second gap, as shown in the Floquet spectrum (Fig.~\ref{subfig:gappedbandstruc}).
\subsection{DC conductance}
\begin{figure}
\subfloat[\label{subfig:gappedbandstruc}]{
\includegraphics[height=5cm,width=.5\columnwidth]{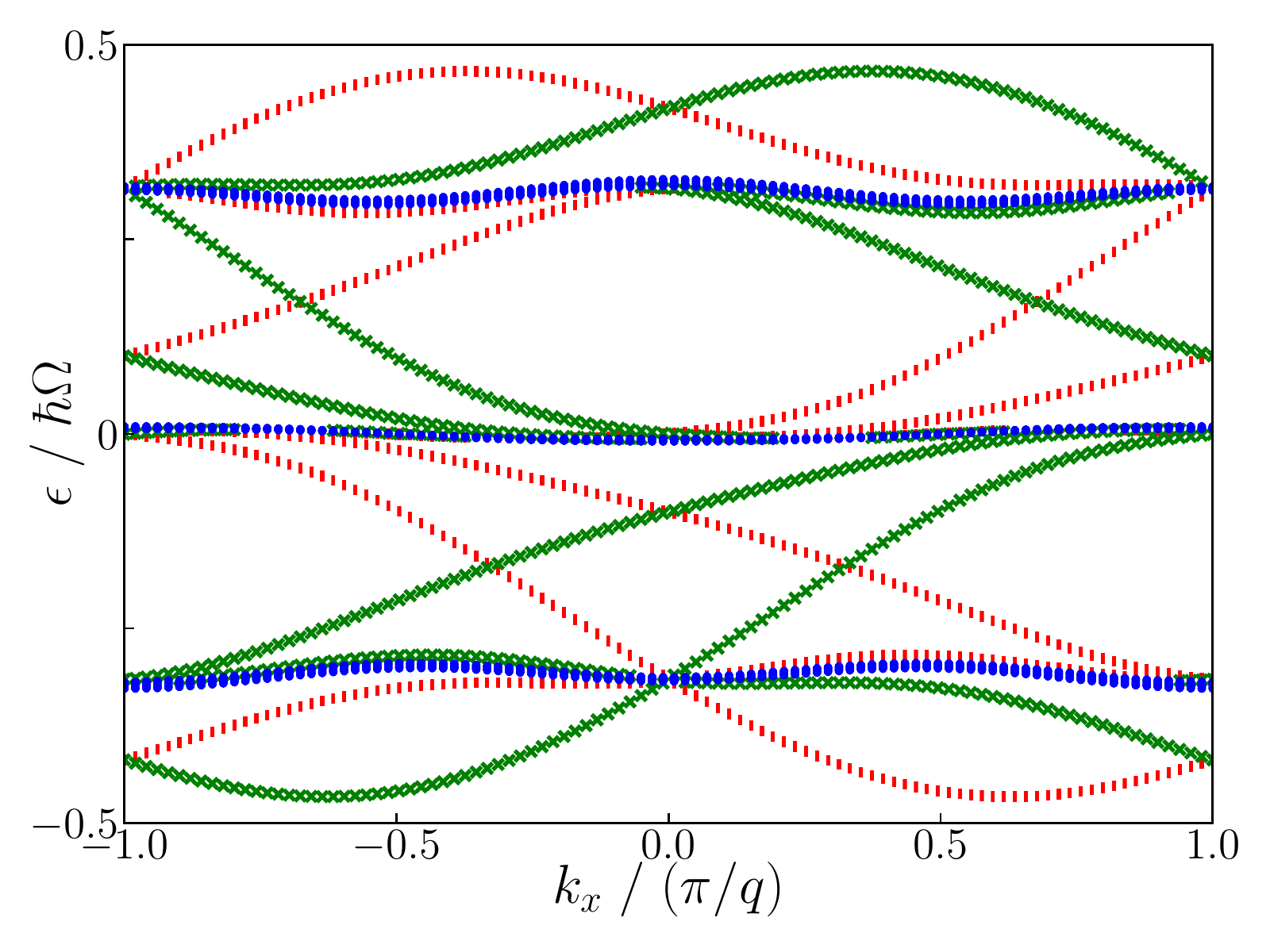}
}
\subfloat[\label{subfig:transmGappedCopropa}]{
\includegraphics[height=5cm,width=.5\columnwidth]{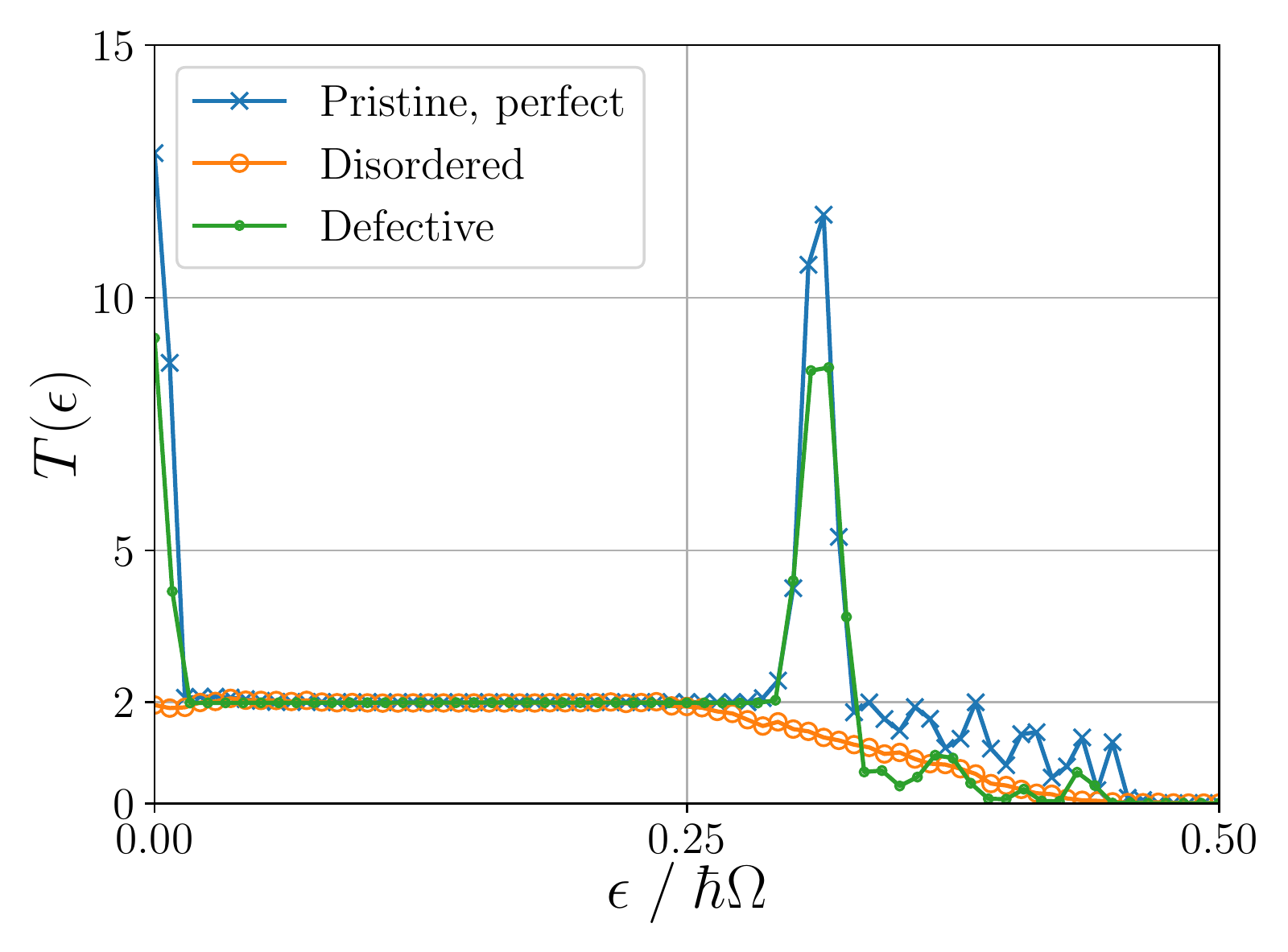}
}
\caption{(color online) (a): Floquet spectrum of the HDHM with parameters $J_y=1.6,\alpha=1/3,\Omega=\pi,$ and $s=0$. showing the existence of gapped edge states on the top and bottom gaps $(q=3)$. (b): DC transmission after the Floquet sum rule. For the disordered curve (cirlces), the disorder strength is $W=1$, penetrating the whole lattice, averaged over {100} realizations. For the defective curve (dots), the defects are introduced identically to the one in Fig.~\ref{subfig:copropaDefectpattern}. System sizes are $N_x=N_y=40,~\nF=7$.}
\end{figure}

We first present the transmission calculation for three cases: (1) perfect, (2) disordered and (3) defective lattices. As shown in Fig.~\ref{subfig:transmGappedCopropa}, be it disorder or defect, the co-propagating (gapless) Floquet edge modes are very robust, whereas the gapped edge states have no quantized conductance even in the perfect, pristine case.

\subsection{Time-averaged LDOS}
We probe the time-averaged local density of states at quasienergy $\epsilon=1.4$ which lies in the second Floquet gap and thus corresponds to gapped edge states.
\begin{figure}
\subfloat[\label{subfig:gappedTLDOS}]{
\includegraphics[height=5cm,width=.5\columnwidth]{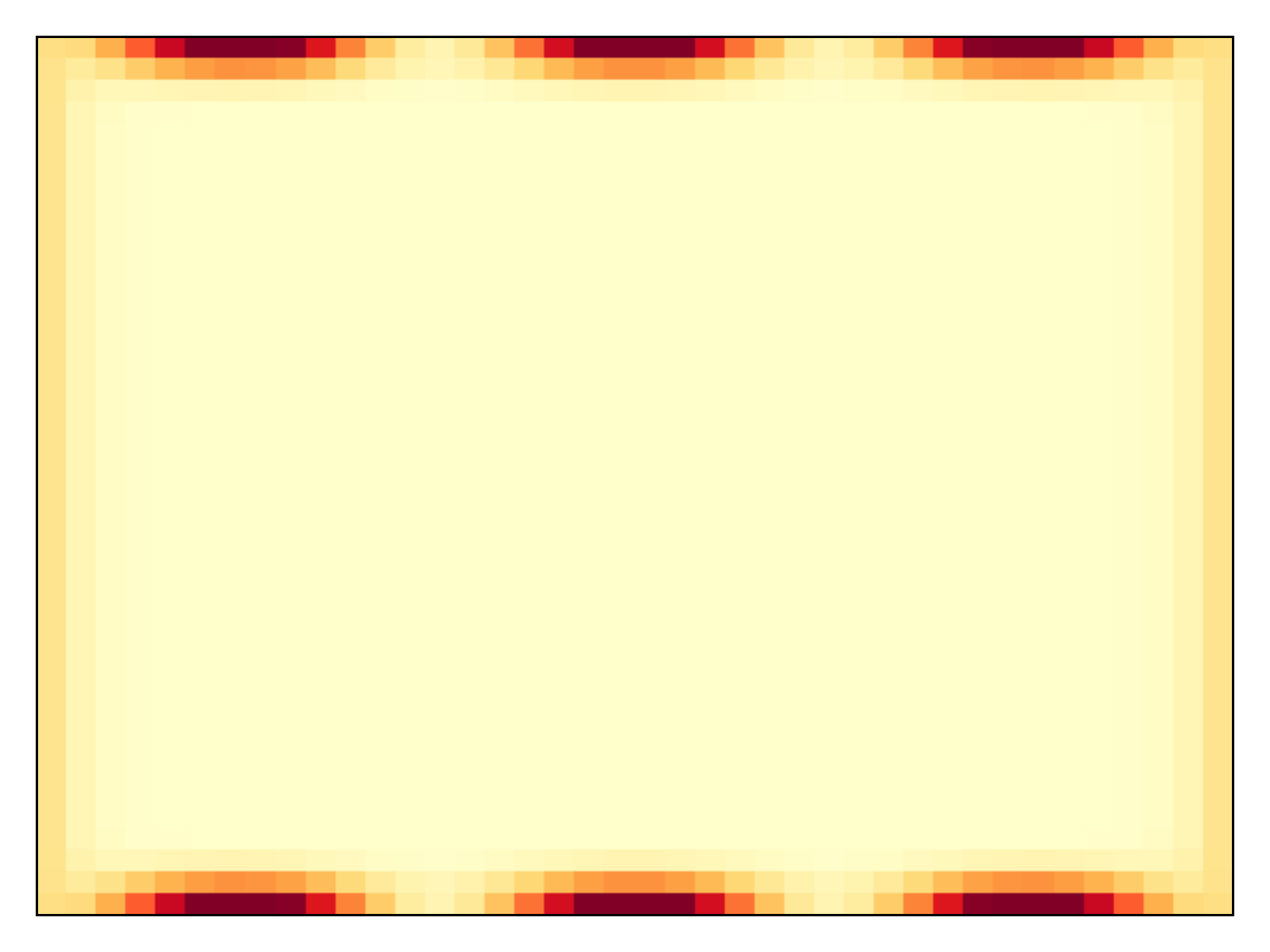}
}
\subfloat[\label{subfig:gappedDisorderedTLDOS}]{
\includegraphics[height=5cm,width=.5\columnwidth]{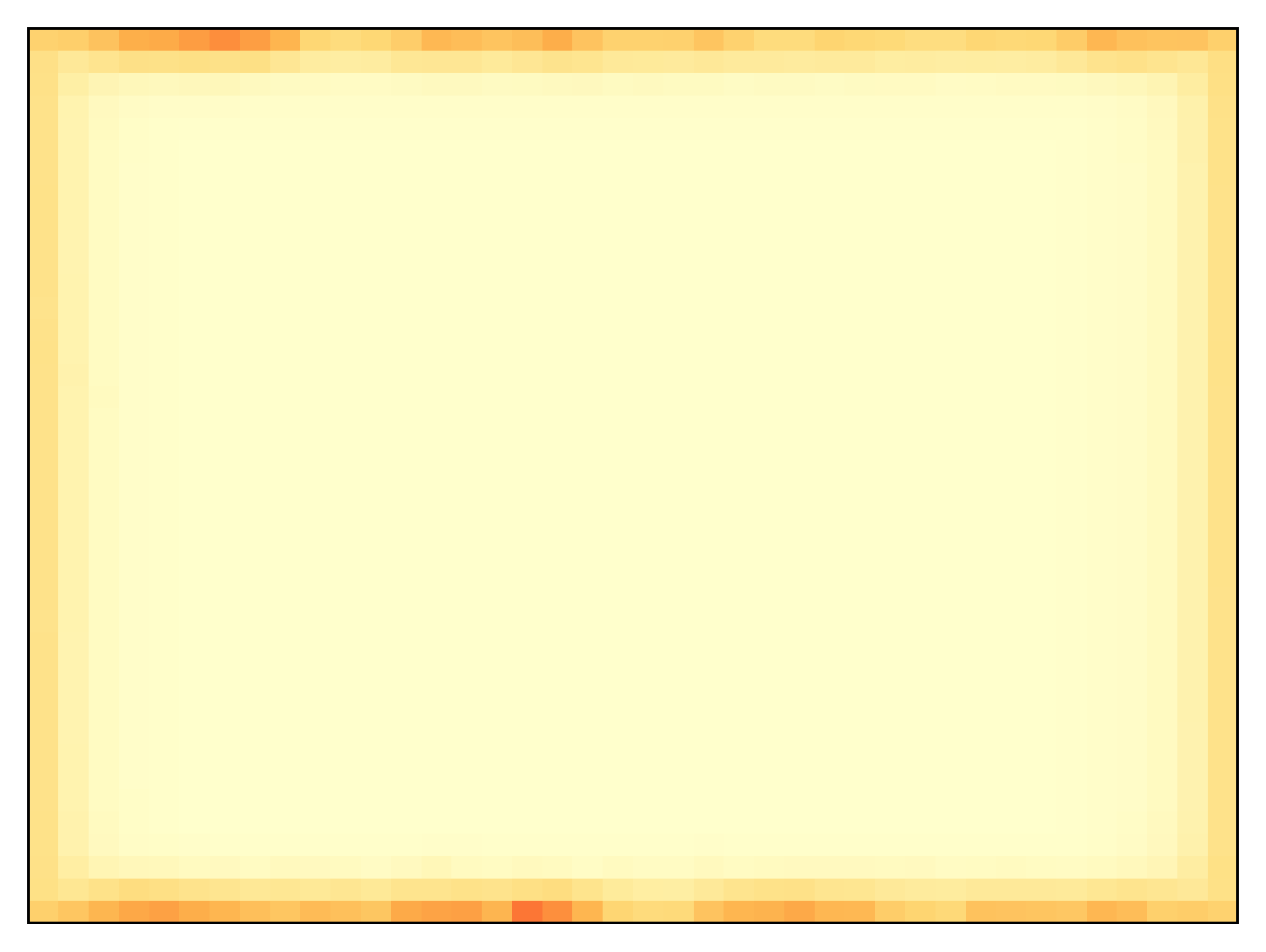}
}
\\
\centering
\subfloat{
\includegraphics[width=.35\columnwidth]{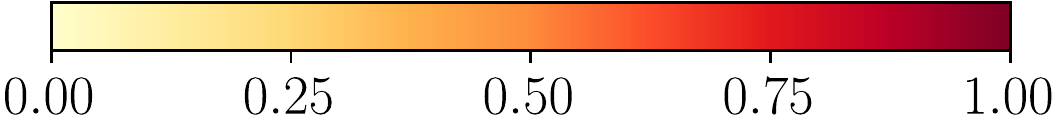}
}
\caption{(color online) Time-averaged LDOS at quasienergy $\epsilon=1.4$ corresponding to a gapped edge state. (a): Pristine. (b): Edge disorder of strength $W=0.5$ penetrate three layers on all four edges. System sizes are $N_x=N_y=40,~\nF=7$. Result shown is averaged over {$200$} disorder realizations.}
\end{figure}
Consistent with the Floquet spectrum (Fig.~\ref{subfig:gappedbandstruc}), at $\epsilon=1.4$, the states are localized on edges, albeit not as uniformly as their gapless counterparts in Fig.~\ref{fig:DOSdisord}. Also, the presence of edge disorder destroys this ``edge islands'' pattern (Fig.~\ref{subfig:gappedDisorderedTLDOS}).

\subsection{Local DC profile}
Here we consider the local current pattern of the gapped and co-propagating edge states, as well as their responses to defects.
\begin{figure}
\subfloat[\label{subfig:gappedPerfectpattern}]{
\includegraphics[height=5cm,width=.5\columnwidth]{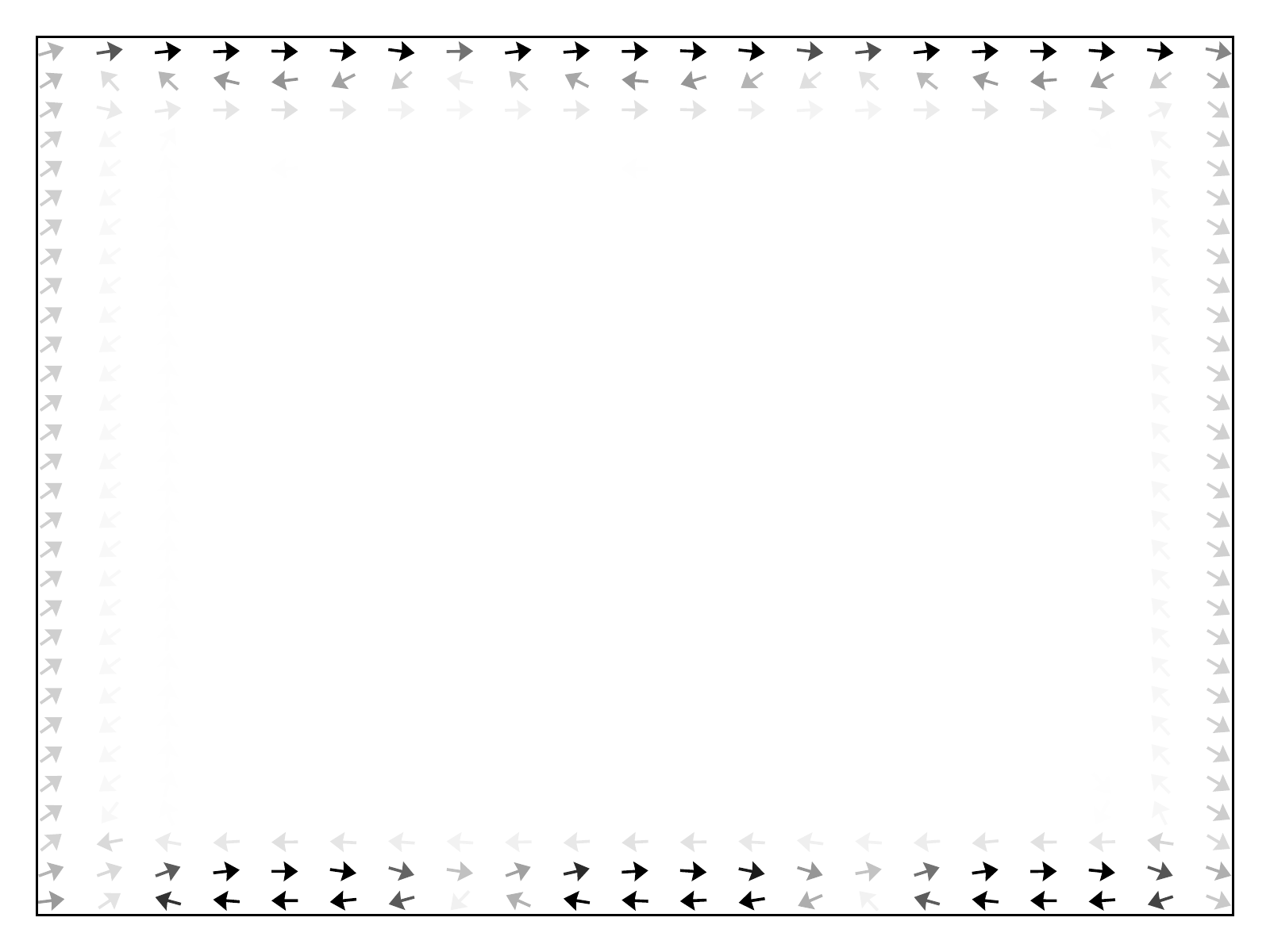}
}
\subfloat[\label{subfig:gappedDefectpattern}]{
\includegraphics[height=5cm,width=.5\columnwidth]{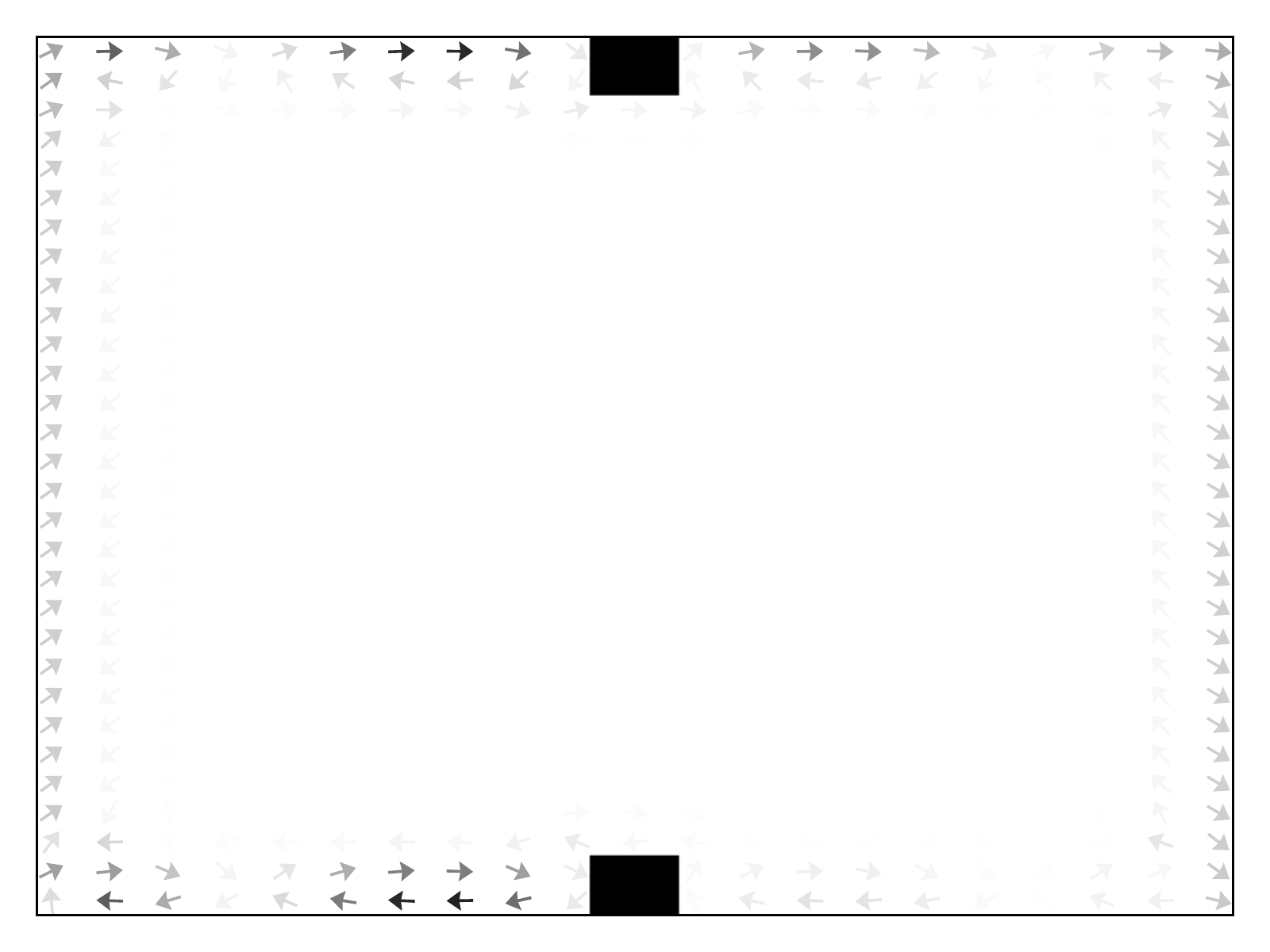}
}
\\
\subfloat[\label{subfig:copropaPerfectpattern}]{
\includegraphics[height=5cm,width=.5\columnwidth]{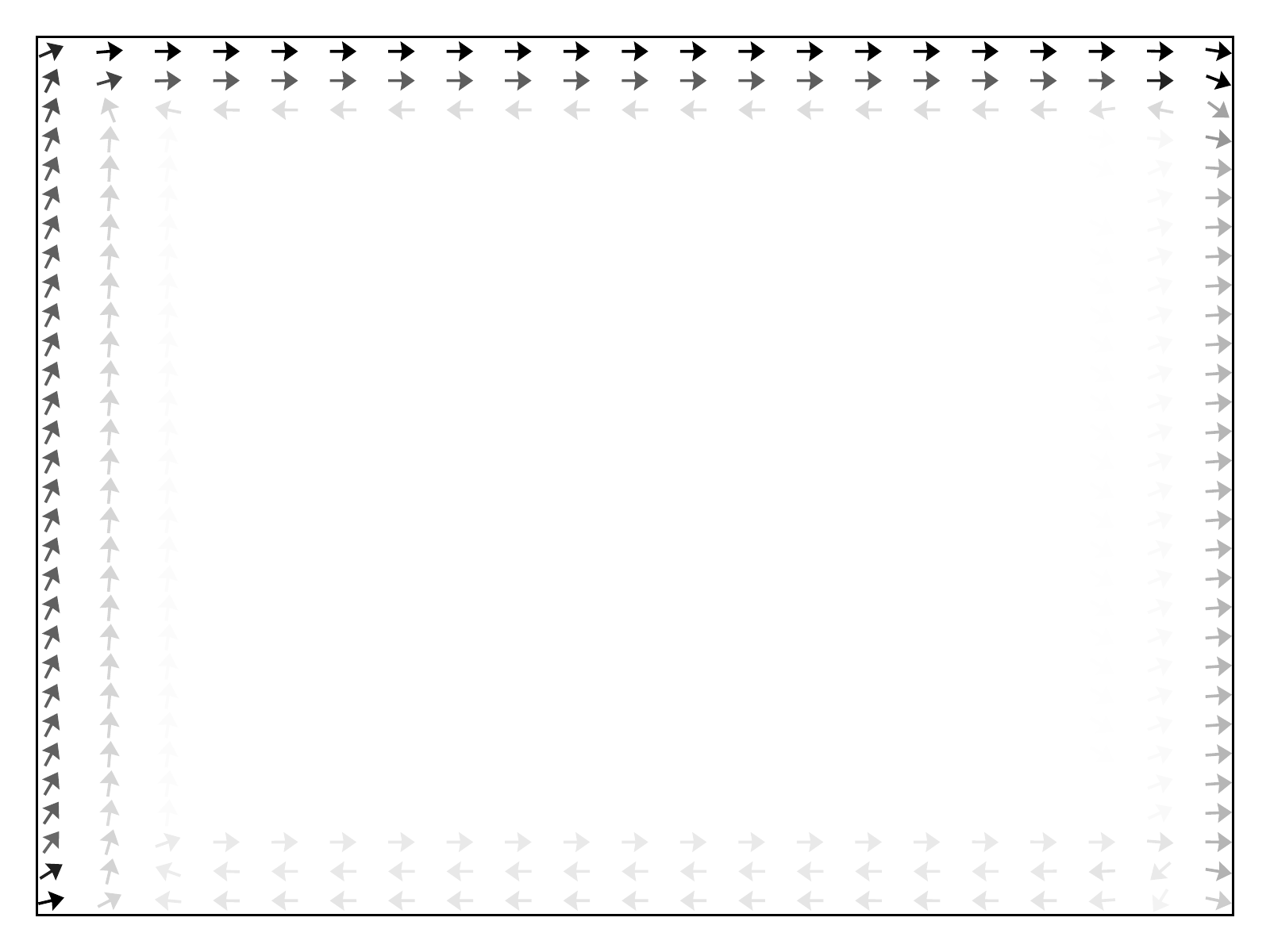}
}
\subfloat[\label{subfig:copropaDefectpattern}]{
\includegraphics[height=5cm,width=.5\columnwidth]{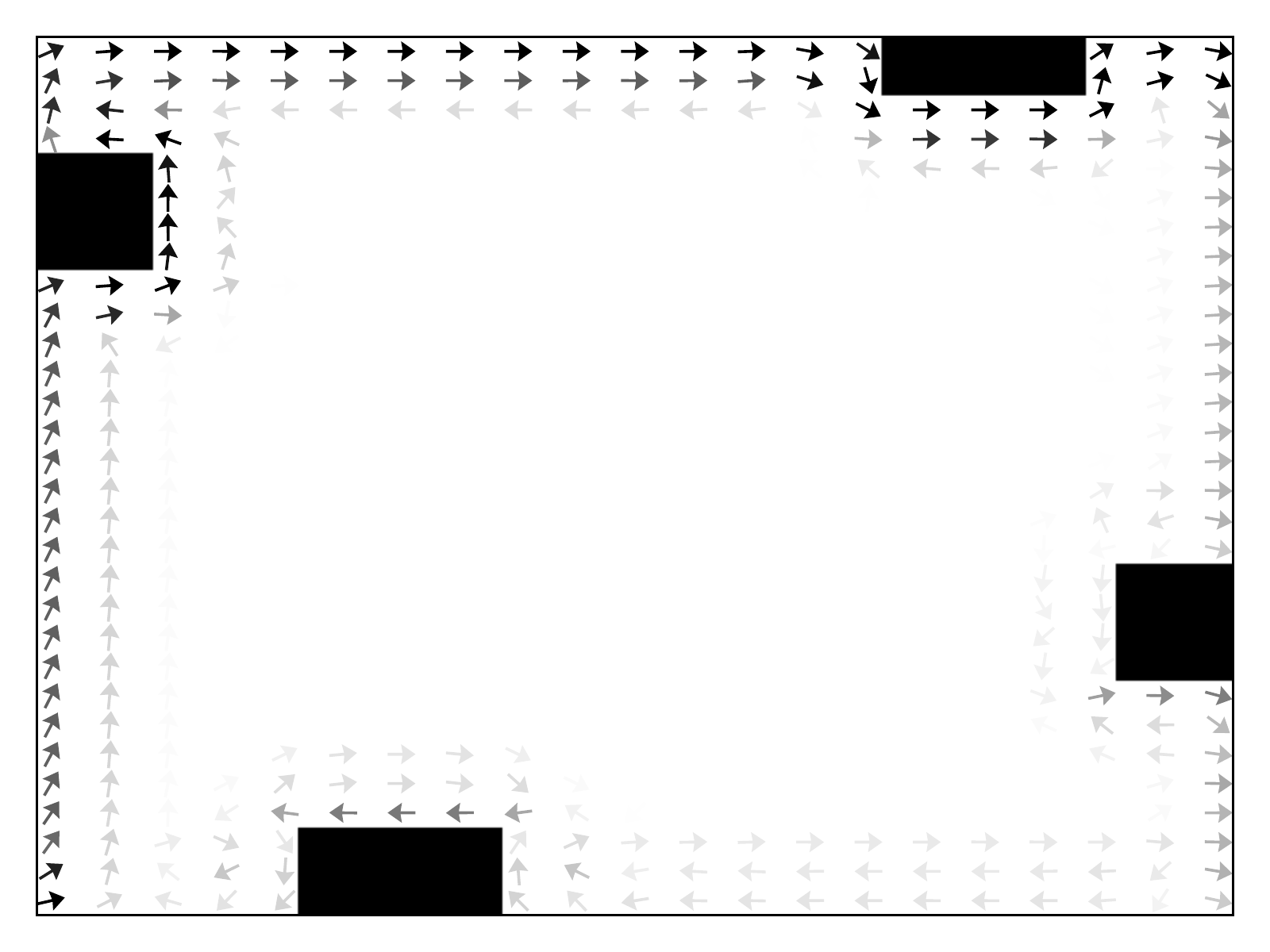}
}
\\
\centering
\subfloat{
\includegraphics[width=.35\columnwidth]{Colorbar.pdf}
}
\caption{DC profile of gapped and co-propagating edge states. Top panels: gapped edge states, with $E=1.4,\mu_L=1.41$, and $\mu_R=1.39$. Bottom panels: co-propagating edge states, with $E=0.5,\mu_L=0.51$, and $\mu_R=0.49$. Panels (a),~(c): perfect lattice. Panels (b),~(d): defective lattices. The Floquet sum rule is applied in all cases. Some arrows are skipped for better display. System sizes: $N_x=41,~N_y=30,~\nF=7$. } \label{fig:TLDOSgapped}
\end{figure}
Unlike their gapless counter-propagating analogs in Fig.~\ref{fig:DCprofileCounterpropa}, the edge current profile is not as evident for gapped edge states (Fig.~\ref{subfig:gappedPerfectpattern}). Upon introduction of defects, the already-weak current gets diminished further. Finally, the chiral feature of the co-propagating (gapped) edge states is evident from Fig.~\ref{subfig:copropaPerfectpattern}, with their topological protection from backscattering exemplified in Fig.~\ref{subfig:copropaDefectpattern}.


\begin{thebibliography}{99}
	\bibitem{Leboeuf1990} P. Leboeuf, J. Kurchan, M. Feingold, and D. P. Arovas, Phys. Rev. Lett. {\bf 65}, 3076 (1990).
	\bibitem{Oka2009} T. Oka and H. Aoki, Phys. Rev. B {\bf 79}, 081406 (2009).
	\bibitem{Kitagawa2011} T. Kitagawa, T. Oka, A. Brataas, L. Fu, and E. Demler, Phys. Rev. B {\bf 84}, 235108 (2011).
	\bibitem{Lindner2011} N. H. Lindner, G. Refael, and V. Galitski, Nat. Phys. {\bf 7}, 490 (2011).
	\bibitem{GongPRL2012} D. Y. H. Ho and J. Gong, Phys. Rev. Lett. {\bf 109}, 010601 (2012).
	\bibitem{Gomezleon2013} {\'A}. G{\'o}mez-Le{\'o}n and G. Platero, Phys. Rev. Lett. {\bf 110}, 200403 (2013).
	\bibitem{Rudner2013} M. S. Rudner, N. H. Lindner, E. Berg, and M. Levin, Phys. Rev. X {\bf 3}, 031005 (2013).
	\bibitem{Cayssol2013} J. Cayssol, B. D{\'o}ra, F. Simon, and R. Moessner, Phys. Status Solidi Rapid Res. Lett. {\bf 7}, 101 (2013).
	\bibitem{Titum2015} P. Titum, N. H. Lindner, M. C. Rechtsman, and G. Refael, Phys. Rev. Lett. {\bf 114}, 056801 (2015).
	\bibitem{Titum2016} P. Titum, E. Berg, M. S. Rudner, G. Refael, and N. H. Lindner, Phys. Rev. X {\bf 6}, 021013 (2016).
	\bibitem{Jiang2011} L. Jiang, T. Kitagawa, J. Alicea, A. R. Akhmerov, D. Pekker, G. Refael, J. I. Cirac, E. Demler, M. D. Lukin, and P. Zoller, Phys. Rev. Lett. {\bf 106}, 220402 (2011).
	\bibitem{Kundu2013} A. Kundu and B. Seradjeh, Phys. Rev. Lett. {\bf 111}, 136402 (2013).
	\bibitem{Gong2013} Q.-J. Tong, J.-H. An, J. Gong, H.-G. Luo, and C. H. Oh, Phys. Rev. B {\bf 87}, 201109 (2013).
	\bibitem{Klinovaja2016} J. Klinovaja, P. Stano, and D. Loss, Phys. Rev. Lett. {\bf 116}, 176401 (2016).
	\bibitem{Thakurathi2017} M. Thakurathi, D. Loss, and J. Klinovaja, Phys. Rev. B {\bf 95}, 155407 (2017).
	\bibitem{WangFWSM2014} R. Wang, B. Wang, R. Shen, L. Sheng, and D. Y. Xing, EPL {\bf 105}, 17004 (2014).
	\bibitem{Raditya2016} R. W. Bomantara, G. N. Raghava, L. Zhou, and J. Gong, Phys. Rev. E {\bf 93}, 022209 (2016).
	\bibitem{Zhou2016} L. Zhou, C. Chen, and J. Gong, Phys. Rev. B {\bf 94}, 075443 (2016).
	\bibitem{Raditya2016b} R. W. Bomantara and J. Gong, Phys. Rev. B {\bf 94}, 235447 (2016).
	\bibitem{Harper2016a} R. Roy and F. Harper, Phys. Rev. B {\bf 94}, 125105 (2016).
	\bibitem{Harper2016b} R. Roy and F. Harper, arXiv:1603.06944.
	\bibitem{Jotzu2014} G. Jotzu, M. Messer, R. Desbuquois, M. Lebrat, T. Uehlinger, D. Greif, and T. Esslinger, Nature {\bf 515}, 237 (2014).
	\bibitem{Wang453} Y. H. Wang, H. Steinberg, P. Jarillo-Herrero, and N. Gedik, Science {\bf 342}, 453 (2013).
	\bibitem{Rechtsman2013} M. C. Rechtsman, J. M. Zeuner, Y. Plotnik, Y. Lumer, D. Podolsky, F. Dreisow, S. Nolte, M. Segev, and A. Szameit, Nature {\bf 496}, 196 (2013).
	\bibitem{Hu2015} W. Hu, J. C. Pillay, K. Wu, M. Pasek, P. P. Shum, and Y. D. Chong, Phys. Rev. X {\bf 5}, 011012 (2015).
	\bibitem{Maczewsky2017} L. J. Maczewsky, J. M. Zeuner, S. Nolte, and A. Szameit, Nat. Commun. {\bf 8}, 13756 (2017).
	\bibitem{Mukherjee2017} S. Mukherjee, A. Spracklen, M. Valiente, E. Andersson, P. Ohberg, N. Goldman, and R. R. Thomson, Nat. Commun. {\bf 8}, 13918 (2017).
	\bibitem{Xiao2015} M. Xiao, G. Ma, Z. Yang, P. Sheng, Z. Q. Zhang, and C. T. Chan, Nat. Phys. {\bf 11}, 240 (2015).
	\bibitem{Susstrunk2015} R. S{\"u}sstrunk and S. D. Huber, Science {\bf 349}, 47 (2015).
	\bibitem{Susstrunk2017} R. S{\"u}sstrunk, P. Zimmermann, and S. D. Huber, New J. Phys. {\bf 19}, 015013 (2017).
	\bibitem{Peano2015} V. Peano, C. Brendel, M. Schmidt, and F. Marquardt, Phys. Rev. X {\bf 5}, 031011 (2015).
	\bibitem{Derek2014} D. Y. H. Ho and J. Gong, Phys. Rev. B {\bf 90}, 195419 (2014).
	\bibitem{Zhou2014} L. Zhou, H. Wang, D. Y. H. Ho, and J. Gong, Eur. Phys. J. B {\bf 87}, 204 (2014).
	\bibitem{Xiong2016} T.-S. Xiong, J. Gong, and J.-H. An, Phys. Rev. B {\bf 93}, 184306 (2016).
	\bibitem{Buttiker1988} M. B\"uttiker, Phys. Rev. B {\bf 38}, 9375 (1988).
	\bibitem{Datta1995} S. Datta, {\it Electronic Transport in Mesoscopic Systems} (Cambridge University Press, UK, 1995).
	\bibitem{Tien1963} P. K. Tien and J. P. Gordon, Phys. Rev. {\bf 129}, 647 (1963).
	\bibitem{Platero2004} G. Platero and R. Aguado, Phys. Rep. {\bf 395}, 1 (2004).
	\bibitem{Kohler2005} S. Kohler, J. Lehmann, and P. H\"anggi, Phys. Rep. {\bf 406}, 379 (2005).
	\bibitem{Jauho1994} A.-P. Jauho, N. S. Wingreen, and Y. Meir, Phys. Rev. B {\bf 50}, 5528 (1994).
	\bibitem{Arrachea2005} L. Arrachea, Phys. Rev. B {\bf 72}, 125349 (2005).
	\bibitem{Tsuji2008} N. Tsuji, T. Oka, and H. Aoki, Phys. Rev. B {\bf 78}, 235124 (2008).
	\bibitem{Bruder1994} C. Bruder and H. Schoeller, Phys. Rev. Lett. {\bf 72}, 1076 (1994).
	\bibitem{Moskalets2002} M. Moskalets and M. B\"uttiker, Phys. Rev. B {\bf 66}, 205320 (2002).
	\bibitem{Kim2004} S. W. Kim, Int. J. Mod. Phys. B {\bf 18}, 3071 (2004).
	\bibitem{Arrachea2006} L. Arrachea and M. Moskalets, Phys. Rev. B {\bf 74}, 245322 (2006).
	\bibitem{Gu2011} Z. Gu, H. A. Fertig, D. P. Arovas, and A. Auerbach, Phys. Rev. Lett. {\bf 107}, 216601 (2011).
	\bibitem{Foatorres2014} L. E. F. Foa Torres, P. M. Perez-Piskunow, C. A. Balseiro, and G. Usaj, Phys. Rev. Lett. {\bf 113}, 266801 (2014).
	\bibitem{Kundu2014} A. Kundu, H. A. Fertig, and B. Seradjeh, Phys. Rev. Lett. {\bf 113}, 236803 (2014).
	\bibitem{Farrell2015} A. Farrell and T. Pereg-Barnea, Phys. Rev. Lett. {\bf 115}, 106403 (2015).
	\bibitem{Farrell2016} A. Farrell and T. Pereg-Barnea, Phys. Rev. B {\bf 93}, 045121 (2016).
	\bibitem{Fruchart2016} M. Fruchart, P. Delplace, J. Weston, X. Waintal, and D. Carpentier, Physica E Low Dimens. Syst. Nanostruct.	{\bf 75}, 287 (2016).
	\bibitem{Fulga2016} I. C. Fulga and M. Maksymenko, Phys. Rev. B {\bf 93}, 075405 (2016).
	\bibitem{Dehghani2015a} H. Dehghani, T. Oka, and A. Mitra, Phys. Rev. B {\bf 91}, 155422 (2015).
	\bibitem{Dehghani2015b} H. Dehghani and A. Mitra, Phys. Rev. B {\bf 92}, 165111 (2015).
	\bibitem{Dehghani2016} H. Dehghani and A. Mitra, Phys. Rev. B {\bf 93}, 205437 (2016).
	\bibitem{AokiRMP2014} H. Aoki, N. Tsuji, M. Eckstein, M. Kollar, T. Oka, and P. Werner, Rev. Mod. Phys. {\bf 86}, 779 (2014).
	\bibitem{GomezLeon_Keldysh} L. Ruocco and \'A. G\'omez-Le\'on, Phys. Rev. B {\bf 95}, 064302 (2017).
	\bibitem{Haug2008} H. Haug and A.-P. Jauho, {\it Quantum kinetics in transport and optics of semiconductors} (Springer, New York, 2008).
	\bibitem{Wang2008} J.-S. Wang, J. Wang, and J. T. L\"u, Eur. Phys. J. B {\bf 62}, 381 (2008).
	\bibitem{Jishi2014} R. A. Jishi, {\it Feynman Diagram Techniques in Condensed Matter Physics} (Cambridge University Press, UK, 2014).
	\bibitem{Wang2014} J.-S. Wang, B. K. Agarwalla, H. Li, and J. Thingna, Front. Phys. {\bf 9}, 673 (2014).
	\bibitem{Cuevas2010} J. C. Cuevas and E. Scheer, {\it Molecular Electronics: An Introduction to Theory and Experiment} (World Scientific, Singapore, 2010).
	\bibitem{footnote1} This equation (continuity equation for local particle number) remains valid on the edges of the central system, with the understanding that, on the top and bottom edges, Dirichlet boundary condition is used, e.g.: $J_{i,i- \haty}=0$ if $y_i=0$, whereas on the left and right edges, scattering boundary condition is used, e.g.: $J_{i,i-\hatx}=t_x$ (the left-center coupling) and $G^<_{i,i-\hatx}=-\frac{\ii}{\hbar}\protect\langle a^\dag_{i-\hatx}c_i \protect\rangle$ if $x_i=0$, where we recall that $a^\dag$ is the creation operator of the left lead.
	\bibitem{Cresti2003} A. Cresti, R. Farchioni, G. Grosso, and G. P. Parravicini, Phys. Rev. B {\bf 68}, 075306 (2003).
	\bibitem{Waintal2008} K. Kazymyrenko and X. Waintal, Phys. Rev. B {\bf 77}, 115119 (2008).
	\bibitem{Lewenkopf2013} C. H. Lewenkopf and E. R. Mucciolo, J. Comput. Electron. {\bf 12}, 203 (2013).
	\bibitem{Mikami2016} T. Mikami, S. Kitamura, K. Yasuda, N. Tsuji, T. Oka, and H. Aoki, Phys. Rev. B {\bf 93}, 144307 (2016).
	\bibitem{Ryndyk2015} D. Ryndyk, {\it Theory of Quantum Transport at Nanoscale: An Introduction} (Springer, New York, 2015).
	\bibitem{Zhao2014} Z. Zhou, I. I. Satija, and E. Zhao, Phys. Rev. B {\bf 90}, 205108 (2014).
	\bibitem{Kolovsky2012} A. R. Kolovsky and G. Mantica, Phys. Rev. B {\bf 86}, 054306 (2012).
	\bibitem{Shirley1965} J. H. Shirley, Phys. Rev. {\bf 138}, B979 (1965).
	\bibitem{Sambe1973} H. Sambe, Phys. Rev. A {\bf 7}, 2203 (1973).
	\bibitem{GomezLeon_counter} \'A. G\'omez-Le\'on, P. Delplace, and G. Platero, Phys. Rev. B {\bf 89}, 205408 (2014).
	\bibitem{Lewenkopf2017} M. M. Odashima and C. H. Lewenkopf, Phys. Rev. B {\bf 95}, 104301 (2017).
	\bibitem{Velicky2010} B. Velick{\'y}, A. Kalvov{\'a}, and V. $\check{{\rm S}}$pi$\check{{\rm c}}$ka, Physica E Low Dimens. Syst. Nanostruct. {\bf 42}, 539 (2010).
	\bibitem{Stefanucci2004} G. Stefanucci and C.-O. Almbladh, Phys. Rev. B {\bf 69}, 195318 (2004).
	\bibitem{footnote2} Counter-propagating edge states may also exist in static systems, e.g. in quantum spin Hall effect (which is spin-degenerate). However, here our system breaks time reversal symmetry and consists of spinless electrons. In absence of driving, such systems are not expected to host counter-propagating edge modes.
	\bibitem{Klitzing1980} K. v. Klitzing, G. Dorda, and M. Pepper, Phys. Rev. Lett. {\bf 45}, 494 (1980).
	\bibitem{Zhang2010} S.-C. Zhang and X.-L. Qi, Phys. Today {\bf 63}, 33 (2010).
	\bibitem{Lababidi2014} M. Lababidi, I. I. Satija, and E. Zhao, Phys. Rev. Lett. {\bf 112}, 026805 (2014).
	\bibitem{Fujita1996} M. Fujita, K. Wakabayashi, K. Nakada, and K. Kusakabe, J. Phys. Soc. Jpn. {\bf 65}, 1920 (1996).
	\bibitem{Fleury2016} R. Fleury, A. B. Khanikaev, and A. Al{\`u}, Nat. Commun. {\bf 7}, 11744 (2016).
	\bibitem{Bandres2016} M. A. Bandres, M. C. Rechtsman, and M. Segev, Phys. Rev. X {\bf 6}, 011016 (2016).
	\bibitem{Caroli1971} C. Caroli, R. Combescot, D. Lederer, P. Nozieres, and D. Saint-James, J. Phys. C: Solid State Phys. {\bf 4}, 916 (1971).
	\bibitem{Khomyakov2005} P. A. Khomyakov, G. Brocks, V. Karpan, M. Zwierzycki, and P. J. Kelly, Phys. Rev. B {\bf 72}, 035450 (2005).
\end{thebibliography}

\end{document}